\documentclass[a4paper,11pt]{article}
\pdfoutput = 1
\usepackage[no-natbib-sort]{jheppub}
\usepackage{amsmath,amssymb,amscd,braket,amsfonts}
\usepackage{color}
\usepackage{graphicx}
\usepackage{subcaption}
\usepackage{slashed}
\usepackage{comment}
\usepackage{amsthm}

\usepackage[normalem]{ulem}
\usepackage{tikz}
\usetikzlibrary{decorations.markings}
\usetikzlibrary{shapes.geometric}
\usetikzlibrary{arrows.meta,positioning,calc}
\usetikzlibrary{decorations.pathmorphing}

\usepackage[most]{tcolorbox}
\newtcolorbox{keyformula}[1][]{
  enhanced,
  breakable,
  colback=gray!4,
  colframe=black!55,
  boxrule=0.5pt,
  arc=1.5mm,
  left=2mm,
  right=2mm,
  top=1mm,
  bottom=1mm,
  fonttitle=\bfseries,
  title={#1},
  before={\par\addvspace{\abovedisplayskip}\noindent},
  after={\par\addvspace{\belowdisplayskip}\noindent\ignorespaces}
}

\usepackage{amstext} 
\usepackage{array}   
\newcolumntype{L}{>{$}l<{$}} 

\usepackage{lettrine, Romantik}

\newcommand{\bea}{\begin{eqnarray}}
\newcommand{\eea}{\end{eqnarray}}
\newcommand{\be}{\begin{equation}}
\newcommand{\ee}{\end{equation}}
\newcommand{\ba}{\begin{align}}
\newcommand{\ea}{\end{align}}

\title{OPE = QNM}

\author[a]{Paolo Arnaudo,}
\emailAdd{p.arnaudo@soton.ac.uk}
\author[b]{Cristoforo Iossa,}
\emailAdd{cristoforo.iossa@unige.ch}
\author[c]{Robin Karlsson,}
\emailAdd{robin.karlsson@maths.ox.ac.uk}
\author[a]{Benjamin Withers}
\emailAdd{b.s.withers@soton.ac.uk}

\affiliation[a]{Mathematical Sciences and STAG Research Centre, University of Southampton, Highfield, Southampton SO17 1BJ, UK}
\affiliation[b]{Section de Math\'ematiques, Universit\'e de Gen\`eve, 1211 Gen\`eve 4, Switzerland}
\affiliation[c]{Mathematical Institute, University of Oxford, Andrew Wiles Building, Radcliffe Observatory
Quarter, Woodstock Road, Oxford, OX2 6GG, UK
}

\abstract{
At microscopic scales the linear response of thermal states of large-$N$ CFTs is governed by a thermal operator product expansion (OPE), while at large scales response is governed by collective excitations known as quasinormal modes (QNM).
We show that the OPE and QNM representations of the retarded correlator in mixed time and spatial momentum coordinates have an overlapping region of convergence in the complex time plane, giving a map between OPE and QNM data.
We show that large-overtone QNM asymptotics are related to OPE singularities, while low-overtone QNM data appear in analytic continuation from short to large times. 
Using this approach we obtain new analytic results for QNM asymptotics, and numerically obtain low-overtone QNMs from OPE data for the Schwarzschild-AdS$_5$ black brane. We further show that the OPE spectrum is intimately related to QNM data through a set of sum rules which we derive in Mellin space.
Finally, using the lightcone OPE, we argue that stress tensor correlators at large spatial momentum thermalise slower as the conformal collider bounds approach saturation. This work points to a new thermal bootstrap programme where OPE and QNM data constrain each other. 
}

\begin{document}
\maketitle

\section{Introduction}

\lettrine{L}{inear response} provides a surgical window into equilibrium states of QFTs at finite temperature, where correlation functions are determined by data intrinsic to the system in question. On one hand this data can be \emph{microscopic}, governed by UV behaviour near coincident points in the form of a thermal operator product expansion (OPE). On the other hand it can be \emph{macroscopic}, in the form of emergent collective phenomena in the IR corresponding to momentum-space poles of correlation functions.

In this work we show how these two classes of data are unified for correlation functions of large-$N$ CFTs at finite temperature. From OPE data in the UV we will recover IR data for collective modes, and vice versa, through analyticity. Indeed, as we will see, due to an overlapping region of convergence, the defining set of data for the OPE is in one-to-one correspondence with the defining set of data for collective excitations, which we will generically refer to as quasinormal modes (QNMs), aligning with the usual notion of QNMs in the case of holographic QFTs. This invites the slogan `OPE = QNM'. 

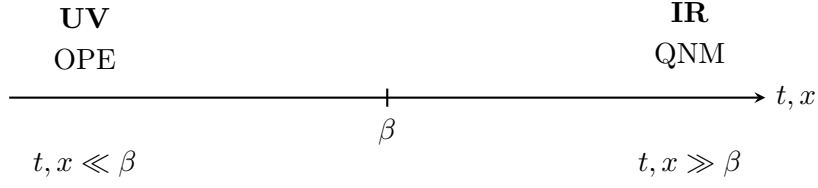
\begin{figure}[h!]
    \centering
    \begin{tikzpicture}[>=stealth]
    \draw[->, thick] (0,0) -- (10,0) node[right] {$t,x$}; 
    \draw[thick] (5,-0.12) -- (5,0.12);
    \node[below=4pt] at (5,0) {$\beta$}; 
    \node[above=7pt, align=center] at (1.0,0) {\textbf{UV}\\[-1pt] OPE}; 
    \node[above=7pt, align=center] at (9.0,0) {\textbf{IR}\\[-1pt] QNM}; 
    \node[below=16pt] at (1.0,0) {$t,x \ll \beta$}; 
    \node[below=16pt] at (9.0,0) {$t,x \gg \beta$}; 
    \end{tikzpicture}
   \caption{At finite temperature $T=\beta^{-1}$, CFT two-point functions are no longer completely determined by conformal symmetry, and exhibit two distinct organisations in the UV and IR. This work relates the defining data in each region to one another.}
    \label{fig:timeline}
\end{figure}

The conformal bootstrap has been a tremendously successful framework to explore CFTs at zero temperature, see e.g.\ \cite{Poland:2018epd}\, for a review. In such cases two- and three-point functions are entirely fixed by conformal symmetry, but become highly nontrivial at four points. It is therefore interesting to extend the bootstrap philosophy to the study of thermal correlators, where already two-point functions are highly nontrivial. There has been various work in this direction, see e.g.\ \cite{El-Showk:2011yvt, Petkou:2018ynm, Iliesiu:2018fao,Iliesiu:2018zlz, Alday:2020eua, Barrat:2025nvu,Buric:2025anb,Buric:2025fye, Niarchos:2025cdg}, but so far finite-temperature bootstrap remains challenging compared to its zero-temperature counterpart. This is largely due to a lack of positivity in thermal OPE coefficients, preventing a direct transfer of zero-temperature approaches to finite temperature (see however \cite{Iliesiu:2018zlz, Barrat:2025wbi} for examples where this has been overcome). Instead, in this work, we aim to make use not only of the OPE as the UV organising principle (for timescales less than $\beta$), but also include IR behaviour (for timescales greater than $\beta$), see Figure \ref{fig:timeline}. Indeed, the IR is where thermal correlators are often studied, in particular, in terms of their late-time exponential decay in the form of QNMs. Our hope is that including this ingredient into the bootstrap programme will enable new progress. Other significant work in this direction include recent work \cite{Dodelson:2024atp,Dodelson:2025rng,Dodelson:2026gak} which obtained QNMs from a small time expansion in the SYK model.

The UV data is defined through the Euclidean thermal two-point function, which admits the following thermal OPE \cite{El-Showk:2011yvt, Iliesiu:2018fao},
\begin{equation}
G_E(\tau,\vec{x})=\frac{1}{(\tau^2+|\vec{x}|^2)^{\Delta_{\mathcal{O}}}}\sum_{(\Delta,J)}a_{\Delta,J}\beta^{-\Delta}C_J^{\left(\frac{d-2}{2}\right)}\left(\frac{\tau}{\sqrt{\tau^2+|\vec{x}|^2}}\right)(\tau^2+|\vec{x}|^2)^{\frac{\Delta}{2}}, \label{introUV}
\end{equation}
where $C_J^{\left(\frac{d-2}{2}\right)}$ are Gegenbauer polynomials. In general the sum \eqref{introUV} converges up to $\sqrt{\tau^2+|\vec{x}|^2} < \beta$. Here the relevant data are the spectrum of exchanged operators appearing in the sum over dimension $\Delta$ and spin $J$, and the coefficient $a_{\Delta,J}$, denoting the product of the zero-temperature OPE coefficient and the thermal one-point function of the exchanged operator. The IR data may be naturally defined in the late-time QNM ringdown of the retarded thermal correlator at fixed spatial momentum $\vec{k}$,
\be
G_R(t, \vec{k}) = \theta(t)\sum_{n=0}^\infty r_n(k) e^{-i \omega_n(k)t},\label{introIR}
\ee
where $k = |\vec{k}|$.
Here the relevant data are the $r_n, \omega_n$ corresponding to $- i\, \times$ residues and frequencies of poles of the frequency space retarded correlator.\footnote{This assumes simple poles, which is the generic behaviour for complex modes in holographic theories. In cases with higher-order poles there may be additional powers of $t$ appearing in \eqref{introIR}, while for branch cuts there can be additional algebraic contributions.}

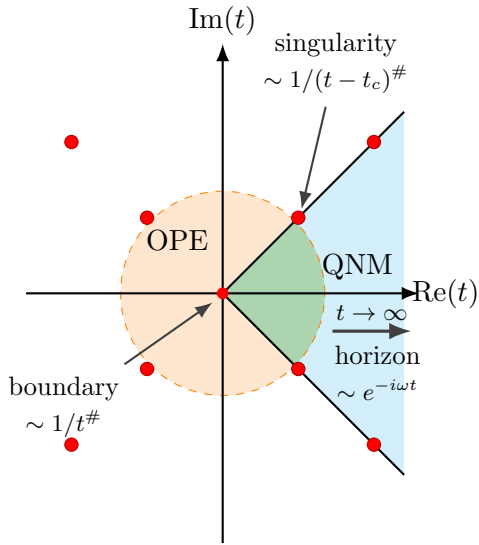
\begin{figure}
    \centering
    \begin{tikzpicture}[
    >=Latex,
    pole/.style={circle, fill=red, draw=red!70!black, minimum size=5pt, inner sep=0pt},
    origin/.style={circle, fill=red, draw=red, minimum size=4pt, inner sep=0pt},
    qnm/.style={fill=cyan!35, draw=none, opacity=0.45},
    opefill/.style={fill=orange!25, draw=none, opacity=0.75},
    overlapfill/.style={fill=orange!50!cyan!70, draw=none, opacity=0.85},
    boundary/.style={black, thick},
    anno/.style={->, thick, black!75, shorten >=1.5pt, shorten <=1.5pt}
]

\def\X{2.6}      
\def\Y{3.3}
\def\A{2.4}      
\def\R{1.35}     

\fill[opefill] (0,0) circle[radius=\R];
\fill[qnm] (0,0) -- (\A,\A) -- (\A,-\A) -- cycle;
\fill[overlapfill] (0,0) -- (45:\R)
    arc[start angle=45,end angle=-45,radius=\R] -- cycle;

\draw[boundary] (0,0) -- (\A,\A);
\draw[boundary] (0,0) -- (\A,-\A);
\draw[orange, dashed] (0,0) circle[radius=\R];

\draw[->, thick] ({-\X},0) -- (\X,0);
\draw[->, thick] (0,{-\Y}) -- (0,\Y);
\node at (0,\Y+0.3) {$\text{Im}(t)$};
\node at (\X+0.35,0) {$\text{Re}(t)$};

\node[origin] at (0,0) {};
\node[pole] at (-2.0, 2.0) {};
\node[pole] at (-1.0, 1.0) {};
\node[pole] at (-1.0,-1.0) {};
\node[pole] at (-2.0,-2.0) {};
\node[pole] at ( 1.0, 1.0) {};
\node[pole] at ( 2.0, 2.0) {};
\node[pole] at ( 1.0,-1.0) {};
\node[pole] at ( 2.0,-2.0) {};

\node at (-0.6,0.7) {OPE};
\node at (1.78,0.35) {QNM};

\node[font=\small, align=center] (bdy) at (-2.1,-1.5)
    {boundary\\[-1pt]{\footnotesize$\sim 1/t^{\#}$}};
\draw[anno] (bdy) -- (-0.09,-0.09);

\node[font=\small, align=center] (sing) at (1.5,3.05)
    {singularity\\[-1pt]{\footnotesize$\sim 1/(t-t_c)^{\#}$}};
\draw[anno] (sing) -- (1.03,1.06);

\draw[anno, very thick] (1.4,-0.5) -- (2.55,-0.5)
    node[midway, above, font=\footnotesize, black]{$t\to\infty$};
\node[font=\small, align=center] at (2.05,-1.05)
    {horizon\\[-1pt]{\footnotesize$\sim e^{-i\omega t}$}};

\end{tikzpicture}
    \caption{
    Structure of the complex time plane of $G_R(t,k)$ in holographic theories. The OPE and the QNM representations in the complex time plane converge respectively in the orange and blue shaded areas. Setting OPE=QNM inside their common domain of convergence in green, we relate IR and UV data to each other. In holographic examples the OPE region is naturally described in the bulk as a near-boundary expansion, the QNM region $t\to\infty$ by the black hole horizon while the point $t\sim |t_c|e^{i\theta}$ is associated to geodesics bouncing off the black hole singularity.}
    \label{fig:opeqnmIntro} 
\end{figure}

Connecting the UV \eqref{introUV} to the  IR \eqref{introIR} is achieved through a series of steps. The first is to compute $G_R(t,\vec{x})$ in the OPE regime by taking the discontinuity of $G_E(\tau, \vec{x})$. Next, one can integrate the OPE of $G_R(t,\vec{x})$ inside the lightcone to recover an OPE for $G_R(t,\vec{k})$, which takes the form
\be
G_R(t,\vec{k})=\theta(t)\,t^{d-1-2\Delta_{\mathcal O}}\sum_{\widehat\Delta}b_{\widehat\Delta,k} \left(\frac{t}{\beta}\right)^{\widehat\Delta}, \label{introGRtk}
\ee
where the spectrum of $\widehat\Delta = \Delta + 2\mathbb{Z}_{\geq 0}$ and the coefficients $b_{\widehat\Delta,k}$ are given by a finite sum of the $a_{\Delta, J}$ from \eqref{introUV}, as detailed in \eqref{abmap}.
Finally the connection to the IR is possible because the OPE converges for $0 < |t| < t_c$ for some $t_c >0$, while the QNM expansion converges for all $t>0$, as we discuss in detail in Section \ref{sec:overlap}. Thus there is an overlap region where both representations are valid and can be matched. This is illustrated in Figure \ref{fig:opeqnmIntro}. In a growing number of holographic examples, the radius of convergence, $t_c$, is set by bouncing singularities \cite{Parisini:2023nbd, Ceplak:2024bja, Buric:2025anb, Buric:2025fye, Barrat:2025twb, Afkhami-Jeddi:2025wra,  Dodelson:2025jff, Ceplak:2025dds, Jia:2025jbi, AliAhmad:2026wem,Giombi:2026kdz, Arnaudo:2026der, Grozdanov:2026cut, Jia:2026ryl, Arnaudo:2026tcy, Grozdanov:2026ktq, Buric:2026qsp}, related to null geodesics which `bounce' from the black hole singularity \cite{Fidkowski:2003nf, Festuccia:2005pi}, and this in particular can happen for $t_c < \beta$.

We find that a useful language for making this connection is the Mellin transform of the retarded correlation function, defined by
\be
\mathcal{M}(s, \vec{k}) \equiv \int_0^\infty dt\,t^{s-1}G_R(t,\vec{k}). \label{MellinCorrelator}
\ee
When evaluated on the QNM sum \eqref{introIR} one sees that $\mathcal{M}(s, \vec{k})$ is a variant of a spectral zeta function,
\be
\mathcal{M}(s, \vec{k}) = \Gamma(s) \sum_{n=0}^\infty r_n(i\omega_n)^{-s},\label{MellinQNM}
\ee
obtained for $\text{Re}(s)>0$ then analytically continued elsewhere.
From the OPE \eqref{introGRtk} one concludes that $\mathcal{M}(s, \vec{k})$ contains simple poles at $s=2\Delta_\mathcal{O}-\widehat{\Delta} -d+1$ with residues $b_{\widehat{\Delta},k}$. This conclusion holds even though the OPE is valid for $|t| < t_c$, since the existence of the pole only comes from knowing the contribution near the lower limit of the Mellin integral in \eqref{MellinCorrelator}.

We present three alternative methods to obtain QNMs from OPE explicitly, and our main example will be retarded correlators dual to scalar perturbations of Schwarzschild-AdS$_5$ black brane.\footnote{We note, however, that our methods are purely field theoretic and only use the bulk as a tool to obtain various input to the OPE and as a numerical oracle for comparing QNM predictions.} The first method involves matching the OPE and QNM at finite but purely real time, similar to what was done for SYK in \cite{Dodelson:2024atp,Dodelson:2025jff}. We start with stress-tensor sector OPE data for scalar perturbations of Schwarzschild-AdS$_5$ black brane, which are known exactly \cite{Fitzpatrick:2019zqz}, and through analytic continuation from $|t|<t_c$ to $t\in \mathbb{R}$ recover the first few QNMs of the black brane, including the fundamental mode. This therefore achieves QNM from OPE+analyticity, and in particular works better at small mode numbers. The result is shown in Tables \ref{table:numericsIntro} and \ref{table:numerics}.

The second method uses a matching procedure at two singular points as described in Figure \ref{fig:opeqnmIntro}. In particular, the non-analytic singular behaviour forces a QNM spectrum which asymptote to a single line at large overtone numbers $n\gg1$, $i\omega_n \sim re^{i\theta} n$, with residues $r_n \sim n^{2\Delta_\mathcal{O} -d}$. By matching subleading terms both at the OPE singularity and the singularity at the edge of convergence, we can systematically improve the large-$n$ tail of QNMs. In particular, from the OPE we obtain the following prediction for the QNMs and the residues $r_n$:
\bea 
i\omega_n  &=& re^{i\theta}n' + \frac{d_{4/3}}{(n')^{4/3}}+\mathcal{O}((n')^{-8/3})= re^{i\theta}n+d_0+d_{4/3}n^{-4/3}+d_{7/3}n^{-7/3}+\mathcal{O}(n^{-8/3}),\cr
r_n&\simeq& C \left(\frac{d}{dn}\frac{i \omega_n}{r e^{i\theta}}\right)\left(\frac{i\omega_n}{r e^{i\theta}} \right)^{2\Delta_\mathcal{O}-4}  \left(1+\mathcal{O}\left((\omega_n)^{-4}\right)\right) ,\label{eq:tailintro}
\eea 
where $n'\equiv n + e^{-i\theta }\frac{d_0}{r}$ and \eqref{coefficientsResults1}-\eqref{coefficientsResults3}. The $n^{-7/3}$ prediction for the QNMs is new as well as the prediction for the residues. The lower order data is known from bulk WKB methods\footnote{Asymptotic expression for QNMs in flat space were first explored in \cite{Motl:2003cd}.}, see \cite{Natario:2004jd,Cardoso:2004up,Dodelson:2023vrw}, and agree with ours. We also see that writing the expansion in terms of $n'= n + e^{-i\theta }\frac{d_0}{r}$ dramatically simplifies the expansion. The expression \eqref{eq:tailintro} predicts even the fundamental $n=1$ mode within a 0.4\% error at $\Delta=11/4$ and quickly converges as $n$ increases, see Figure \ref{fig:tailqnms}. 

The third and last method to obtain QNMs from OPE uses Mellin sum rules. These are given by\footnote{These are divergent sum rules that we zeta-regularise as described in Section \ref{sec:DTSumRules}.}
\be 
r_1 (i \omega_1)^q +\sum_{n=2}^\infty r_n (i \omega_n)^q = 0,\qquad q=0,1,2,\cdots, 
\ee 
which follow from the absence of analytic terms in the retarded correlator (double-trace operators). We take the following strategy: for the sum from $n=2,3,\cdots,\infty$ we input the asymptotic tail expression from the OPE and read off the prediction for the fundamental mode. In this way we obtain an expression for the fundamental mode in terms of QNM moments with $n>1$.  Including more terms in the asymptotic approximation of the moments systematically improves the prediction for the fundamental mode. 

\begin{table}[t!]
\begin{center}
\begin{tabular}{ |c||c|c|c| } 
\hline
mode & OPE analytic continuation & exact (numerics) & tail formula \eqref{eq:qnmTail}\\
\hline\hline 1,2 & $\pm 1.97041 - 1.51923 i$ & $\pm 1.97046 - 1.51917 i$ & $\pm 1.97115 - 1.52957 i$\\
\hline 39,40 & $\pm 40.5110 - 39.0443 i$ & $\pm 39.9706 - 39.5294 i$ & $\pm 39.9706 - 39.5294 i$\\
\hline
\end{tabular}
\end{center}
\caption{Here we show a few modes obtained from the first and second OPE methods as compared to numerical values obtained from the bulk dual. For an extended list, see Table \ref{table:numerics}. The OPE analytic continuation column is obtained from matching at real $t>0$, together with a Pad\'e continuation, while the last column comes from the analytical large-mode number expansion obtained by matching at singular points. The two methods converge from opposite sides, i.e.\ for small mode number $n$ vs large mode number allowing us to obtain any QNM consistently from the OPE.}
\label{table:numericsIntro}
\end{table}

Beyond explicit computations of OPE and QNM described above, we study useful analytic examples such as scalar correlators in 2d CFT dual to the BTZ black hole, R-current correlators at $k=0$ in $\mathcal{N}=4$ SYM \cite{Myers:2007we}, and correlators of fundamental scalars in the large-$N$ O($N$) model. In the O($N$) model the correlator is governed by two undamped QNMs (i.e. normal modes). A general consequence of having only a finite number of QNMs is that the sum \eqref{MellinQNM} cannot generate new Mellin poles. This means that in such examples, the only poles of $\mathcal{M}(s, \vec{k})$ are just those of $\Gamma(s)$ and hence \eqref{introGRtk} takes the form of a Taylor series in $t$ with no OPE singularity.

We further consider the OPE$=$QNM relation in the limit of large spatial momentum $k$. In this limit, the spatial Fourier transform singles out the lightcone limit and the OPE is organised in terms of the twist $\tau=\Delta-J$ instead of the scaling dimension $\Delta$. Due to the universality of low-twist spectrum, we argue that lightcone modes with $\omega\sim\pm k+\cdots$ have a universal scaling behaviour of $\mathcal{O}(k^{-\frac{d-2}{d+2}})$ consistent with WKB in the bulk dual \cite{Festuccia:2008zx}. In particular, while in this work we mainly explore external scalar operators, in the context of stress tensor thermal two-point functions, we argue that the universal correction to the lightcone modes is controlled by the conformal collider bounds and in $d=4$ the conformal anomalies $(a,c)$. As an example of this, we consider Gauss-Bonnet gravity which has been explored in the past in this context \cite{Brigante:2007nu,Brigante:2008gz,Buchel:2009sk,Fidkowski:2003nf} and recently in \cite{Buchel:2026rep} in more detail. We argue from the OPE, see also \cite{Kulaxizi:2010jt,Esper:2023jeq}, that the leading lightcone mode correction vanishes as we approach the critical value where the conformal collider bounds become saturated.
\\\\
The layout of the paper is as follows. In Section \ref{sec:analytic}, we discuss properties of thermal correlators at fixed spatial momentum. We introduce both the OPE and QNM representation of retarded correlators and argue that there is an overlapping regime of convergence. We further introduce a dispersive representation of the Wightman correlator in terms of the retarded correlator. In Section \ref{sec:opeExtended}, we derive QNMs from OPE by analytically extending the OPE beyond its region of convergence and match the OPE and QNM representations at finite $t$. In Section \ref{sec:QNMasyfromOPE}, we derive QNM asymptotics from OPE data by matching the two representations at singular points. In Section \ref{sec:mellin}, we introduce the Mellin space representation of retarded correlators and derive various sum rules that we further explore. We further discuss how to analytically extend the Mellin representation. In Section \ref{sec:lightcone}, we consider the OPE$=$QNM relation in a lightcone limit with large spatial momentum $k$ and furthermore study stress-tensor correlators. We conclude with a discussion in Section \ref{sec:discussion}. The appendices include further technical details, the $O(N)$ model and introduce Mellin space in the bulk. 
\\\\
\emph{\textbf{Note added}: While finalising this work we became aware of the related work \cite{Barrat:2026toAppear}, which appears simultaneously with ours.}

\section{Analytic structure at fixed spatial momentum}\label{sec:analytic}

In this work we develop a general framework relating thermal OPE data to QNM data. The central object in this relation is the mixed $t,\vec{k}$ retarded correlator, $G_R(t, \vec{k})$, which admits both OPE and QNM decompositions. In this section we provide correlator definitions, the OPE and QNM decompositions, and discuss the region in the complex $t$ plane where these decompositions overlap.

We begin with the key definitions. Let $\mathcal{O}$ be a bosonic scalar primary operator\footnote{It is straightforward to extend to operators with spin, we discuss in particular stress tensor correlators in Section \ref{sec:lightcone}.} of scaling dimension $\Delta_{\mathcal{O}}$ in a thermal state at inverse temperature $\beta$ on a spatial manifold $\mathbb{R}^{d-1}$. The Euclidean, Wightman, two-sided, and  retarded correlators are defined as follows, 
\bea
G_E(\tau,\vec{x})&=&\langle \mathcal{O}(\tau,\vec{x})\mathcal{O}(0,0)\rangle_\beta,\\
G_>(t,\vec{x}) &=& \langle \mathcal{O}(t,\vec{x})\mathcal{O}(0,0)\rangle_\beta,\\
G_<(t,\vec{x}) &=& \langle \mathcal{O}(0,0)\mathcal{O}(t,\vec{x})\rangle_\beta,\\
G_{12}(t, \vec{x}) &=& \langle \mathcal{O}(t-i\beta/2,\vec{x})\mathcal{O}(0,0)\rangle_\beta,\\
G_R(t,\vec{x})&=&-i\theta(t)\langle[\mathcal{O}(t,\vec{x}),\mathcal{O}(0,0)]\rangle_\beta.
\eea
The Euclidean and Wightman correlators are related through analytic continuation,
\bea
G_>(t,\vec{x}) &=& G_E(it + 0^+, \vec{x}),\\
G_<(t,\vec{x}) &=&  G_E(it - 0^+, \vec{x}),
\eea
where the second line follows from the KMS condition for bosonic operators. By combining the above relations, $G_R(t,\vec{x})$ can be expressed directly as a discontinuity of the Euclidean correlator,
\bea
G_R(t,\vec{x}) &=& -i \theta(t)\,\text{disc}\,G_E(it,\vec{x}), \label{GRdisc}\\
\text{disc}\,G_E(it,\vec{x})&\equiv&G_E(it+0^+,\vec{x})-G_E(it-0^+,\vec{x}).\label{discdef}
\eea

The retarded Green's function in mixed time-momentum variables is given by the following spatial Fourier transform,
\begin{equation}
G_R(t,\vec{k})=\int \mathrm{d}^{d-1}x\,e^{-i\vec{k}\cdot\vec{x}}G_R(t,\vec{x}).\label{GRmixed}
\end{equation}
Causality implies that the retarded correlator vanishes outside the lightcone,
\begin{equation}
G_R(t,\vec{x})=0\qquad\text{for}\qquad |\vec{x}|>t,
\end{equation}
which limits the integration range in the Fourier transform \eqref{GRmixed}. Thus, combined with \eqref{GRdisc} we arrive at a key formula for the OPE=QNM setup,
\begin{equation}
G_R(t,\vec{k})= -i\theta(t)\int_{|\vec{x}|<t} \mathrm{d}^{d-1}x\,e^{-i\vec{k}\cdot\vec{x}}\,\text{disc}\,G_E(it,\vec{x}).
\label{GRGE}
\end{equation}
On the left hand side the real-time correlator can be represented in terms of a QNM residue sum, while on the right hand side the Euclidean correlator can be expressed using the thermal OPE. These have an overlapping region of convergence in the $t$ plane.

\subsection{OPE}
The Euclidean thermal two-point function admits the OPE \eqref{introUV}, copied here for convenience,
\begin{equation}
G_E(\tau,\vec{x})=\sum_{(\Delta,J)}a_{\Delta,J}\beta^{-\Delta}C_J^{\left(\frac{d-2}{2}\right)}\left(\frac{\tau}{\sqrt{\tau^2+|\vec{x}|^2}}\right)(\tau^2+|\vec{x}|^2)^{\frac{\Delta}{2}-\Delta_\mathcal{O}}, \label{OPEcopy}
\end{equation}
where the sum runs over primary operators of scaling dimension $\Delta$ and even-spin $J$. In this section we use this OPE to compute the small $t$ expansion of $G_R(t,\vec{k})$ using \eqref{GRGE}. 

The first step is computing the discontinuity \eqref{discdef}. 
For real operators we evaluate it using the relation
\be
\text{disc}\,G_E(it,\vec{x}) = 2 i \theta(t>|\vec{x}|)\,\text{Im} G_E(it+0^+,\vec{x}). \label{discIm}
\ee
Using this, we obtain
\begin{equation}
G_R(t,\vec{x})=2\theta(t>|\vec{x}|)\sum_{\substack{(\Delta,J)}}\sin\left[\pi\left(\frac{\Delta}{2}-\Delta_\mathcal{O}\right)\right]\frac{a_{\Delta,J}}{\beta^{\Delta}}C_J^{\left(\frac{d-2}{2}\right)}\left(\frac{t}{\sqrt{t^2-|\vec{x}|^2}}\right)(t^2-|\vec{x}|^2)^{\frac{\Delta}{2}-\Delta_\mathcal{O}}.\label{GRtx}
\end{equation}
Note that whenever $\frac{\Delta}{2}-\Delta_\mathcal{O} \in \mathbb{Z}_{\geq 0}$ there is no contribution to this sum, and hence this sum only contains single trace exchanges. Thus, with \eqref{GRtx} the spatial Fourier transform \eqref{GRmixed} becomes a weighted sum of spatial Fourier transforms of the blocks, 
\bea\label{eq:GRtk}
G_R(t,\vec{k}) &=& \sum_{(\Delta,J)} 2\sin\left[\pi\left(\frac{\Delta}{2}-\Delta_\mathcal{O}\right)\right]\frac{a_{\Delta,J}}{\beta^\Delta}\mathcal{I}_{\Delta,J}(t,\vec{k}),\\
\mathcal{I}_{\Delta,J}(t,\vec{k})&\equiv &\int_{|\vec{x}|<t}\mathrm{d}^{d-1}x\,e^{-i\vec{k}\cdot\vec{x}}\,C_J^{\left(\frac{d-2}{2}\right)}\left(\frac{t}{\sqrt{t^2-|\vec{x}|^2}}\right)(t^2-|\vec{x}|^2)^{\frac{\Delta}{2}-\Delta_\mathcal{O}}.\label{eq:IDeltaJtk}
\eea
We evaluate the integral $\mathcal{I}_{\Delta,J}(t,\vec{k})$ explicitly in Appendix \ref{app:FTblock}. With the result for $\mathcal{I}_{\Delta,J}(t,\vec{k})$ in an expansion in $t$ given in \eqref{eq:IDeltaJexpansion}, we arrive at the final expression for the OPE for $G_R(t,\vec{k})$, given in \eqref{abmap}.
\begin{keyformula}[Mixed correlator OPE]
\begin{equation}
\begin{aligned}
G_R(t,\vec{k})=&\,\theta(t)\,t^{d-1-2\Delta_{\mathcal O}}\sum_{\widehat\Delta}b_{\widehat\Delta,k} \left(\frac{t}{\beta}\right)^{\widehat\Delta} \\
b_{\widehat\Delta,k} = &\sum_{\substack{(\Delta,J),\ell:\\\Delta+2\ell=\widehat\Delta}}\, 2\sin\left[\pi\left(\frac{\Delta}{2}-\Delta_{\mathcal O}\right)\right]\,
 \left(k\beta\right)^{2\ell} \mathcal{I}_{\Delta, J, \ell}\;a_{\Delta,J}
\end{aligned}
\label{abmap}
\end{equation}
\end{keyformula}
The coefficients $\mathcal{I}_{\Delta, J, \ell}$ appearing in \eqref{abmap} are given in \eqref{IDeltaJell}, and $k\equiv |\vec{k}|$. Here the sum runs over every operator $(\Delta, J)$ in the original OPE sum \eqref{OPEcopy}, and every $\ell \in \mathbb{Z}_{\geq 0}$ such that $\Delta + 2\ell = \widehat{\Delta}$. The general expectation is that this thermal OPE has a finite radius of convergence. Note that the $\sin$ factor removes analytic terms (double-trace operators) due to these having zero discontinuity, see also \cite{Manenti:2019wxs}. For $\Delta_\mathcal{O}\in \mathbb{Z}$ and $J\in2\mathbb{Z}$ the coefficients $\mathcal{I}_{\Delta, J, \ell}$ have a simple pole leading to a finite non-vanishing contribution. 

In the above expression \eqref{abmap} we have provided a map from the original OPE data in position space, $a_{\Delta, J}$, to the OPE data in mixed time-momentum domain, $b_{\widehat{\Delta}, k}$. It is interesting to ask if this relation can be inverted. When $k=0$ only the $\ell = 0$ term in the sum in \eqref{abmap} survives, and this means we cannot invert the relation in general. However, by turning on $k$ we can attempt the inversion. Let us denote the coefficient of $(\beta k)^{2\ell}$ in $b_{\widehat{\Delta}, k}$ as 
\be
b_{\widehat{\Delta};\ell} = \left[(\beta k)^{2\ell}\right]b_{\widehat{\Delta},k}.
\ee
We now fix a $\Delta$ in the OPE spectrum, then 
\eqref{abmap} is the linear transformation from the $J$ basis to the $\ell$ basis, 
\be
b_{\Delta + 2\ell;\ell} = \sum_J M_{\ell,J}^{(\Delta)}a_{\Delta, J},
\ee
where
\be
M_{\ell,J}^{(\Delta)} = 2\sin\left[\pi\left(\frac{\Delta}{2}-\Delta_{\mathcal O}\right)\right]\, \mathcal{I}_{\Delta, J, \ell}.
\ee
For a given $\Delta$ there are finitely many values for $J = J_1, J_2, \ldots J_{N_\Delta}$, and by taking equally many values for $l = 0, 1, \ldots N_{\Delta} -1$ we have a square linear system which may then be invertible. This is the case for the large-$N$ O($N$) model, where the inverted relation is given in \eqref{btoaON}.

\subsection{QNM}
The retarded correlator may also be represented as a sum over QNMs \eqref{introIR},
\begin{keyformula}[Mixed correlator QNM]
\begin{equation}
G_R(t,\vec{k})=\theta(t)\sum_n r_n(k)e^{-i\omega_n(k)t}. \label{QNMsumcopy}
\end{equation}
\end{keyformula}
This assumes that the frequency space Green's function is meromorphic, admitting simple poles, whose sum is complete in the time-domain. This is the generic behaviour in large-$N$ holographic theories. We include normal modes in this definition of QNMs as a special case corresponding to poles at real frequencies.

Let us comment on the assumptions used when writing \eqref{QNMsumcopy}, in particular what can happen by relaxing the presence of simple poles only or meromorphy. If the frequency-space correlator has higher-order poles, the corresponding terms in time are polynomials in $t$ multiplying the exponential factors. If branch cuts are present, as can happen at finite $N$ or at zero temperature, the pole sum must be supplemented by branch-cut integrals. These cases are not included in the simple form \eqref{QNMsumcopy}, although the following discussions can be generalised to cover them as well. In holographic theories, the retarded correlator is meromorphic and one can prove using scattering theory that the poles are simple poles \cite{Dodelson:2023vrw}, barring the case of purely imaginary modes which could be of higher order.

\subsection{Overlapping regime of convergence}\label{sec:overlap}

Let us now discuss in detail the analytic structure of $G_R(t,\vec{k})$ in the complex $t$ plane. We will show that the convergence regions for the OPE and the QNM expansion overlap.

To gain some intuition we start with the simple case of a $2d$ thermal correlator. It is useful to start with the two-sided correlator whose closed form (taking $\beta=2\pi$ and $\Delta_\mathcal{O}=2$ for simplicity) expression is:
\begin{equation}
G_{12}(t,x)=\frac{1}{8\pi\cosh^2\left(\frac{t+x}{2}\right)\cosh^2\left(\frac{t-x}{2}\right)} \,.
\end{equation}
Here $G_{12}$ is a simple meromorphic function of $x$ with double poles at $x=\pm t \pm i \pi$ plus thermal images. After integrating over space the analytic structure gets much richer. Consider the $k=0$ correlator given by, 
\begin{equation}
    G_{12}(t,k=0)= \int_\mathbb{R}dx\, G_{12}(t,x) \,. \label{BTZG12k0int}
\end{equation}
If we give $t$ an imaginary part, the poles of $G_{12}(t,x)$ in $x$ will move and they will cross the contour of integration in \eqref{BTZG12k0int} as soon as $\text{Im}\,t= \beta/2=\pi$. This produces a branch cut in the $(t,k=0)$ correlator, with the discontinuity given by $G_R(t,k=0)$,
\begin{equation}
    G_R(t,k=0) = -i\theta(t)\text{disc} \, G_{12}(t,k=0) = \theta(t) \, \frac{2\cosh t}{\sinh^3 t},
\end{equation}    
where now the appropriate discontinuity in the case of $G_{12}$ is given by $\text{disc} \, G_{12}(t,k=0) =  G_{12}(t + i \beta/2-i\epsilon,k=0)- G_{12}(t - i \beta/2+i\epsilon,k=0)$\footnote{This is is most clearly seen as the discontinuity on the cylinder, see Section \ref{sec:disp}.}.
The retarded correlator has poles at $t_b=\pm i n \pi = \pm i n \beta/2$. In the context of holography, these poles are the 2d analogues of the bouncing singularities appearing in higher dimensional cases, which show up in this case even though no curvature invariant is divergent at $r=0$ for BTZ. 

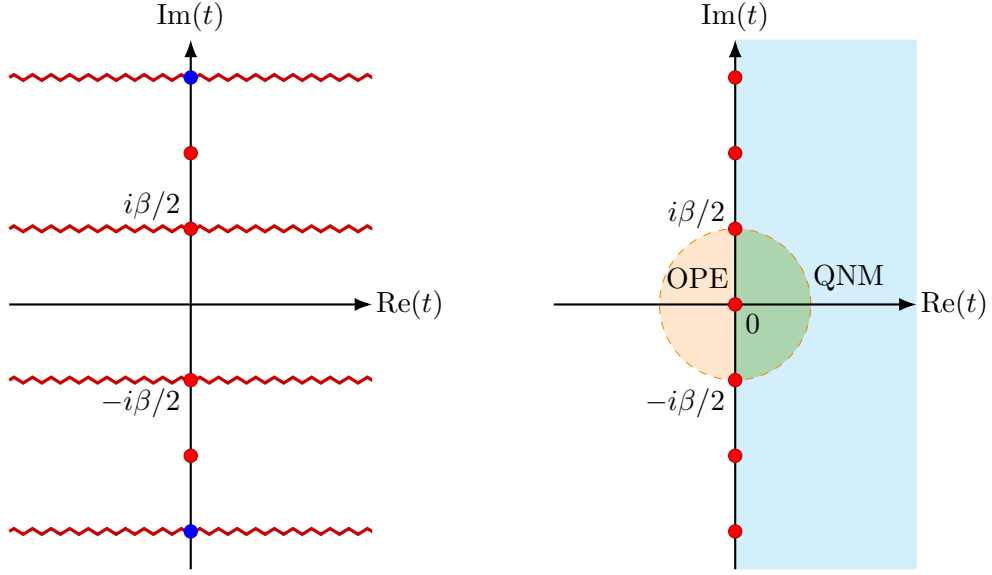
\begin{figure}[t]
\centering
    \begin{tikzpicture}[
    >=Latex,
    cut/.style={
        red!80!black,
        very thick,
        decorate,
        decoration={zigzag, segment length=7pt, amplitude=1.1pt}
    },
    pole/.style={
        circle,
        fill=red,
        draw=red!70!black,
        minimum size=5pt,
        inner sep=0pt
    },
    poleg/.style={
        circle,
        fill=blue,
        draw=blue,
        minimum size=5pt,
        inner sep=0pt
    },
    overlapfill/.style={
        fill=orange!50!cyan!70,
        draw=none,
        opacity=0.85
    },
    qnm/.style={
        fill=cyan!35,
        draw=none,
        opacity=0.45
    },
    guide/.style={gray!45, densely dashed}
]

\def\X{2.4}
\def\piY{1}
\def\Y{3.5}
\def\panelSep{6.2}

\pgfmathsetmacro{\topLabel}{\piY + 0.2}
\pgfmathsetmacro{\botLabel}{-\piY - 0.2}
\pgfmathsetmacro{\titleY}{-\Y - 0.35}

\begin{scope}

\draw[->, thick] ({-\X},0) -- (\X,0);
\draw[->, thick] (0,{-\Y}) -- (0,\Y);

\draw[guide] ({-\X},\piY) -- (\X,\piY);
\draw[guide] ({-\X},{-\piY}) -- (\X,{-\piY});

\draw[cut] ({-\X},\piY) -- (\X,\piY);
\draw[cut] ({-\X},{-\piY}) -- (\X,{-\piY});
\draw[cut] ({-\X},3*\piY) -- (\X,3*\piY);
\draw[cut] ({-\X},-3*\piY) -- (\X,-3*\piY);

\node[pole] at (0,\piY) {};
\node[pole] at (0,2*\piY) {};
\node[poleg] at (0,3*\piY) {};
\node[pole] at (0,-\piY) {};
\node[pole] at (0,-2*\piY) {};
\node[poleg] at (0,-3*\piY) {};

\node[left] at (0,\topLabel+0.1) {$i\beta/2$};
\node[left] at (0,\botLabel-0.1) {$-i\beta/2$};

\node[align=center] at (0,3.8) {$\text{Im}(t)$};
\node[align=center] at (2.9,0.0) {$\text{Re}(t)$};

\end{scope}
\hspace{10mm}
\begin{scope}[shift={(\panelSep,0)}]


\fill[orange!25, opacity=0.75] (0,0) circle[radius=1];

\fill[qnm] (0,0) -- (0,3.5) -- (2.4, 3.5) -- (2.4, -3.5) -- (0, -3.5) -- cycle;

\fill[overlapfill] (0,0) -- (90:1)
    arc[start angle=90,end angle=-90,radius=1] -- cycle;
    
\draw[orange, dashed] (0,0) circle [radius={1}];

\node[left] at (0,\topLabel) {$i\beta/2$};
\node[left] at (0,\botLabel-0.1) {$-i\beta/2$};
\node[below right] at (0,0) {$0$};
\node[align=center] at (-0.5,0.35) {OPE};
\node[align=center] at (1.5,0.35) {QNM};

\draw[->, thick] ({-\X},0) -- (\X,0);
\draw[->, thick] (0,{-\Y}) -- (0,\Y);

\node[align=center] at (0,3.8) {$\text{Im}(t)$};
\node[align=center] at (2.9,0.0) {$\text{Re}(t)$};

\node[pole] at (0,\piY) {};
\node[pole] at (0,0) {};
\node[pole] at (0,{-\piY}) {};
\node[pole] at (0,2*\piY) {};
\node[pole] at (0,3*\piY) {};
\node[pole] at (0,-2*\piY) {};
\node[pole] at (0,-3*\piY) {};

\end{scope}

\end{tikzpicture}
\caption{Here we plot $G_{12}(t,k=0)$ (left) and $G_R(t,k=0)$ (right) with the $\theta(t)$ factor removed (right) for the BTZ case, in the complex $t$ plane. The two-sided correlator is analytic in the strip, with the discontinuity across the cut given by $G_R$. When crossing the cut, poles of $G_R$ show up on the second sheet. The OPE of $G_R$ converges in the orange (punctured) disk, while the QNM expansion converges in the right half plane, leading to a region of overlapping convergence.}
\label{fig:btz}
\end{figure}

All in all, the $k=0$ two-sided correlator as a function of $t$ is analytic in the strip, with cuts at the edges of the strip. Crossing the cuts we pick up a discontinuity given by $G_R$, and its \textit{bouncing-}like singularities will show up in the second sheet, as illustrated in Figure \ref{fig:btz}. Note that, contrary to what happens for $G_{12}(t,x)$ here, in order to make $G_{12}(t,k=0)$ periodic in imaginary time we have to give up analyticity.

The retarded correlator can be now expanded at small (OPE) and large (QNM) times:
\begin{equation}
\begin{aligned}
    G_R(t,k=0)&=  \theta(t)\left(\frac{2}{t^3}+\sum_{n\ge1} \frac{(2n-1)(2n-2)2^{2n}B_{2n}}{(2n)!}t^{2n-3} \right)\,\qquad &\text{OPE},\\
    &= 8 \theta(t)\sum_{n}n^2 e^{-2nt} \qquad &\text{QNM},
\end{aligned}
\end{equation}
where $B_{2n}$ are Bernoulli numbers. The OPE converges until we meet the first singularity at $|t|=\pi$, while the QNM sum is a geometric series in $e^{-2t}$, which is convergent in the half plane $\text{Re}\,t>0$. We note that the convergence of the QNM sum to $t=0$ in the case of BTZ was previously pointed out in \cite{Arnaudo:2025uos}. In this case we see that OPE$=$QNM in a common domain of convergence, illustrated in Figure \ref{fig:btz}. In this figure we show $G_R(t,k=0)$ with the $\theta(t)$ factor removed -- in other words, consider $G_R(t,k=0)$ at $t>0$ and then analytically continue to $t\in \mathbb{C}$.

\begin{figure}[t]
\centering
\begin{tikzpicture}[
    >=Latex,
    cut/.style={
        red!80!black,
        very thick,
        decorate,
        decoration={zigzag, segment length=7pt, amplitude=1.1pt}
    },
    pole/.style={
        circle,
        fill=red,
        draw=red!70!black,
        minimum size=5pt,
        inner sep=0pt
    },
    bluepole/.style={
        circle,
        fill=red,
        draw=red!70!black,
        minimum size=5pt,
        inner sep=0pt
    },
    violetpole/.style={
        circle,
        fill=blue,
        draw=blue,
        minimum size=5pt,
        inner sep=0pt
    },
    origin/.style={
        circle,
        fill=red,
        draw=red,
        minimum size=4pt,
        inner sep=0pt
    },
    qnm/.style={
        fill=cyan!35,
        draw=none,
        opacity=0.45
    },
    opefill/.style={
        fill=orange!25,
        draw=none,
        opacity=0.75
    },
    overlapfill/.style={
        fill=orange!50!cyan!70,
        draw=none,
        opacity=0.85
    },
    boundary/.style={
        black,
        thick
    }
]

\def\X{2.4}
\def\Y{3.5}
\def\panelSep{6.2}


\begin{scope}


\draw[->, thick] ({-\X},0) -- (\X,0);
\draw[->, thick] (0,-\Y) -- (0,\Y);

\draw[cut] ({-\X},0.8) -- (\X,0.8);
\draw[cut] ({-\X},-0.8) -- (\X,-0.8);
\draw[cut] ({-\X},2.4) -- (\X,2.4);
\draw[cut] ({-\X},-2.4) -- (\X,-2.4);

\node[pole] at (0,-0.8) {};

\node[bluepole] at (0,0.8) {};

\node[bluepole] at (-0.75,1.6) {};
\node[bluepole] at ( 0.75,1.6) {};

\node[bluepole] at (-1.5,2.4) {};
\node[violetpole] at (0,2.4) {};
\node[bluepole] at ( 1.5,2.4) {};

\node[bluepole] at (-0.75,-1.6) {};
\node[bluepole] at ( 0.75,-1.6) {};

\node[bluepole] at (-1.5,-2.4) {};
\node[violetpole] at (0,-2.4) {};
\node[bluepole] at ( 1.5,-2.4) {};

\node[align=center] at (0,3.8) {$\text{Im}(t)$};
\node[align=center] at (2.9,0.0) {$\text{Re}(t)$};

\end{scope}

\hspace{10mm}

\begin{scope}[shift={(\panelSep,0)}]

\def\A{2.4}
\def\R{1.25}

\fill[opefill] (0,0) circle[radius=\R+0.1];

\fill[qnm] (0,0) -- (\A,\A) -- (\A,-\A) -- cycle;

\fill[overlapfill] (0,0) -- (45:\R+0.1)
    arc[start angle=45,end angle=-45,radius=\R+0.1] -- cycle;

\draw[boundary] (0,0) -- (\A,\A);
\draw[boundary] (0,0) -- (\A,-\A);

\draw[orange, dashed] (0,0) circle[radius=\R+0.1];

\draw[->, thick] ({-\X},0) -- (\X,0);
\draw[->, thick] (0,{-\Y}) -- (0,\Y);

\node[origin] at (0,0) {};

\node[pole] at (-2.0, 2.0) {};
\node[pole] at (-1.0, 1.0) {};

\node[pole] at (-1.0,-1.0) {};
\node[pole] at (-2.0,-2.0) {};

\node[pole] at ( 1.0, 1.0) {};
\node[pole] at ( 2.0, 2.0) {};

\node[pole] at ( 1.0,-1.0) {};
\node[pole] at ( 2.0,-2.0) {};

\node[align=center] at (-0.55,0.75) {OPE};
\node[align=center] at (1.7,0.35) {QNM};

\node[align=center] at (0,3.8) {$\text{Im}(t)$};
\node[align=center] at (2.9,0.0) {$\text{Re}(t)$};

\end{scope}

\end{tikzpicture}
\caption{$G_{12}(t,k=0)$ (left) and $G_R(t,k=0)$ with the $\theta(t)$ factor removed (right) for the R-current correlator in the complex $t$ plane. The two-sided correlator is analytic in the strip, with a discontinuity given by $G_R$. When crossing the cut, poles of $G_R$ show up on the second sheet. The OPE of $G_R$ converges in the orange (punctured) disk, while the QNM expansion converges in the right wedge, leading to a region of overlapping convergence. }
\label{fig:rcurrent_tplane}
\end{figure}
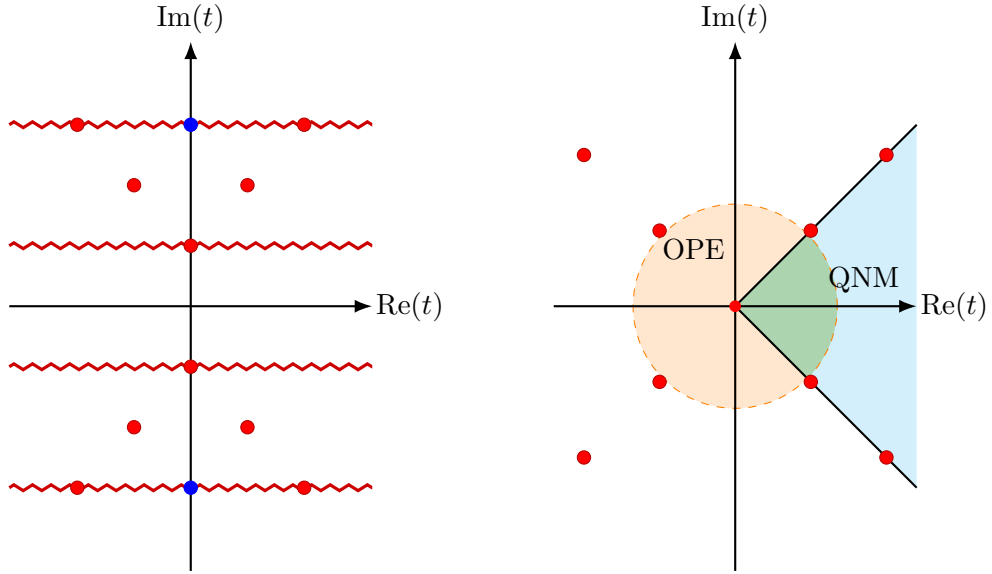

Let us now consider thermal correlators in higher dimensions. A simple example is the R-current correlator in $\mathcal N=4$ SYM with $k=0$, known in closed form from the bulk dual \cite{Myers:2007we}. We start with the expression for the two-sided Wightman function in \cite{Dodelson:2023vrw}, in units in which $\beta = 2\pi$ and arbitrary normalisation,
\begin{equation}
G_{12}(\omega,\vec{k} = 0)=\frac{\omega^2}{\sinh\left(\frac{\pi\omega(1+i)}{2}\right)\sinh\left(\frac{\pi\omega(1-i)}{2}\right)}.
\end{equation}
The two-sided correlator in the time domain is therefore
\begin{equation}
G_{12}(t,\vec{k} = 0)=\frac{1}{2\pi}\int_{\mathbb R}d\omega\,\frac{\omega^2e^{-i\omega t}}{\sinh\left(\frac{\pi\omega(1+i)}{2}\right)\sinh\left(\frac{\pi\omega(1-i)}{2}\right)}.
\end{equation}
Closing the frequency contour in the lower half-plane, the retarded correlator becomes 
\begin{equation}\label{RcurrentGR}
G_R(t,\vec{k} = 0)=\theta(t)\left(-\frac{1-i}{\pi}\frac{\cosh\left(\frac{1+i}{2}t\right)}{\sinh^3\left(\frac{1+i}{2}t\right)}-\frac{1+i}{\pi}\frac{\cosh\left(\frac{1-i}{2}t\right)}{\sinh^3\left(\frac{1-i}{2}t\right)}\right). 
\end{equation}
We remark that the Fourier transform is only convergent inside the strip $|\text{Im}(t)|<\beta/2$: equation \eqref{RcurrentGR} can only strictly be obtained in the strip, where no singularity shows up. The situation is then analogous to the BTZ case: the analytic continuation of $G_{12}(t,k=0)$ is not periodic in imaginary time, and to make it periodic we introduce cuts at the edges of the strip of analyticity. The discontinuity across the cuts is again $G_R(t,k=0)$, whose singularities in complex $t$ precisely correspond to bouncing singularities. The situation is now depicted in Figure \ref{fig:rcurrent_tplane}.

From \eqref{RcurrentGR} we can now expand in different regions and read off OPE and QNM data. We obtain the two QNM towers
\bea
\omega_n^{+} &=& (1-i)n,\qquad r_n^{+}=-\frac{4(1-i)}{\pi}n^2,\qquad n\in\mathbb Z_{>0},\\
\omega_n^{-} &=& (-1-i)n,\qquad r_n^{-}=-\frac{4(1+i)}{\pi}n^2,\qquad n\in\mathbb Z_{>0}.
\eea
Expanding \eqref{RcurrentGR} around $t=0$ we recover the $k=0$ OPE, with spectrum
\be\label{Rcurrent-OPE1}
\widehat{\Delta} = 4m,\qquad m\in \mathbb{Z}_{\geq 0}
\ee
and coefficients 
\begin{equation}\label{Rcurrent-OPE2}
b_{0,0}=\frac{8}{\pi}, \qquad b_{4m,0}=\frac{(-1)^{m-1}2^{4-2m}}{\pi^{1+4m}}(8m^2-6m+1)\zeta(4m).
\end{equation}

The OPE has a radius of convergence set by bouncing singularities. These occur at zeros of the denominator appearing in \eqref{RcurrentGR}, i.e.
\be
t = (1\pm i)\pi q \qquad q \in \mathbb{Z}\setminus\{0\},
\ee
corresponding to a diagonal lattice set by the QNM frequencies. Thus the OPE converges in the annular region $0<t<\sqrt{2}\pi = \beta/\sqrt{2}$. Meanwhile, the QNM expansion is the sum of two Taylor series around $t\to \infty$ in the form,
\be
G_R(t,0) = \sum_{n>0} r_n^{-}(X^{-})^n + \sum_{n>0} r_n^{+}(X^{+})^n
\ee
with $X^{\pm} = e^{(-1\mp i)t}$. Since $r_n^{\pm}\propto n^2$, each sum is a geometric series which converges in the disk $|X^{\pm}|<1$. The convergence region $\{|X^{+}| < 1\} \cap \{|X^{-}| < 1\}$ in the $t$ plane is the wedge $\text{Re}(t) > 0$ and $|\text{Im}(t)| < \text{Re}(t)$, along the edge of which lie the OPE and lattice of bouncing singularities. This is illustrated in Figure \ref{fig:rcurrent_tplane}. This provides a complementary view on the origin of the bouncing singularity lattice, as the line of singularities limiting the QNM convergence in the appropriate $X$ variable, with each asymptotic QNM line giving an edge of a wedge. 

More generally in higher dimensional cases in holography we have no explicit expression for boundary correlators. However it has been shown that two-sided correlators in momentum space have no zeroes in the complex $\omega-$plane, and hence can be written as a product over QNMs \cite{Dodelson:2023vrw}. Let us assume we have a single line of poles asymptotically behaving as
\begin{equation}\label{eq:prodqnm}
    \omega_n\simeq -i\,r e^{i\theta}n+s e^{i\phi} \,,
\end{equation}
where in our convention $\theta\in(0,\pi/2)$. Then 
\begin{equation}
    \widetilde{G}_{12}(\omega, \vec{k})=\prod_n \left(1-\frac{\omega^2}{\omega_n^2}\right)^{-1}\left(1-\frac{\omega^2}{(\omega_n^*)^2}\right)^{-1} \simeq \prod_{\pm,\pm}\Gamma\left(\frac{e^{\pm i\theta}}{r}\left( \pm \omega+e^{-i\phi}s\right)+1\right) \,.
\end{equation}
Following \cite{Dodelson:2025jff} we throw away all perturbative terms at large $\omega$ while keeping all the non-perturbative terms, which gives the simple expression
\begin{equation}
    \widetilde{G}_{12}(\omega, \vec{k}) \simeq\frac{\omega^{2\Delta_\mathcal{O}-d}}{\sin\left(\frac{\pi e^{-i\theta}}{r}(\omega-e^{i\phi}s)\right)\sin\left(\frac{\pi e^{i\theta}}{r}(\omega-e^{-i\phi}s)\right)} \,. \label{TPFG12wk}
\end{equation}
Moreover, in \cite{Dodelson:2023vrw} a relation for the asymptotic QNM parameters was established as follows 
\begin{equation}
    \beta=\frac{4\pi\sin\theta}{r} \,, \quad 2\Delta_\mathcal{O} - d = \frac{4s\cos(\theta-\phi)+2r}{r}  \,,\label{TPFG12wkparams}
\end{equation}
where $\beta$ is the inverse temperature. 
From \eqref{TPFG12wk} we may read off the residues for the asymptotic lines of QNMs in the lower half plane, through (see \cite{Dodelson:2023vrw})
\be
r_n = -i\, \text{res}\left(\widetilde{G}_R(\omega, \vec{k}), \omega = \omega_n\right) = 2\sinh\left(\frac{\beta\omega_n}{2}\right) \text{res}\left(\widetilde{G}_{12}(\omega, \vec{k}), \omega = \omega_n\right).
\ee
Moreover, using \eqref{TPFG12wkparams} we find the asymptotic behaviour of residues,
\be
r_n \simeq 2\sinh\left(\frac{\beta\omega_n}{2}\right)\frac{(-1)^n e^{i\theta}r}{\pi} \frac{\omega_n^{2\Delta_\mathcal{O} -d}}{\sin\left(\pi e^{2i\theta} n\right)} \sim \text{const}\times n^{2\Delta_\mathcal{O} -d}.
\ee
Thus this asymptotic QNM line contributes, up to the constant,
\be
G_R(t, \vec{k}) \supset \sum_{n=1}^\infty n^{2\Delta_\mathcal{O} -d} e^{-\sigma n t} = \text{Li}_{-2\Delta_{\mathcal O}+d}\left(e^{-\sigma t}\right),
\ee
where $\sigma =  i r e^{i\theta}$. The polylog has singularities at $e^{-\sigma t} = 1$ only, and so the region of convergence is $\left|e^{-\sigma t}\right| < 1$, bounded by both the $t=0$ OPE singularity and bouncing singularities at complex values of $t$. For this single line of poles, these are located at $-\sigma t = 2\pi m i$ with $m\in \mathbb{Z}$, i.e. $t = -m \frac{2 \pi}{r} e^{-i\theta}$. With another asymptotic line of QNMs, such as the one under $\omega \to -\omega^*$, one recovers another line of bouncing singularities and a wedge convergence region.

Another way to see the wedge region of convergence is through the Fourier transform of \eqref{TPFG12wk} to get $G_R(t, \vec{k})$ in a slightly different way. Expanding the denominator into geometric series and integrating term by term we get
\begin{equation}\label{eq:manyb}
\begin{aligned}
    &G_{12}(t,k=0)=\sum_{n,m=0}^{\infty}
\frac{e^{\frac{2\pi i s}{r}}
\left(m e^{i(\phi-\theta)} - n e^{i(\theta-\phi)}\right)}
{\left(i(t-t_{nm})\right)^{2\Delta_\mathcal{O}-d+1}}
+ (t \to -t) \,, \\ &t_{nm}=i\frac{\beta}{2}+n \frac{2\pi}{r}e^{i\theta}-m \frac{2\pi}{r}e^{-i\theta} \,.
\end{aligned}
\end{equation}
As argued in \cite{Dodelson:2025jff}, both perturbative $\omega$ corrections and subleading corrections to the asymptotic behaviour of $\omega_n$ will contribute to subleading orders close to each singularity at $t_{nm}$. These singularities are precisely the bouncing singularities of black branes in $AdS_{d+1}$. As was the case in both the $2d$ correlator and the R-current correlator discussed above, in general we see that the Fourier transform is only convergent inside the fundamental strip $|\text{Im}(t)|<\beta/2$ as discussed in Figure \ref{fig:rcurrent_tplane}. Note that while at any finite $|t|$ the sum over QNMs converges inside the wedge region due to the asymptotic behaviour of QNMs, the behaviour as $|t|\to\infty$ inside this wedge is governed by the minimal angle QNM, and in particular, the correlator might only exponentially decay inside this smaller wedge.

Here we have used holographic examples to argue for the overlapping regime of convergence. Beyond holography, a piece of evidence for the overlap is given by the O($N$) model studied in Appendix \ref{app:ON}. There, $G_R(t,\vec{k})$ is an entire function of $t$ and both the OPE and QNM representations converge for all $t$.
Another example is the large $N$ SYK model where QNMs can be recovered from a small $t$ expansion \cite{Dodelson:2024atp}.\footnote{Note also that the convergence of the QNM expansion for the retarded bulk-to-bulk Green's function in asymptotically flat Schwarzschild is also set by lines of bouncing singularities \cite{Arnaudo:2026tcy}, a mechanism that appears to be universal across all examples discussed here. However, because of the bulk separation, the OPE singularity lies beyond the QNM radius of convergence in that example.}

In summary, the above evidence points to a common region of overlap in which the two expansions converge, where we can write the following expression as the basis of the OPE=QNM relation,
\begin{keyformula}[Mixed correlator overlap]
\be
t^{d-1-2\Delta_{\mathcal O}}\sum_{\widehat\Delta}b_{\widehat\Delta,k} \left(\frac{t}{\beta}\right)^{\widehat\Delta} = \sum_n r_n(k)e^{-i\omega_n(k)t}, \label{overlap}
\ee
\end{keyformula}
with only OPE data on the left hand side, and only QNM data on the right hand side. Here and throughout we will often omit the $\theta(t)$ factors by considering $t>0$ then analytically continuing the expressions to $t\in \mathbb{C}$. 

\subsection{A dispersive representation at fixed $k$}\label{sec:disp}
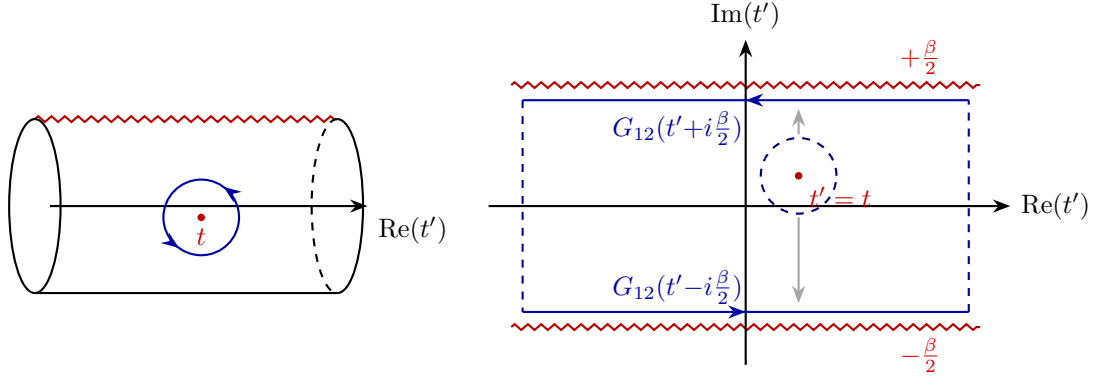
\begin{figure}[t]
\centering
\begin{tikzpicture}[>=Stealth, line width=0.8pt,
  cut/.style   ={orange!85!black, thick,
                 decorate, decoration={zigzag, amplitude=1.1pt, segment length=5pt}},
  cont/.style  ={blue!65!black, thick},
  pole/.style  ={red!75!black},
  defarr/.style={gray!70, ->, shorten >=1pt}]

\begin{scope}[shift={(0,0)}]
  \def\r{1.15}\def\L{4.0}
  \draw[cut,red!75!black] (0,\r) -- (\L,\r);
  \draw[black] (0,-\r) -- (\L,-\r);
  \draw[black] (\L,\r) arc (90:-90:0.34 and \r);
  \draw[black, dashed] (\L,\r) arc (90:270:0.34 and \r);
  \draw[black] (0,0) ellipse (0.34 and \r);
  \draw[->,black] (0.2,0) -- (\L+0.4,0) node[below right]{\small $\text{Re}(t')$};
  \coordinate (Tc) at (2.2,-0.15);
  \fill[pole] (Tc) circle (1.4pt);
  \node[pole, below] at (Tc) {\small $t$};
  \draw[cont, postaction={decorate, decoration={markings,
        mark=at position 0.15 with {\arrow{Stealth}},
        mark=at position 0.65 with {\arrow{Stealth}}}}]
        (Tc) circle (0.5);
\end{scope}

\begin{scope}[shift={(9.4,0)}]
  \def\W{3.1}\def\H{1.6}
  \draw[cut,red!75!black] (-\W,\H)  -- (\W,\H);
  \draw[cut,red!75!black] (-\W,-\H) -- (\W,-\H);
  \draw[->,black] (-\W-0.3,0) -- (\W+0.4,0) node[right]{\small $\text{Re}(t')$};
  \draw[->,black] (0,-\H-0.5) -- (0,\H+0.6) node[above]{\small $\text{Im}(t')$};
  \node[red!85!black, above right] at (\W-1.2,\H) {\small $+\tfrac{\beta}{2}$};
  \node[red!85!black, below right] at (\W-1.2,-\H){\small $-\tfrac{\beta}{2}$};
  
  \coordinate (T) at (0.7,0.4);
  \fill[pole] (T) circle (1.4pt);
  \node[pole, below right] at (T) {\small $t'=t$};
  \draw[blue!45!black, dashed] (T) circle (0.5);

  \draw[defarr] (0.7,0.95) -- (0.7,\H-0.28);
  \draw[defarr] (0.7,-0.15) -- (0.7,-\H+0.28);

  \draw[cont, postaction={decorate, decoration={markings,
        mark=at position 0.5 with {\arrow{Stealth}}}}]
        (\W-0.15,\H-0.2) -- (-\W+0.15,\H-0.2);
  \draw[cont, postaction={decorate, decoration={markings,
        mark=at position 0.5 with {\arrow{Stealth}}}}]
        (-\W+0.15,-\H+0.2) -- (\W-0.15,-\H+0.2);

  \draw[cont, dashed] (\W-0.15,-\H+0.2)  -- (\W-0.15,\H-0.2);
  \draw[cont, dashed] (-\W+0.15,\H-0.2)  -- (-\W+0.15,-\H+0.2);
 
  \node[blue!65!black, below] at (-0.9,\H-0.2) {\small $G_{12}(t'{+}i\frac{\beta}{2})$};
  \node[blue!65!black, below] at (-0.9,\H-0.2-2.1) {\small $G_{12}(t'{-}i\frac{\beta}{2})$};
\end{scope}

\end{tikzpicture}
\caption{Unfolding the Cauchy contour of \eqref{eq:disprel} from the thermal cylinder onto the strip. (left) On the cylinder the contour is a small loop around $t'=t$, enclosing the single image pole of the coth kernel in the first strip. (right) Cutting the cylinder along $\text{Im}(t)=\pm\beta/2$ and deforming the contour gives the dispersive representation for $G_{12}(t,\vec{k})$. The vertical segments at $\text{Re}(t')\to\pm\infty$ vanish by quasinormal-mode decay.}
\label{fig:unfold}
\end{figure}

In this section we obtain a dispersive representation of the fixed-$k$ two-sided (Wightman) correlator in the fundamental strip. The fact that the analytic continuation of the fixed $k$ correlators  $G_{12}(t, \vec{k})$, $G_{>}(t, \vec{k})$, or $G_{<}(t, \vec{k})$ in the strip are not periodic functions in imaginary time translates to them not being single valued on the cylinder, see Figure \ref{fig:unfold}.

Let us set $z=e^{\frac{2\pi}{\beta}t}$ and write down a Cauchy representation on the cylinder. It is convenient to work with the two-sided correlator so that $t=0$ is not a singular point:
\begin{equation}
\begin{aligned}
    G_{12}(t, \vec{k}) &= \frac{1}{2\pi i} \oint_{C_t} dz'\, \frac{G_{12}(t(z'), \vec{k})}{z'-e^{\frac{2\pi}{\beta}t}}= \frac{1}{2\pi i} \oint_{C_t} dt'\, \frac{\pi}{\beta} G_{12}(t', \vec{k}) \left(\coth\left(\frac{\pi(t'-t)}{\beta}\right)+1\right)= \\ &=\frac{1}{2\pi i} \oint_{C_t} dt'\, \frac{\pi}{\beta} \coth\left(\frac{\pi(t'-t)}{\beta}\right)G_{12}(t', \vec{k}) \,,
\end{aligned}
\end{equation}
where $C_t$ is a small contour around $z'=e^{\frac{2\pi}{\beta}t}$, see Figure (\ref{fig:unfold}), in the first step we have changed variable $z'=e^{{\frac{2\pi}{\beta}t'}}$, and in the second step we have used the assumption that $G(t', \vec{k})$ is analytic inside the contour. The Cauchy kernel can be obtained as a sum over images from the usual one from
\begin{equation}
    \frac{\pi}{\beta}\coth \frac{\pi x}{\beta}= \sum_{n\in\mathbb{Z}} \frac{1}{x+i\beta n} \,.\label{eq:coth}
\end{equation}
Now we can blow up the contour of integration and pick up the discontinuity along the cut in Figure \ref{fig:unfold}
\bea
    G_{12}(t, \vec{k})&=&\frac{1}{2\pi i} \int_\mathbb{R} dt'\, \frac{\pi}{\beta} \tanh\left(\frac{\pi(t'-t)}{\beta}\right) \text{disc}\,G_{12}(t'+i\frac{\beta}{2}, \vec{k}).
    \label{eq:disprel}
\eea
There are no contributions from the vertical lines at $\text{Re}(t)\to\pm\infty$ due to quasinormal mode decay. The kernel, hyperbolic tangent, has poles at $t'=t+i\beta(n+\frac{1}{2})$, hence as soon as $t=i\frac{\beta}{2}$ one simple pole crosses the contour. Picking up the residue produces a discontinuity given by $\text{disc} \, G_{12}(t'+i\frac{\beta}{2}, \vec{k})$, following the same mechanism we have discussed for the BTZ correlator. In particular, the representation \eqref{eq:disprel} makes it manifest that $G_{12}$ will remain analytic in the strip, since the kernel is, even if $\text{disc}\,G_{12}$ might not be. 

The dispersion relation \eqref{eq:disprel} is a convolution and can therefore naturally be written in momentum space. To this end consider
\bea 
&\displaystyle\int& dt\, e^{i\omega t}\text{disc}\,G_{12}(t+i\frac{\beta}{2}, \vec{k}) = -2\sinh \big(\frac{\beta\omega}{2}\big)G_{12}(\omega),\cr 
&\displaystyle\int& dt \, e^{i\omega t}\frac{\pi}{\beta}\tanh\big(\frac{\pi t}{\beta}\big) = \frac{i\pi}{\sinh(\frac{\beta\omega}{2})},\qquad \omega\neq 0.
\eea 
In particular, we see that the Fourier transform of the kernel introduces poles at the Matsubara frequencies $\omega_n = \frac{2\pi in}{\beta}$ for $n\in \mathbb{Z}$ which, however, are cancelled by the prefactor in the discontinuity and therefore we see that the representation \eqref{eq:disprel} consistently reproduce $G_{12}(t,k)=\int \frac{d\omega}{2\pi}e^{-i\omega t}G_{12}(\omega,k)$ in momentum space as expected. 

Moreover, let us consider splitting the integral over $\mathbb{R}$ into $(-\infty,-\delta)$, $(-\delta,\delta)$ and $(\delta,\infty)$ and set $t=t_p+i\frac{\beta}{2}$. The discontinuity is regular everywhere except $t'\to 0$ corresponding to the OPE singularity.  Near $t'=0$ the discontinuity is fixed by the OPE and is schematically of the form
\begin{equation}
\text{disc}\,G_{12}(t'+i\frac{\beta}{2}, \vec{k})\;\simeq\; t'^{-2\Delta_\mathcal{O}+d-1}\sum_{\Delta'} d_{\Delta'}\,\left(\frac{t'}{\beta}\right)^{\Delta'},
\label{eq:discOPE}
\end{equation}
with no analytic terms. The region $(-\delta,\delta)$ contains both the simple pole from the kernel at
$t'=t_{p}$ and the singularity of the discontinuity at $t'=0$. By the Sokhotski--Plemelj theorem,
\bea
\frac{1}{2\pi i}\int_{-\delta}^{\delta}dt'\frac{\text{disc}\,G_{12}(t'+i\frac{\beta}{2}, \vec{k})}{t'-t_{p}}
&=&\tfrac12 \text{disc}\,G_{12}(t_p+i\frac{\beta}{2}, \vec{k})
\cr
&+&\frac{1}{2\pi i}\,\mathrm{PV}\!\int_{-\delta}^{\delta}dt'
\frac{\text{disc}\,G_{12}(t'+i\frac{\beta}{2}, \vec{k})}{t'-t_{p}},
\label{eq:sokhotski}
\eea
where $\text{PV}$ refers to the principal value. This is the region which contributes non-analytic terms in the OPE. In particular, using 
\be 
\frac{1}{2\pi i}\,\mathrm{PV}\!\int_{-\delta}^{\delta}
dt'\frac{(t')^{\lambda}}{t'-t_{p}} = \left[\frac{1}{2i\sin(\pi \lambda)}-\frac{1}{2}\right]t_p^{\lambda}+\cdots,
\ee 
we see that non-analytic terms \eqref{eq:discOPE} give the following contribution to the OPE of $G_{12}(t'+i\frac{\beta}{2},k)$
\bea 
\frac{1}{2\pi i}\int_{-\delta}^{\delta}dt'\frac{\text{disc}\,G_{12}(t'+i\frac{\beta}{2}, \vec{k})}{t'-t_{p}} &=&\sum_{\Delta'} \frac{-d_{\Delta'}(\frac{t_p}{\beta})^{-2\Delta_\mathcal{O}+d-1+\Delta'}}{2i\sin(\pi(2\Delta_{\mathcal O}-d+1-\Delta'))}+\cdots,\label{eq:nonanalytic}  
\eea
up to analytic terms. Here we see how the inverse sine factor appears, consistent with the discontinuity of \eqref{eq:nonanalytic} reproducing \eqref{eq:discOPE} if we take the discontinuity.

In the remaining region we can expand the kernel inside the integral for small $t_p$ which leads to manifestly analytical terms
\begin{equation}
\frac{1}{2\pi i}\Big[\int_{-\infty}^{-\delta}dt'+\int_{\delta}^{\infty}dt'\Big]\,
\frac{\pi}{\beta}\coth\tfrac{\pi(t'-t_{p})}{\beta}\,\text{disc}\,G_{12}(t'+i\frac{\beta}{2}, \vec{k})=\sum_{m\ge0}a_m\,(\frac{t_{p}}{\beta})^m
\label{eq:outer}
\end{equation}
The $\delta$-dependence cancels between the middle and outer regions, since the full integral is
$\delta$-independent. We therefore end up with the expected expansion of $G_{12}(t+i\frac{\beta}{2})$ as $t\to0$:
\begin{equation}
G_{12}(t+i\frac{\beta}{2},\vec{k})\simeq t^{-(2\Delta_O-d+1)}\sum_{\Delta'}d'_{\Delta'}\,\left(\frac{t}{\beta}\right)^{\Delta'}
\;+\sum_{m\ge0}a_m\,\left(\frac{t}{\beta}\right)^m,
\label{eq:twotowers}
\end{equation}
with both analytical and non-analytical terms and $(d_{\Delta'},d'_{\Delta'})$ related by the sin factor \eqref{eq:nonanalytic}. This further makes it clear that the non-analytic terms in $G_{12}$ can be obtained from the OPE of the discontinuity while the analytic terms can not and require knowledge about the correlator for all values of $t$. Note that, crucially, being at fixed spatial momentum $\vec{k}$ allowed us to safely drop the contribution from the vertical segments at infinity.

\section{Analytic continuation of OPE and extracting QNMs}\label{sec:opeExtended}
In the frequency space correlator, 
\be
\widetilde{G}_R(\omega, \vec{k}) = \int G_R(t,\vec{k})e^{i\omega t} dt,
\ee
QNMs appear as poles in $\omega$, so if one could construct $\widetilde{G}_R(\omega, \vec{k})$ using OPE then QNMs can be read off from the resulting expression. The difficulty of course is that the Fourier transform of the OPE \eqref{introGRtk} requires evaluating the OPE sum outside its radius of convergence, and so analytic continuation is required. Nevertheless, for intuition, we can naively evaluate the Fourier transform term by term in the OPE sum,\footnote{See however \cite{Bzowski:2014qja} for a discussion of the subtleties of OPE in momentum space.}
\be
\widetilde{G}_R(\omega, \vec{k}) \sim  \sum_{\widehat\Delta} \beta^{-\widehat{\Delta}} b_{\widehat{\Delta},k} \Gamma(d+\widehat{\Delta}-2\Delta_\mathcal{O}) (-i\omega)^{-d-\widehat{\Delta}+2\Delta_\mathcal{O}}. \label{GRtildeasy}
\ee
The problem of reconstructing QNMs can then be stated as follows,
\begin{keyformula}[QNM from OPE]
\be
\begin{aligned}
r_n &= -i\,\text{res}\left[\widetilde{G}_R(\omega, \vec{k}),\, \omega = \omega_n\right]\\
\widetilde{G}_R(\omega, \vec{k}) &\sim \sum_{\widehat\Delta} \beta^{-\widehat{\Delta}} b_{\widehat{\Delta},k} \Gamma(d+\widehat{\Delta}-2\Delta_\mathcal{O}) (-i\omega)^{-d-\widehat{\Delta}+2\Delta_\mathcal{O}}. 
\end{aligned}\label{qnmfromope}
\ee
\end{keyformula}
Here it is to be understood that $\widetilde{G}_R(\omega, \vec{k})$ is to be reconstructed from the OPE data which appears in its asymptotic large $\omega$ expansion. This mirrors a similar expression we will derive in a later section for constructing OPE from QNM in Mellin space \eqref{OPEreconstruction}. There the roles of OPE and QNM are reversed, with terms in the OPE appearing as poles in Mellin while the QNMs appear as asymptotics. The reconstruction in \eqref{qnmfromope} makes obtaining general results about this direction more difficult. In particular, because of the $\Gamma$ factor, \eqref{GRtildeasy} is a divergent sum that receives non perturbative contributions, as explored in holographic examples in \cite{Afkhami-Jeddi:2025wra, Giombi:2026kdz}.

One way around this problem is to analytically continue the OPE beyond its radius of convergence. We now carry out this exercise in the case of the CFT dual to the Schwarzschild-AdS$_5$ black brane. We will start with the OPE, extend its domain of convergence using a Pad\'e approximant, and fit the resulting function to damped exponentials. In doing so we are able to extract accurate results for many QNM frequencies directly from OPE data alone.

Here the relevant OPE data appearing as in \eqref{introUV} correspond to the stress-tensor sector, with dimensions $\Delta = 4m$ with $m=0,1,2,\ldots$. These can be computed analytically using the method in \cite{Fitzpatrick:2019zqz}, with the first few given by
\begin{equation}
\begin{aligned}
&a_{0,0}=\mathcal{N},\qquad a_{4,2}=\mathcal{N}\frac{\pi ^4 \Delta_{\mathcal O} }{120},\qquad a_{8,0}=\mathcal{N}\frac{\pi ^8 \Delta_{\mathcal O}  (\Delta_{\mathcal O}  (\Delta_{\mathcal O}  (\Delta_{\mathcal O}  (7 \Delta_{\mathcal O} -45)+100)-80)+48)}{201600 (\Delta_{\mathcal O} -4) (\Delta_{\mathcal O} -3) (\Delta_{\mathcal O} -2)},\\
&a_{8,2}=\mathcal{N}\frac{\pi ^8 \Delta_{\mathcal O}  (\Delta_{\mathcal O} (\Delta_{\mathcal O}  (7 \Delta_{\mathcal O} -23)+22)+12)}{201600 (\Delta_{\mathcal O} -3) (\Delta_{\mathcal O} -2)},\qquad a_{8,4}=\mathcal{N}\frac{\pi ^8 \Delta_{\mathcal O}  (\Delta_{\mathcal O}  (7 \Delta_{\mathcal O} +6)+4)}{201600 (\Delta_{\mathcal O} -2)},
\end{aligned}
\end{equation}
where we keep an arbitrary normalisation factor $\mathcal{N}$. The first step is to translate this to the data $b_{\widehat\Delta,k}$ that appears in the OPE for $G_R(t,\vec{k})$ using \eqref{abmap}. We work at $k=0$, then the relevant coefficients are
\be
b_{\widehat{\Delta}, 0} = \sum_{J,\Delta} 2 \sin\left[\pi\left(\frac{\Delta}{2} - \Delta_{\mathcal{O}}\right)\right] \mathcal{I}_{\Delta, J, 0}\, a_{\Delta, J},
\ee
where recall $\mathcal{I}_{\Delta, J, \ell}$ correspond to the Fourier transform of the blocks.
From this formula we see that the double-trace OPE sector with $\Delta = 2\Delta_{\mathcal{O}} + 2n$ do not contribute, due to the sine prefactor. Thus we have only stress-tensors contributing, and hence
\be
b_{4m, 0} = -2\sin\left(\pi \Delta_{\mathcal{O}}\right)\sum_{p}  \mathcal{I}_{4m, 2p, 0}\, a_{4m, 2p},
\ee
giving the first few coefficients,
\bea
b_{0,0} &=& -\mathcal{N}\frac{2 \pi^{3/2} \Gamma(1 - \Delta_{\mathcal{O}})\, \sin(\pi \Delta_{\mathcal{O}})}
        {\Gamma\!\left(\frac{5}{2} - \Delta_{\mathcal{O}}\right)}, \label{b00} \\
b_{4,0} &=& \mathcal{N}\frac{\pi^{11/2} (-4 + \Delta_{\mathcal{O}})\, \Delta_{\mathcal{O}}\, \Gamma(2 - \Delta_{\mathcal{O}})\, \sin(\pi \Delta_{\mathcal{O}})}
       {20\, \Gamma\!\left(\frac{9}{2} - \Delta_{\mathcal{O}}\right)}, \\
b_{8,0} &=& -\mathcal{N}\frac{\pi^{21/2}     \left(7\Delta_{\mathcal O}^6-17\Delta_{\mathcal O}^5-340\Delta_{\mathcal O}^4+1100\Delta_{\mathcal O}^3+1008\Delta_{\mathcal O}^2+432\Delta_{\mathcal{O}}\right)}{100800\, \Gamma\left(\frac{13}{2} - \Delta_{\mathcal{O}}\right)\Gamma( \Delta_{\mathcal{O}}-1)}.\label{b80}
\eea

To extract QNMs accurately we will need more $b_{4m,0}$ terms than the first few. The process above can in principle be extended arbitrarily, however a more efficient way to obtain them is to work directly at fixed $k$ where they can be obtained from a simple bulk procedure. This was carried out in \cite{Arnaudo:2026der} (see also \cite{Afkhami-Jeddi:2025wra}), where the coefficients are given by
\begin{equation}
b_{4m,0} = \frac{\pi\,\left(2\pi\right)^{4m-2\Delta_{\mathcal O} +3}}{\cos(\pi\Delta_{\mathcal O})\Gamma(4m+4-2\Delta_{\mathcal O})}\,  c_m,
\end{equation}
and the first 70 $c_m$ can be found in the ancillary file attached to \cite{Arnaudo:2026der}, in the normalisation $\mathcal{N} = - \frac{(\Delta_\mathcal{O} - 2)^2(\Delta_\mathcal{O}-1)}{\pi^2 \cos(\pi\Delta_\mathcal{O})}$. 

\begin{figure}[h!]
\centering
    \includegraphics[width=0.9\textwidth]{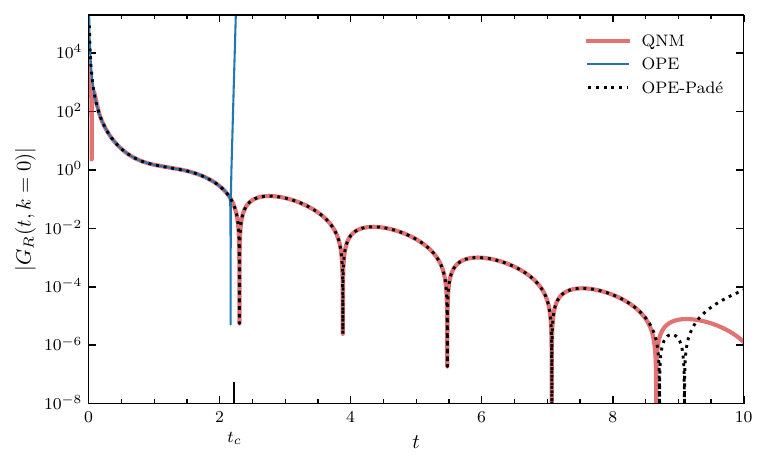}
    \caption{OPE=QNM in $d=4$ holography at $\Delta=11/4$ with $\beta = \pi$. 
    In blue we show the first 100 terms in the OPE for $G_R(t,k=0)$, whose convergence radius is $t_c$ indicated, as set by the nearest bouncing singularity.
    In red we show the QNM mode sum representation of $G_R(t,k=0)$ using the first 20 pairs of QNMs by residue sums, which converges for any $t>0$. 
    Thus in the common region of overlap there is a good agreement between OPE and QNM. In addition, we show the analytic continuation of the OPE sum using a diagonal Pad\'e approximant (black dotted), extending the overlap region beyond $t=t_c$.}
    \label{fig:pade}
\end{figure}

As a concrete example, we turn to $\Delta_\mathcal{O} = 11/4$. We compute the diagonal Pad\'e approximant of the first 100 terms in the OPE for $G_R(t,k=0)$. This extends the OPE beyond its radius of convergence set by the first bouncing singularity, $t_c = \frac{\beta}{\sqrt{2}}$. This is shown in Figure \ref{fig:pade}, shown alongside the QNM sum, demonstrating the agreement as a function of $t$.

To extract QNMs from the Pad\'e extended OPE, we then use Prony's method to fit to a sum of damped exponentials in some interval $[t_a, t_b]$. This method gives rise to several unphysical modes, but these are highly sensitive to the interval. Thus repeating the exercise for a second interval $[t_a', t_b']$ and keeping only those modes which don't change significantly reveals the correct QNM spectrum.  We use $[t_a, t_b] = [10^{-5}, 4]\times \beta/\pi$ and $[t_a', t_b'] = [10^{-3}, 6]\times \beta/\pi$ with 300 sample points for Prony's method. The results for the QNM frequencies are shown in Table \ref{table:numerics}, displaying excellent agreement with numerics.
\begin{table}[h!]
\begin{center}
\begin{tabular}{ |c||c|c|c| } 
\hline
mode & OPE analytic continuation & exact (numerics) & tail formula \eqref{eq:qnmTail}\\
\hline\hline 1,2 & $\pm 1.97041 - 1.51923 i$ & $\pm 1.97046 - 1.51917 i$ & $\pm 1.97115 - 1.52957 i$\\
\hline 3,4 & $\pm 3.97054 - 3.52892 i$ & $\pm 3.97070 - 3.52841 i$ & $\pm 3.97083 - 3.52943 i$\\
\hline 5,6 & $\pm 5.97156 - 5.52972 i$ & $\pm 5.97069 - 5.52908 i$ & $\pm 5.97075 - 5.52940 i$\\
\hline 7,8 & $\pm 7.97147 - 7.52734 i$ & $\pm 7.97068 - 7.52925 i$ & $\pm 7.97071 - 7.52939 i$\\
\hline 9,10 & $\pm 9.96701 - 9.52826 i$ & $\pm 9.97067 - 9.52930 i$ & $\pm 9.97069 - 9.52938 i$\\
\hline 19,20 & $\pm 20.0512 - 19.4878 i$ & $\pm 19.9707 - 19.5294 i$ & $\pm 19.9707 - 19.5294 i$\\
\hline 39,40 & $\pm 40.5110 - 39.0443 i$ & $\pm 39.9706 - 39.5294 i$ & $\pm 39.9706 - 39.5294 i$\\
\hline
\end{tabular}
\end{center}
\caption{QNM frequencies, $\omega_n$, extracted from the first 100 terms in the OPE after analytic continuation beyond $t = t_c$. Here we focus on $\Delta_\mathcal{O} = 11/4$ scalars in Schwarzschild-AdS$_5$ holography, in units where $\beta = \pi$. The fundamental mode is the most accurate, with quality of the method deteriorating deeper into the complex frequency plane. For comparison we also provide the result from an analytic asymptotic formula \eqref{eq:qnmTail}, which we obtain from OPE in Section \ref{sec:QNMtail}. This latter approach improves deeper into the complex frequency plane, but is already very accurate for the fundamental mode. We therefore conclude that by combining the methods in this section with those in Section \ref{sec:QNMtail}, we can use the OPE to obtain accurate predictions for any QNMs in the complex frequency plane.}
\label{table:numerics}
\end{table}

\section{QNM asymptotics from OPE}\label{sec:QNMasyfromOPE}
In this section we show how to constrain the asymptotic QNM spectrum exploiting the fundamental relation \eqref{overlap}. The basic observation is that a finite sum of exponentials is an entire function of time, therefore all singularities at finite $t$ must be reproduced by the asymptotic QNM spectrum. The OPE is sensitive to the singularities at $t=0$ and to the first complex singularity at $t=t_b$. Matching the QNM asymptotics against these singularities constrains the structure of the QNM spectrum.

We elucidate this in the example of the black brane correlator in $d=4$. Setting $k=0$ for simplicity and suppressing the explicit $k$ dependence, equation \eqref{overlap} becomes
\begin{equation}
    G_R(t)=t^{3-2\Delta_\mathcal{O}} \sum_{m\ge0} b_{4m} \left(\frac{t}{\beta}\right)^{4m} = \sum_n r_n e^{-i\omega_n t} \,.
\end{equation}
The behaviour close to the singularity at $t=0$ is controlled by the low lying OPE coefficients, which we derived earlier and present in \eqref{b00}-\eqref{b80}. In this section we work with normalisation
\be
\mathcal{N} = -\frac{\cot(\pi \Delta_\mathcal{O})\Gamma\left(\frac{5}{2}-\Delta_\mathcal{O}\right) \Gamma\left(2\Delta_\mathcal{O}-3\right)}{2\pi^\frac{5}{2} \Gamma(1-\Delta_\mathcal{O})},
\ee
to match the one used in \cite{Afkhami-Jeddi:2025wra}, so that
\begin{equation}
    b_0= \frac{1}{\pi} \Gamma\left(2\Delta_\mathcal{O}-3\right) \cos \pi \Delta_\mathcal{O} \,, \quad b_4= \frac{2\pi^3}{5} \left(\Delta_\mathcal{O}-4\right)_5 \,\,\Gamma\left(2\Delta_\mathcal{O}-7\right)  \cos \pi \Delta_\mathcal{O} \,, \quad \dots
\end{equation}
On the other hand the expansion close to the first singularity in the first quadrant, $t_b$, is controlled by the large $m$ behaviour of $b_{4m}$ \cite{Afkhami-Jeddi:2025wra}
\begin{equation}\label{eq:asyopech}
    b_{4m} \simeq \beta^{4m} t_b^{-4m} m^{2(\Delta_\mathcal{O}-2)} \left(B_0+B_1 m^{-1}+B_{4/3} m^{-4/3}+B_2 m^{-2}+B_{7/3} m^{-7/3}+\dots\right) \,,
\end{equation}
with
\begin{equation}
\begin{aligned}
    &t_b=\frac{\beta}{\sqrt{2}} e^{i \frac{\pi}{4}} \,, \quad B_0=\frac{1}{\pi}{4^{2\Delta_\mathcal{O}-3}} \,, \quad\frac{B_1}{B_0}=-\frac{\left(2\Delta_\mathcal{O}-4\right)\left(2\Delta_\mathcal{O}-3\right)}{8} \,, \\
    &B_{4/3} = 4^{2\Delta_\mathcal{O}-\frac{16}{3}}(2\pi)^{\frac{1}{3}} \frac{\pi(24+7\Delta_\mathcal{O}(\Delta_\mathcal{O}-4))\Gamma\left(\frac{2}{3}\right)}{2\times 18^{\frac{1}{3}}\Gamma\left(\frac{1}{6}\right)\Gamma\left(\frac{13}{6}\right)}  \,, \,\dots
\end{aligned}
\end{equation}
Summing over $m$ we get an expansion close to the first bouncing singularity 
\begin{equation}\label{eq:firstbounce}
\begin{aligned}
    G_R(t)\simeq& B_0\Gamma\left(2\Delta_\mathcal{O}-3\right)(4(t-t_b))^{-2\Delta_\mathcal{O}+3}+ \\&+B_{4/3}\Gamma\left(2\Delta_\mathcal{O}-\frac{13}{3}\right) (4(t-t_b))^{-2\Delta_\mathcal{O}+3+4/3} \,t_b^{-4/3} + \dots
\end{aligned}
\end{equation}

On the QNM side, let us assume we have two lines of asymptotically linearly spaced QNM as it is the case for the black brane:
\begin{equation}\label{eq:1lpa}
\begin{aligned}
&i \omega_n^+= r e^{i \theta} n + d_0 + d_{\delta_1} n^{-\delta_1}+\dots \\
&r_n^{+} = C n^\alpha \left(1+\ell_{\eta_1} n^{-\eta_1}+\ell_{\eta_2}n^{-\eta_2} +\dots \right)\\
&i \omega_n^-=\left(i \omega_n^+\right)^* \,, \quad r_n^- = \left(r_n^+\right)^* \,,
\end{aligned}
\end{equation}
where $0<\delta_2-\delta_1<1$ and similarly for $\eta_{1,2}$. The goal is to relate the coefficients appearing in the QNM ansatz to OPE data.

This QNM spectrum will produce singularities at
\begin{equation}\label{eq:predictedsing}
e^{-r e^{\pm i \theta}t^*} = 1 \Longrightarrow t^*_{m,\mp}= \frac{2 \pi i m}{r} e^{\mp i \theta} \,.
\end{equation}
The singularities accessible through OPE data correspond to $m=1$ (first bounce) and $m=0$ (OPE singularity). We now match the QNM expansion close to these singular points against the OPE predictions.

\subsection{QNM vs OPE singularity}\label{subsec:opevsqnmsing}
Let us start with the OPE singularity. Adding up the two lines and approximating the QNM sum with an integral we get
\begin{equation}
\begin{aligned}
    G_R(t) &=2\theta(t)\, \text{Re}\,\sum_n r_n^+ e^{-i\omega_n^+ t} \simeq 2\, \text{Re}\,\int dn\, r_n^+ e^{-i\omega_n^+ t} \simeq \\ &\simeq2\, \text{Re}\, \int d \left(i\omega_n^+\right) \frac{r_n^+}{d \left(i\omega_n^+\right)/dn} e^{-i\omega_n^+ t} \,,
\end{aligned}
\end{equation}
to be matched against  
\begin{equation}
G_R(t) = \theta(t)\,t^{3-2\Delta_\mathcal{O}}\left(\frac{1}{\pi} \Gamma\left(2\Delta_\mathcal{O}-3\right) \cos \pi \Delta_{\mathcal{O}}+ \mathcal{O}\left(t^4\right)\right) \,.
\end{equation}
Since 
\begin{equation}
    \cos \pi \Delta_\mathcal{O}=\text{Re}\, e^{-i\pi\Delta_{\mathcal{O}}} \,,
\end{equation}
it is tempting to just forget about the real part and get double the amount of constraints. Let us argue that this is in fact allowed. The retarded correlator is given by
\begin{equation}
    G_R(t)=-i \theta(t) \,\left( G_>(t) - G_<(t) \right) \,,
\end{equation}
which in momentum space reads
\begin{equation}
G_> (\omega) = i \frac{G_R(\omega)-G_A(\omega)}{1-e^{-\beta \omega}} \,, \quad  G_< (\omega) = e^{-\beta \omega} \, G_>(\omega)
\end{equation}
where $G_A$ is the advanced correlator. Closing to $\omega-$contour we get a QNM representation for $G_>(t)$ at fixed $k$. For $t>0$ we close the contour on the LHP and this only gets contribution from QNMs of $G_R$. Since $\beta \,\omega_n^\pm$ have respectively positive/negative real part, only the $\omega^+_n$ line will contribute to singular terms. Similarly, only the $\omega_n^-$ line will contribute to the non analytic terms in $G_<$. The $\cos \pi\Delta_\mathcal{O}$ is reconstructed by the difference of $G_{>/<}$ since
\begin{equation}
G_{>/<}(t) \sim (\pm i t)^{-2\Delta_{\mathcal{O}}+3} \,,
\end{equation}
therefore matching singular terms in $G_{>/<}(t)$ before taking the difference resolves the mixing of QNM lines, and allowing us to set 
\begin{equation}\label{eq:opeqnmg<>}
    2\int^{e^{i\pi/4}\infty} d \left(i\omega_n^+\right) \frac{r_n^+}{d \left(i\omega_n^+\right)/dn} e^{-i\omega_n^+ t} = t^{3-2\Delta_{\mathcal{O}}} \left(\frac{1}{\pi}\Gamma\left(2\Delta_\mathcal{O}-3\right)e^{-i\pi \Delta_\mathcal{O}}+\mathcal{O}(t^{4})\right) \,,
\end{equation}
where the equality is only understood at the level of singular terms. The lower extremum of  integration only affects analytic terms. The previous equation forces
\begin{equation}\label{eq:idd}
r_n^+\simeq C \left(i\omega_n^+ \right)^\alpha  \left(1+\mathcal{O}\left((\omega_n^+)^{-4}\right)\right) \,\,\frac{d}{dn}i \omega_n^+  \,, \quad \alpha=2\Delta_\mathcal{O}-4 \,,
\end{equation}
since 
\begin{equation}
    \int^{e^{i\pi/4}\infty} d \left(i\omega_n^+\right)  \left(i\omega_n^+\right)^\alpha e^{-i\omega_n^+ t} = \Gamma\left(\alpha+1\right)\, t^{-\alpha-1} + \text{analytic terms} \,.
\end{equation}
The relation \eqref{eq:idd} relates the residues to the QNM, and from \eqref{eq:opeqnmg<>} and \eqref{eq:idd} we get
\begin{equation}\label{eq:CDeltafirst}
C   =  \frac{1}{2\pi}e^{-i \pi \Delta_\mathcal{O}} \,.
\end{equation}
Assuming that the QNM sum correctly reproduces the OPE singularity has halved the number of unknowns. 

\subsection{QNM vs first bouncing singularity}
The next step is to impose that the QNM sum correctly reproduces the first bouncing singularity. Up to the relevant order the 2 lines ansatz reads
\begin{equation}\label{eq:1lpass}
\begin{aligned}
&i \omega_n^+= r e^{i \theta} n + d_0 + d_{\delta_1} n^{-\delta_1}+\dots \\
&r_n^{+}\simeq C r e^{i\theta} \left(r e^{i\theta}n\right)^\alpha \left(1+ \frac{\alpha d_0}{r e^{i\theta}} \frac{1}{n}+\left(\alpha-\delta_1\right) \frac{d_{\delta_1}}{r e^{i\theta}}n^{-\delta_1-1}+\dots\right)\\
&i \omega_n^-=\left(i \omega_n^+\right)^* \,, \quad r_n^- = \left(r_n^+\right)^* \,
\end{aligned}
\end{equation}
with $\alpha=2\Delta_\mathcal{O}-4$. Imposing $t_{1,-}=t_b$ in \eqref{eq:predictedsing} yields
\begin{equation}\label{coefficientsResults1}
r=\frac{2\pi}{|t_b|}=\frac{2\sqrt{2}\,\pi}{\beta} \,, \quad \theta=\frac{\pi}{4} \,,
\end{equation}
Note that contrarily to what happens at the OPE singularity, only the + line contributes to the first bounce. In fact since $-i \omega_n^- t_b \simeq -2\pi n$ this line gets exponentially suppressed and only contributes to analytic terms close to $t\sim t_b$. We also remark that a single asymptotic QNM line would be incompatible with the black brane OPE as it would force $\theta=\pi/2$.

To expand close to the first bounce we set $z=t_b-t$ where $r e^{i\theta}t_b=2\pi i$. Expanding the singular part of the plus line yields
\begin{equation}
    G_R(t)\simeq C e^{-d_0 t_b} \left(-z\right)^{-\alpha-1} \bigg( \Gamma(\alpha+1)-t_b d_{\delta_1} \Gamma\left(\alpha-\delta_1+1\right) \left(-re^{i\theta}z\right)^{\delta_1} + \dots  \bigg) \,.
\end{equation}
Matching against \eqref{eq:firstbounce} gives $\delta_1=4/3$ and
\begin{equation}\label{eq:firstbounceconstraintsnomel}
   C e^{-d_0t_b}=\frac{1}{\pi}\left(-1\right)^{\alpha+1} \,, \quad -t_b d_{4/3}=\frac{B_{4/3}}{B_0}\left(\frac{4}{-r e^{i\theta}t_b}\right)^{4/3} \,.
\end{equation}
Solving for $d_0,d_{\delta_1}$ gives (using the branch $(-1)=e^{-i\pi}$)
\begin{equation}\label{coefficientsResults3}
d_0
=
r e^{i\pi/4}
\left(
\frac{\Delta_\mathcal{O}-3}{2}
+
\frac{i\log2}{2\pi}
\right) \,, \quad d_{4/3}
= e^{-\frac{7\pi}{12}i}
\frac{\sqrt{2}}{\beta}
\frac{
\pi\left(24+7\Delta_\mathcal{O}(\Delta_\mathcal{O}-4)\right)
\Gamma\left(\frac{2}{3}\right)
}{
16\,18^{1/3}
\Gamma\left(\frac{1}{6}\right)
\Gamma\left(\frac{13}{6}\right)
} \,,
\end{equation}
in agreement with \eqref{TPFG12wkparams}. It is curious to note that as a function of $\Delta$, $d_{4/3}(\Delta_0)=0$ for $\Delta_0=\frac{2}{7} \left(7\pm\sqrt{7}\right)\approx \{1.24,2.76\}$ for which $i\omega_n^+=r^{i\theta}n'+\mathcal{O}((n')^{-8/3})$.

\subsection{Subleading orders}
\begin{figure}
    \centering
    \includegraphics[width=0.7\linewidth]{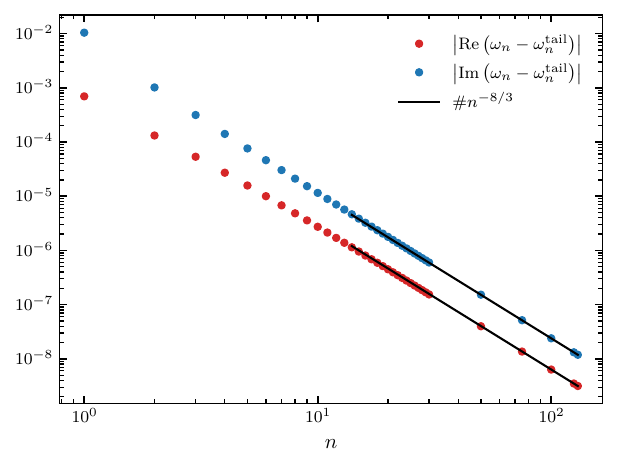}
    \caption{Residual error for the analytic tail formula for QNM frequencies in $d=4$ holography, derived from OPE. Here $\omega_n^\text{tail}$ refers to the expansion in \eqref{eq:qnmTail} up to and including $n^{-7/3}$, while the exact $\omega_n$ is obtained from numerics. Here we use $\Delta_\mathcal{O} = 11/4$ in units where $\beta = \pi$. The residual scales as $n^{-8/3}$ at large $n$ which is the expected next correction, confirming the result.}
    \label{fig:tailqnms}
\end{figure}

Let us comment on subleading orders in the QNM expansion and on the generic structure of the series. The next corrections to the OPE asymptotics in \eqref{eq:asyopech} read \cite{Afkhami-Jeddi:2025wra}
\begin{equation}
B_2= \frac{B_0}{96}(-2+\Delta_\mathcal{O})(-5+2\Delta_\mathcal{O})(-3+2\Delta_\mathcal{O})(-5+3\Delta_\mathcal{O}) \,, \quad
B_{7/3}=-\frac{B_{4/3}}{8} \left(\frac{80}{9}-14\Delta_\mathcal{O}+4\Delta_\mathcal{O}^2\right) \,.
\end{equation}
These corrections cancel out close to the first bounce, making \eqref{eq:firstbounce} valid up to $\mathcal{O}((t-t_b)^{7/3})$. Matching against the QNM expansion predicts the next exponent to arise at $\mathcal{O}(n^{-7/3})$, with 
\begin{equation}\label{eq:73qnm}
\begin{aligned}
&d_{7/3}=-\frac{4}{3} \frac{d_0 d_{4/3}}{r e^{i\pi/4}}  \\ &= -\frac{2}{3}\left(\Delta_{\mathcal{O}}-3+\frac{i\log 2}{\pi}\right) e^{-\frac{7\pi}{12}i}
\frac{\sqrt{2}}{\beta}
\frac{
\pi\left(24+7\Delta_\mathcal{O}(\Delta_\mathcal{O}-4)\right)
\Gamma\left(\frac{2}{3}\right)
}{
16\,18^{1/3}
\Gamma\left(\frac{1}{6}\right)
\Gamma\left(\frac{13}{6}\right)
} \,.
\end{aligned}
\end{equation}
This is new data for the QNM asymptotic. The first equality in \eqref{eq:73qnm} suggests that the large $n$ series rearranges in powers of $n'=n+e^{-i\theta}d_0/r$. In fact expanding at large $n$
\begin{equation}
\begin{aligned}
i\omega_n^+ &= r e^{i\theta} n'+d_{4/3} (n')^{-4/3} +\dots \\&= r e^{i\theta}n + d_0 + d_{4/3}n^{-4/3}-\frac{4}{3}\frac{d_0 d_{4/3}}{re^{i\theta}}n^{-7/3}+\dots
\end{aligned} \label{eq:qnmTail}
\end{equation}
consistently with our findings. We numerically confirm \eqref{eq:qnmTail}, including the newly derived asymptotic data, in Figure \ref{fig:tailqnms}.

Let us now investigate the generic form of the large $n'$ series. Equation \eqref{eq:idd} at all orders reads
\begin{equation}
r_n^+=C(i\omega_n^+)^{\alpha} \frac{d(i \omega_n^+)}{dn} g^{(0)}(i \omega_n^+) \,, \quad b_0\,g^{(0)}(i\omega_n^+) \equiv\sum_{m\ge0} b_{4m} \frac{\Gamma(\alpha+1)}{\Gamma(\alpha+1-4m)} (i \beta \omega_n^+)^{-4m} \,.
\end{equation}
Note that $g^{(0)}(\omega) \propto G_R(\omega)$ is a factorially divergent series. 

In order to get an all order relation, we can set again $z=t_b-t$. Using that $r e^{i\theta}t_b=2\pi i$ we have
\begin{equation}
e^{-i\omega_n^+ t} = e^{-(i \omega_n^+-r e^{i\theta}n)t_b} e^{i\omega_n^+ z} \,.
\end{equation}
Converting the residue sum to an integral we get
\begin{equation}\label{eq:integralexpb}
\sum_n r_n^+ e^{-i\omega_n^+ t} \simeq C \int d(i\omega_n^+)(i\omega_n^+)^\alpha g^{(0)}(i\omega_n^+) e^{-(i \omega_n^+-r e^{i\theta}n)t_b} e^{i\omega_n^+ z} \,.
\end{equation}
This expression can be now expanded at small $z$. Let us denote the expansion close to the first bounce as
\begin{equation}\label{eq:firstbounceall}
    G_R(t)\simeq \sum_mD_{4m/3} t_b^{-4m/3}\Gamma\left(2\Delta_\mathcal{O}-3-\frac{4m}{3}\right)(4(t-t_b))^{-2\Delta_\mathcal{O}+3+4m/3}
\end{equation}
and matching against the small $z$ expansion of the integral expression in \eqref{eq:integralexpb} we get
\begin{equation}
C g^{(0)}(i\omega_n^+) e^{-(i \omega_n^+-r e^{i\theta}n)t_b} = (-4)^{-2\Delta_\mathcal{O}+3} \sum_m D_{4m/3} \left( -\frac{4}{i\omega_n^+ t_b} \right)^{4m/3} \equiv g^{(1)}(i\omega_n^+) \,.
\end{equation}
All in all this gives
\begin{equation}\label{eq:allorder}
C e^{-(i \omega_n^+-r e^{i\theta}n)t_b} =\frac{g^{(1)}(i\omega_n^+)}{g^{(0)}(i\omega_n^+)}
\end{equation}
Since $g^{(1)}$ reconstructs the expansion close to the first bouncing singularity, we can think of it as the perturbative series attached to the first nonperturbative sector in the large $\omega$ expansion of $G_R(\omega)$. It is suggestive to view the previous formula as a way to determine QNM data from the interference between the first two nonperturbative sectors in the large frequency expansion.

Expanding now $n'=n'(\omega_n^+)$ at large argument we can solve order by order. Schematically one finds
\begin{equation}\label{eq:kappacoeff}
    i \omega_n^+-r e^{i\theta}n' = \sum_m \kappa_{4m/3} (i\omega_n)^{-4m/3}
\end{equation}
where the coefficients $\kappa_{4m/3}$ are determined by equation \eqref{eq:allorder}. Expanding $i \omega_n^+$ at large $n'$ we get
\begin{equation}\label{eq:higherordqnm}
\begin{aligned}
    i \omega_n^+ = re^{i\theta}n'+&d_{4/3}(n')^{-4/3}+d_{8/3}(n')^{-8/3}+\\ &\quad\quad+d_{11/3}(n')^{-11/3}+d_{4}(n')^{-4}+d_{5}(n')^{-5}+\dots
\end{aligned}
\end{equation}
Note that solving \eqref{eq:kappacoeff} generated new powers other than just $4m/3$. We also remark that we expect the large $n'$ expansion to be asymptotic series, since the RHS of \eqref{eq:kappacoeff} is. This is inherited directly from the asymptotic structure of $G_R(\omega)$.

All in all this provides a concrete map between OPE data and QNM asymptotics. We can summarize the procedure as follows:
\begin{itemize}
    \item compute a large number of OPE coefficients $b_{4m}$,
    \item from the large order behavior of $b_{4m}$ reconstruct the local expansion close to the first bouncing singularity,
    \item derive the coefficients $\kappa_{4m/3}$ in equation \eqref{eq:kappacoeff} and invert the relation to compute QNM asymptotics.
\end{itemize}
We report numerical values for the first coefficients appearing in \eqref{eq:higherordqnm} in Table \ref{table:numericsdasy}.

\begin{table}[t!]
\begin{center}
\begin{tabular}{ |c|c|c|c| } 
\hline $d_{8/3}/(r e^{i\theta})$ &$d_{11/3}/(r e^{i\theta})$ & $d_4/(r e^{i\theta})$ & $d_5/(r e^{i\theta})$  \\
\hline $-1.88005 \,\,e^{-\frac{i\pi}{6}}\times 10^{-3}$ & $-3.74111 \,\,e^{\frac{i\pi}{3}} \times 10^{-8}$ & $6.59557 \,i \times 10^{-7}$ & $1.25968 \times 10^{-6}$ \\ \hline
\end{tabular}
\end{center}
\caption{Higher order corrections to the large $n'$ expansion of black brane QNMs as in equation \eqref{eq:higherordqnm}. Here $\Delta_\mathcal{O}=11/4$.}
\label{table:numericsdasy}
\end{table}

In general spacetime dimension $d$ the holographic results takes the same form 
\be
i\omega_n = r_de^{i\theta_d}n'+ \frac{d_{\frac{d}{d-1}}}{(n')^{\frac{d}{d-1}}}+\cdots
\ee 
with $d_{\frac{d}{d-1}}$ given in \cite{Dodelson:2023vrw} up to $\mathcal{O}(n^{-\frac{d}{d-1}})$. Assuming that the expansion again organises in terms of $n'$, this would predict the $\mathcal{O}(n^{-\frac{d}{d-1}-1})$ correction. It would be interesting to verify this in general.

\begin{figure}[t]
    \centering
    \begin{subfigure}[t]{0.32\textwidth}
        \centering
        \includegraphics[width=\linewidth]{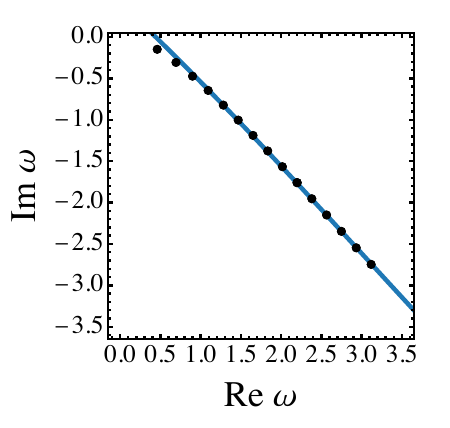}
        \caption{}
        \label{fig:P1}
    \end{subfigure}
    \hfill
    \begin{subfigure}[t]{0.32\textwidth}
        \centering
        \includegraphics[width=\linewidth]{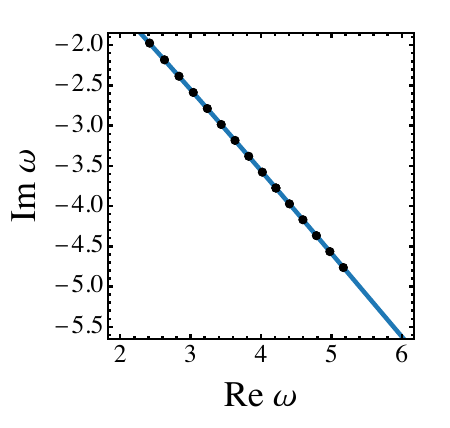}
        \caption{}
        \label{fig:P2}
    \end{subfigure}
    \hfill
    \begin{subfigure}[t]{0.32\textwidth}
        \centering
        \includegraphics[width=\linewidth]{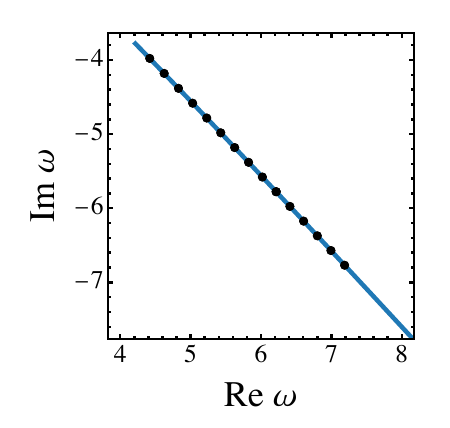}
        \caption{}
        \label{fig:P3}
    \end{subfigure}
    \caption{$\omega_{n}(\Delta)$ for $n=1$ (a), $n=2$ (b) and $n=3$ (c) comparing exact (numerics) and the asymptotic formula \eqref{eq:qnmTail} (blue line). Different dots correspond to uniformly distributed values of $\Delta\in(1,4)$.}
    \label{fig:three-column}
\end{figure}

\section{Mellin space and sum rules}\label{sec:mellin}

The Mellin transform of $G_R(t,\vec{k})$ provides a clean way to compare the two expansions appearing in the overlap \eqref{overlap}. While the Mellin transform is widely used for CFT correlators at zero temperature, to our knowledge it has not been explored extensively at finite temperature (see e.g.\ \cite{Alday:2020eua} for some work in this direction).

The Mellin transform of $G_R(t,\vec{k})$ and its inverse are defined as follows, 
\bea
\mathcal{M}(s, \vec{k}) &\equiv& \int_0^\infty dt\,t^{s-1}G_R(t,\vec{k}), \label{MellinCorrelatorCopy}\\
G_R(t, \vec{k}) &=& \int_{c-i\infty}^{c+i\infty}\frac{ds}{2\pi i}t^{-s}\mathcal{M}(s, \vec{k}),
\eea
where the constant $c$ must be chosen appropriately.
In \eqref{MellinCorrelatorCopy} $G_R(t,\vec{k})$ can either be evaluated using the OPE as in on left hand side of \eqref{overlap}, or the QNM expansion as on the right hand side of \eqref{overlap}. These different forms of the correlator give different information about the properties of $\mathcal{M}(s, \vec{k})$ in the $s$-space, in a way that is convenient for making identifications between the data. We further discuss the notion of  Mellin space transform in the bulk in Appendix \ref{app:bulkmellin}.

By evaluating \eqref{MellinCorrelatorCopy} on the QNM sum, one finds the following \emph{Mellin--QNM relation}
\be
\mathcal{M}(s, \vec{k}) = \Gamma(s) \sum_{n} r_n(i\omega_n)^{-s}.\label{MellinQNMcopy}
\ee
It will also sometimes be convenient to work with a version without the $\Gamma(s)$ factor,
\begin{equation}
\mathcal{Z}(s,\vec{k})\equiv\frac{\mathcal{M}(s,\vec{k})}{\Gamma(s)}=\sum_{n} r_n(i\omega_n)^{-s}.\label{calZdef}
\end{equation}
The poles of $\mathcal{M}(s,\vec{k})$ extend towards the left half-plane and determine the small-$t$ (OPE) expansion of $G_R(t,\vec{k})$. This OPE has a finite radius of convergence, $|t|<t_c$. For $|t|>t_c$ there has to be another contribution such that the result is finite, which is given by an arc contribution in the inverse Mellin transform. These considerations suggest 
\begin{equation}
\mathcal{M}(s+i \, \epsilon,\vec{k}) \sim t_c^{-|s|} \,, \quad s\to-\infty
\end{equation}
where the $i \epsilon$ shift is there to avoid the OPE poles on the real negative axis.

The opposite asymptotic direction is sensitive to the IR and explores the QNM spectrum. From the representation \eqref{calZdef}, each term in $\mathcal{Z}(s,\vec{k})$ is weighted by $(i\omega_n)^{-s}$ so that along the positive real $s$-axis it is controlled by the fundamental mode
\begin{equation}
    \mathcal{M}(s,\vec{k}) \simeq \Gamma(s) \sum_{n\in\text{fund}} r_n \left(i\omega_n\right)^{-s} \,,
\end{equation}
where the sum runs over modes with the minimal modulus. When the QNM representation converges the asymptotics on vertical lines is controlled by QNM angle from the real axis. In fact for each summand in the Mellin transform we find
\begin{equation}
\left|\Gamma(s)r_n(i\omega_n)^{-s}\right|\sim\sqrt{2\pi}|r_n||\text{Im}(s)|^{\text{Re}(s)-\frac{1}{2}}|\omega_n|^{-\text{Re}(s)}\exp\left(-\frac{\pi}{2}|\text{Im}(s)|+\arg(i\omega_n)\text{Im}(s)\right).
\end{equation}
In particular, the limit $\text{Im}(s)\to+\infty$ is controlled by the largest QNM angle, while $\text{Im}(s)\to-\infty$ is controlled by the smallest one. The vertical exponential growth of $\mathcal{M}(s,\vec{k})$ thus probes the extremal angular directions occupied by the QNM spectrum.

If $\mathcal{M}(s,\vec{k})$ has a pole at $s=s_*$ with residue $B_*$, then the inverse Mellin transform contains a term $B_* t^{-s_*}$. 
Comparing with a single OPE term from the left hand side of \eqref{overlap}, $\beta^{-\widehat{\Delta}}b_{\widehat\Delta,k} t^{d-1-2\Delta_{\mathcal{O}}+\widehat{\Delta}}$,
one obtains the following \emph{Mellin--OPE relation},
\begin{equation}
\begin{aligned}
\text{res}\left[\mathcal{M}(s,\vec{k}),s=s_*\right] &= \beta^{-\widehat{\Delta}} b_{\widehat\Delta,k},\\
s_* &= 2\Delta_{\mathcal{O}}-d+1-\widehat{\Delta}.
\end{aligned}\label{MellinOPE}
\end{equation}

Extracting OPE data from a QNM sum is the more straightforward of the two directions. Through Mellin space, we can formalise this relation in a very simple formula by combining the Mellin--OPE relation \eqref{MellinOPE} with the Mellin--QNM relation \eqref{MellinQNMcopy}; given an operator with dimension $\widehat{\Delta}$ in the OPE spectrum, its OPE coefficient $b_{\widehat\Delta,k}$ is given purely in terms of QNM data as follows,
\begin{keyformula}[OPE from QNM]
\be
b_{\widehat\Delta,k} = \beta^{\widehat{\Delta}}\; \text{res}\left[\Gamma(s) \sum_{n} r_n(i\omega_n)^{-s},\,s=2\Delta_{\mathcal{O}}-d+1-\widehat{\Delta}\right]. \label{OPEreconstruction}
\ee
\end{keyformula}

Using \eqref{OPEreconstruction} we can write down constraints on QNM data based on the OPE.
The poles of $\Gamma(s)$ in \eqref{OPEreconstruction} at $s=-q$ with $q\in \mathbb{Z}_{\geq 0}$ correspond to terms in the OPE of the form $b_{2\Delta_\mathcal{O} - d + 1 + q,k}t^q$. Consider theories in which any of these terms vanish. In such cases, the pole of $\Gamma(s)$ must be cancelled by a zero in the QNM sum in the residue formula, \eqref{OPEreconstruction}, so that $b_{2\Delta_\mathcal{O} - d + 1 + q,k} = 0$. Thus, we arrive at the following sum rules
\begin{keyformula}[Double trace sum rules]
\be
\sum_{n} r_n(i\omega_n)^{q} = 0, \label{sumrule}
\ee
\end{keyformula}
for every $q$ where $b_{2\Delta_\mathcal{O} - d + 1 + q,k} = 0$. Such terms correspond to double-trace contributions in holographic examples, which are all absent. Thus in such cases we have an infinite set of sum rules, \eqref{sumrule} for all $q\in \mathbb{Z}_{\geq 0}$ in generic $\Delta_\mathcal{O}$ cases, on holographic QNM data. 

Let us note that the sum rules \eqref{OPEreconstruction} and \eqref{sumrule} are in general divergent expressions (when there is an OPE singularity) that need to be regularised. We will see explicitly in Section \ref{sec:QNMtail} how this can be done using zeta regularisation to obtain non-analytic terms in the OPE. 

Finally, in holographic cases we have explored the idea of computing correlators directly in Mellin space, by working in Mellin space in the bulk, and this is presented in Appendix \ref{app:bulkmellin}.

\subsection{QNM asymptotics revisited}\label{sec:QNMtail}
Evidently, the relation \eqref{OPEreconstruction} is especially simple in cases where there are finitely many QNMs in the spectrum. Then, the sum in \eqref{OPEreconstruction} is an entire function of $s$, leaving the only poles at $s = -q$, $q\in \mathbb{Z}_{\geq 0}$. Thus OPE spectrum data is completely fixed as,
\bea
b_{\widehat{\Delta},k} &=& \beta^{\widehat\Delta} \frac{(-i)^q}{q!}\sum_{n=1}^{N}r_n\omega_n^q,\\
\widehat\Delta &=& 2\Delta_{\mathcal O}-d+1+q, \qquad q\in\mathbb{Z}_{\ge0}.
\eea
A consequence is that there is no OPE singularity for finitely many QNMs, with the leading term in the OPE $\sim t^0$. In time domain this is also clear: a finite sum of exponentials cannot produce a singularity. Thus, an OPE singularity implies the presence of infinitely many QNMs. Note that any singular contribution to the OPE is associated with a pole at $s>0$.

Following the same logic of Section \ref{sec:QNMasyfromOPE} we now show that the asymptotic behaviour of QNMs is set by the singular terms in the OPE. We sketch the procedure for the black brane correlator for simplicity and to have an immediate comparison with previous results. Again we assume two asymptotic lines
\begin{equation}
\begin{aligned}
&i \omega_n^+ = r e^{i\theta} n+d_0 +\dots \,,  \\ &r_n^+ = C \, n^\alpha \left(1+ \ell_1 n^{-1}+\dots\right) \,, \\
&i\omega_n^+ = (i\omega_n^-)^{*} \,, \quad r_n^+ = (r_n^-)^{*} \,,
\end{aligned}
\end{equation}
where by matching with the first bouncing singularity location we can read off again $(r,\theta)$ as in \eqref{coefficientsResults1}. Summing up the large $n$ tail gives
\begin{equation}
\sum_n r_n(i\omega_n)^{-s}
= C \, r^{-s}e^{ i\pi s/4}
\sum_{j\ge 0} A_j(s)\zeta(s-\alpha+j) \,,
\end{equation}
where for example $A_0(s)=1$, $A_1(s)=\ell_1-s a$ and so on. The crucial point is that only finitely many terms in the large $n$ asymptotics of QNMs and residues will contribute to the first poles. In particular, only the leading asymptotics will contribute to the identity exchange. We can then truncate the asymptotic expansion and match with the low lying terms in the OPE. The first Mellin pole will appear at
\begin{equation}
    s^*=1+\alpha=2\Delta_\mathcal{O}-3 \Longrightarrow \alpha=2\Delta_\mathcal{O}-4 \,,
\end{equation}
consistently with previous findings. From \eqref{OPEreconstruction} we get 
\begin{equation}
\begin{aligned}
 &\Gamma(\alpha+1)r^{-(\alpha+1)} 2 \,\text{Re}\,\left( C e^{ i\theta(\alpha+1)} \right) = \frac{1}{\pi}\Gamma\left(2\Delta_\mathcal{O}-3\right) \text{Re} \left( e^{-i\pi \Delta_\mathcal{O}}\right) \,,\\
&2 \,r^{-\alpha}  \Gamma(\alpha)\,\text{Re}\,\left(
 C e^{\sigma i\pi\alpha/4}
\left(\ell_1-\alpha d_0\right) \right) =0\,.
\end{aligned}
\end{equation}
Resolving the two lines matching before taking the discontinuity as we did in Section \ref{subsec:opevsqnmsing} we get back the same constraints for $C$ (see equation \eqref{eq:CDeltafirst}) and the same relation between residues and poles as in equation \eqref{eq:idd} at this order.

In order to recover the constraints coming from the first bouncing singularity in Mellin space it is convenient to perform the Mellin transform with respect to the variable $z=t_b-t$, where $t_b$ is the location of the first bouncing singularity. One has 
\begin{equation}
    G_R(t_b-z)= \sum_n r_n^+ e^{-i\omega_n^+ t_b} e^{i \omega_n^+ z} +  \sum_n r_n^- e^{-i\omega_n^- t_b} e^{i \omega_n^- z} \,.
\end{equation}
This effectively rescales residues by $e^{-i\omega_n^\pm t_b}$. Choosing an appropriate contour, the corresponding Mellin transform reads (we denote by $u$ the Mellin variable conjugate to $z$)
\begin{equation}
    \frac{\tilde{\mathcal{M}}(u)}{\Gamma(u)}=\sum_n r_n^+ e^{-i\omega_n^+ t_b} (-i \omega_n^+)^{-u}+\sum_n r_n^- e^{-i\omega_n^- t_b} (-i \omega_n^-)^{-u} \,.
\end{equation}
Now since
\begin{equation}
    e^{-i\omega_n^+ t_b} \simeq 1 \,, \quad e^{-i \omega_n^- t_b} \simeq e^{-2\pi n} \,,
\end{equation}
the residues in the minus-line come with an exponential suppression. This makes the sum convergent and hence unable to generate singularities: as we have found in section \ref{subsec:opevsqnmsing}, only the plus-line can generate singularities close to $t\sim t_b$. Comparing against \eqref{eq:firstbounce} we expect the first poles at 
\begin{equation}
    u^*= \bigg\{ 2\Delta_\mathcal{O}-3, 2\Delta_\mathcal{O}-3-\frac{4}{3} \,, \dots \bigg\}
\end{equation}
with residues
\begin{equation}
\begin{aligned}
    &(-4)^{-2\Delta_\mathcal{O}+3} \Gamma\left(2\Delta_\mathcal{O}-3\right)B_0=\text{res}\left[ \Gamma(u) \sum_n r_n^+ e^{i\omega_n^+ t_b} (-i\omega_n^+)^{-u},u=2\Delta_\mathcal{O}-3\right] \,, \\
    &(-4)^{-2\Delta_\mathcal{O}+13/3}\Gamma\left(2\Delta_\mathcal{O}-\frac{13}{3}\right) B_{4/3} = \text{res}\left[ \Gamma(u) \sum_n r_n^+ e^{i\omega_n^+ t_b} (-i\omega_n^+)^{-u},u=2\Delta_\mathcal{O}-13/3 \right] \,.
\end{aligned}
\end{equation}
Here by $\zeta-$regulating the sums we obtain again the constraints in equation \eqref{eq:firstbounceconstraintsnomel}. 
 
\subsection{An exact solution resembling $d=4$ holography}
As discussed, holographic examples must solve all sum rules \eqref{sumrule} when $\Delta_\mathcal{O}$ is generic. In this section, we consider a toy model where we solve the double-trace sum rules \eqref{sumrule}. Let us first try to solve the sum rules by keeping only the leading asymptotic behaviour of QNMs as the exact spectrum,
\begin{equation}\label{asymptoticModel}
\omega_{n}^{+}=r e^{-\frac{\pi}{4}i} n,\qquad \omega_{n}^{-}=r e^{-\frac{3\pi}{4}i} n,\qquad r_{n}^{\pm}=\mathcal R_\pm n^{2\Delta_\mathcal{O}-d}.
\end{equation}
For this spectrum, the left hand side of the $q$th sum rule then reads
\begin{equation}
\sum_{n} r_n(i\omega_n)^{q} = r^{q}\left(\mathcal R_+e^{iq\frac{\pi}{4}}+\mathcal R_-e^{-iq\frac{\pi}{4}}\right)\zeta(-q-2\Delta_\mathcal{O}+d).
\end{equation}
The sum rule at $q=0$ would require $\mathcal{R}_+ = -\mathcal{R}_-$, giving a $\sin(\pi q/4)$ factor in the remaining cases, and thus the remaining sum rules are not solved by this exact QNM spectrum. Thus the leading QNM asymptotics alone are not sufficient.

We can however build completions of the QNM asymptotics of $d=4$ holography such that all sum rules are obeyed by adding $1/n$ corrections. Of course one such completion is the exact holographic result itself. However, there is another completion which is simple and bears much similarity to holography,
\bea
\omega_{n}^{+} &=& e^{-\frac{\pi}{4}i}(2\Delta_\mathcal{O}-3 + 2n),\qquad r_{n}^{\pm} = e^{\pm\frac{(2\Delta_\mathcal{O}-3) \pi}{4}i} \frac{(2\Delta_\mathcal{O}-3)_n}{n!},\label{eq:toyQNM}
\eea
with $\omega_{n}^{-}=-(\omega_{n}^{+})^*$, $n \in \mathbb{Z}_{\geq 0}$ and where we pick $r=2$ for convenience, corresponding to an inverse temperature of $\beta = \sqrt{2}\pi$. Summing this QNM spectrum we find,
\be \label{almostholography}
G_R(t) = \theta(t) \left(\frac{e^{\frac{(2\Delta_\mathcal{O}-3) \pi}{4}i}}{(2 \sinh(e^{\frac{\pi}{4}i}t))^{2\Delta_\mathcal{O}-3}} + \frac{e^{-\frac{(2\Delta_\mathcal{O}-3)\pi}{4}i}}{(2 \sinh(e^{-\frac{\pi}{4}i}t))^{2\Delta_\mathcal{O}-3}}\right),
\ee
whose OPE is
\be
G_R(t) \sim t^{3-2\Delta_\mathcal{O}} \times (\text{power series in $t^4$}).
\ee
Thus this QNM spectrum solves the double-trace sum rules \eqref{sumrule}, and the resemblance to $d=4$ holography is as follows:
\begin{itemize}
    \item Two lower-half plane asymptotic lines of QNMs at 45 degree angle which are linearly spaced in $n$, and associated residues that asymptote to $r_n \sim n^{2\Delta_\mathcal{O}-4}$.
    \item An OPE spectrum $\widehat\Delta = 4m$ with $m\in \mathbb{Z}_{\geq 0}$ corresponding to the correct identity contribution at generic $\Delta_\mathcal{O}$ and stress-tensor exchanges.
    \item $G_R(t)$ has two 45 degree lines of bouncing singularities at $t = i m \pi \, e^{\pm\frac{\pi}{4}i}$. 
\end{itemize}
Since this example appears similar but not identical to $d=4$ holography, it begs an `OPE=QNM bootstrap' question as to whether it is an allowed point in the space of theories. Indeed, the OPE spectrum of operators is the same as holography, and the QNM spectrum is causal and stable. Is there another physical criterion that this model fails to meet? 

A first question one could ask is whether the above spectrum can be reproduced by a holographic theory. The answer is negative, since the model is not consistent with the thermal product formula \cite{Dodelson:2023vrw}.
Indeed, if one takes only its pole locations and inserts them in the thermal product formula, the large $n$-scaling of the residues, which is completely fixed by the frequencies only, does not agree with the one of the model. Indeed, assuming the thermal product formula with the modes \eqref{eq:toyQNM} would lead to residues
\begin{equation}
\tilde{r}_{n,+}=-\frac{i\,G_{12}(0)\,\sinh\left(\frac{\beta\omega_n}{2}\right)\omega_n}{\left(1-\frac{\omega_n^2}{(\omega_n^*)^2}\right)\prod_{m\neq n}\left(1-\frac{\omega_n^2}{\omega_m^2}\right)\prod_m\left(1-\frac{\omega_n^2}{(\omega_m^*)^2}\right)},
\end{equation}
which can be checked to be different from \eqref{asymptoticModel}.

Even though the above spectrum and residues cannot arise from a holographic thermal correlator, can it still represent a sensible CFT? Looking closer, and assuming the correlator \eqref{almostholography} describes $k=0$, the first two OPE coefficients read off from \eqref{almostholography} are,
\bea
b_{0,0} &=& 2^{4-2\Delta_\mathcal{O}},\cr
b_{4,0} &=& - \frac{2^{1-2\Delta_\mathcal{O}}(2\Delta_\mathcal{O}-3)(10\Delta_\mathcal{O}-13)}{45}\beta^4.
\eea
We can then invert \eqref{abmap} to obtain the Euclidean OPE coefficients, 
\bea
a_{0,0} &=& -\frac{2^{3-2\Delta_\mathcal{O}}\csc(\pi \Delta_\mathcal{O}) \Gamma\left(\frac{5}{2}-\Delta_\mathcal{O}\right)}{\pi^\frac{3}{2}\Gamma(1-\Delta_\mathcal{O})},\cr
a_{4,2} &=& \frac{4^{-\Delta_\mathcal{O}}\left(-39 + 4(14-5\Delta_\mathcal{O})\Delta_\mathcal{O}\right)\csc(\pi \Delta_\mathcal{O}) \Gamma\left(\frac{9}{2}-\Delta_\mathcal{O}\right)}{135 \pi^\frac{3}{2}(\Delta_\mathcal{O}-4)\Gamma(2-\Delta_\mathcal{O})}\beta^4.
\eea
Thus, we find the following first stress-tensor behaviour normalised by the identity and temperature,
\be
\frac{a_{4,2}}{a_{0,0}\beta^4} = -\frac{(2\Delta_\mathcal{O}-7)(2\Delta_\mathcal{O}-5)(39 + 4\Delta_\mathcal{O}(5\Delta_\mathcal{O}-14))}{4320(\Delta_\mathcal{O}-4)(\Delta_\mathcal{O}-1)}.
\ee
However, in physical theories Ward identities fix the stress-tensor thermal coefficient to be linear in $\Delta_\mathcal{O}$ \cite{Iliesiu:2018fao}, which is not the case here. Thus the model considered in this subsection is ultimately ruled out on these grounds.

\subsection{Finitely many QNMs}
In this section we consider the case where there are finitely many QNMs in the spectrum. In this case one cannot satisfy the sum rules \eqref{sumrule} for all $q \in \mathbb{Z}_{\geq 0}$, since then the only solution is the trivial one, with $r_n = 0$ for all $n$.

Instead, motivated by holographic examples, suppose that the OPE spectrum takes the form
\begin{equation}
\widehat\Delta_m=hm, 
\qquad
m\in\mathbb{Z}_{\ge0}, \label{equalspacedOPEansatz}
\end{equation}
for some fixed integer spacing $h\in\mathbb{Z}_{>0}$.
Then from \eqref{MellinOPE} the Mellin poles are located at
\begin{equation}
s_*=s_0-hm,
\qquad
s_0=2\Delta_{\mathcal O}-d+1.
\end{equation}
Therefore, for this spectrum of poles to be compatible with finitely many QNMs, $s_0$ must be an integer. Then, a pole of the $\Gamma$ function at $s=-q$ is allowed only if
\begin{equation}
q\equiv -s_0 \pmod{h},
\end{equation}
and so at the other values of $q$ we have sum rules to ensure that the poles of $\Gamma(s)$ must cancel. Let us denote with $q_0$ the congruence class of $-s_0$. Then we have the following sum rules,
\begin{equation}
\sum_{n=1}^N r_n \omega_n^q = 0,\qquad q\not\equiv q_0 \pmod{h}
\end{equation}

These sum rules strongly constrain the finite set of frequencies. We group the QNMs according to the value of their $h$th power (recall that $h$ is a fixed positive integer for this discussion, corresponding to the spacing of the OPE spectrum). For each value $x$, define
\begin{equation}
\mathcal{B}_x=\{n:\omega_n^h=x\}.
\end{equation}
Writing $q=q'+hm$, the sum rules become
\begin{equation}
\sum_x x^m
\left(
\sum_{n\in\mathcal{B}_x}r_n\omega_n^{q'}
\right)=0,
\qquad
m\in\mathbb{Z}_{\ge0},
\qquad
q'\not\equiv q_0 \pmod{h}
\end{equation}
Since there are only finitely many distinct values of $x$, one can write the above system of equations as a Vandermonde matrix, and conclude that each coefficient must vanish separately:
\begin{equation}
\sum_{n\in\mathcal{B}_x}r_n\omega_n^{q'}=0, \qquad q'\not\equiv q_0 \pmod{h}.\label{blockconstraints}
\end{equation}

A nontrivial block therefore consists of frequencies with the same $h$th power. Let us label each block by $j$ and within each block we label each QNM mode by $\ell$, so that the QNM frequencies are given by
\begin{equation}
\omega_{j,\ell}=(\zeta_h)^\ell\Omega_j,
\qquad
\ell=0,1,\ldots,h-1,
\qquad
\zeta_h=e^{2\pi i/h}, \label{constrained_frequencies}
\end{equation}
and the residues correspondingly by $r_{j,\ell}$.
Then the block constraints \eqref{blockconstraints} become
\begin{equation}
\sum_{\ell=0}^{h-1}r_{j,\ell}(\zeta_h)^{\ell q}=0,
\qquad
q\not\equiv q_0 \pmod{h}.
\end{equation}
The only nontrivial solution is of the form
\begin{equation}
r_{j,\ell}=c_j\,(\zeta_h)^{-q_0\ell}. \label{constrained_residues}
\end{equation}
As a check of these findings, we can substitute the frequencies \eqref{constrained_frequencies} and residues \eqref{constrained_residues} in the Mellin transform \eqref{MellinQNMcopy}, to get
\begin{equation}
\mathcal{M}(s, \vec{k})=\Gamma(s)\sum_j c_j (i\Omega_j)^{-s}\sum_{\ell=0}^{h-1}\exp\left[-\frac{2\pi i\ell}{h}(q_0+s)\right],
\end{equation}
and indeed, the finite sum over $\ell$ vanishes whenever $s+q_0 \not\equiv 0 \pmod{h}$, cancelling poles of $\Gamma(s)$, as required.

This construction shows why finite QNM spectra are highly constrained by the structure of the OPE.
Each nontrivial block contains the frequencies,
\begin{equation}
\omega_{j,\ell}=(\zeta_h)^\ell\Omega_j, \qquad \ell=0,1,\ldots,h-1.
\end{equation}
If for example, $h>2$, then these frequencies appear in both the lower and upper half $\omega$-planes. At real values of spatial momentum $\vec{k}$ this is incompatible with causality, since causality implies that $\widetilde{G}_R(\omega, \vec{k})$ is analytic in $\text{Im}(\omega) > |\text{Im}(k)|$. Thus, OPE spectra of the form \eqref{equalspacedOPEansatz} with $h>2$ cannot generally be reproduced by finitely many QNMs. Notably, this includes examples such as scalar correlators holographically described by Schwarzschild-AdS$_5$, where the OPE spectrum takes the form \eqref{equalspacedOPEansatz} with $h=4$. Examples with finitely many QNMs are the large-$N$ O($N$) model discussed in Appendix \ref{app:ON} which has $h = 2$. In the case $h=1$ there are no sum rules.

\subsection{Rigidity of the QNM spectrum}
Assume that one starts from a QNM spectrum obeying the sum rules \eqref{sumrule} for all $q \in \mathbb{Z}_{\geq 0}$, and then changes only finitely many $\omega_n$ and $r_n$. Let the difference between the new and old spectral zeta functions be
\bea
\Delta \mathcal{Z}(s,\vec{k}) &=& \sum_{n \in Q'} r'_n (i \omega'_n)^{-s} - \sum_{n \in Q} r_n (i \omega_n)^{-s}=\sum_{a=1}^{A} c_a\mu_a^{-s},
\eea
where $Q$ denotes the set of indices associated with QNMs in the original spectrum that get modified, $Q'$ denotes the set of indices associated with QNMs in the new spectrum that were not in the original one, and when passing to the second line we denote with $A$ the cardinality $|Q|+|Q'|$, and we relabel the indices with $a=1,\dots,A$, the residues $r'_n$ and $-r_n$ with $c_a$, and the frequencies $i\omega'_n$ and $i\omega_n$ with $\mu_a$.
In particular, the complex numbers $\mu_a$ are distinct and collect all old and new frequencies appearing in the finite modification. The deformed spectrum obeys the sum rules only if
\begin{equation}
\Delta\mathcal{Z}(-q,\vec{k})=\sum_{a=1}^{A} c_a\mu_a^q=0,\qquad q\in\mathbb{Z}_{\ge 0}. \label{deltaZsumrules}
\end{equation}
This set of equations gives a finite Vandermonde system. Its determinant is
\begin{equation}
\det V=\prod_{1\leq a<b\leq A}(\mu_b-\mu_a).
\end{equation}
Since the $\mu_a$ are distinct, this determinant is non-zero. Hence from \eqref{deltaZsumrules} the only way for the new spectrum to satisfy the sum rules would be if $c_a=0$ for all $a=1,\dots,A$.

In particular, one cannot move the position of a single QNM, or of any finite number of QNMs, while preserving zeros of the terms $t^q$ in the OPE. Such a deformation necessarily violates at least one of the OPE sum rules.
Equivalently, a finite exponential polynomial cannot have all non-negative integer moments equal to zero unless it is identically zero. Given this rigidity, it is interesting to consider the possibility of writing a (subtracted) dispersive representation for $\omega_n=\omega(n)$ that lets us analytically continue in $n$.

\subsection{Asymptotic subtraction for double trace sum rules}\label{sec:DTSumRules}
We now investigate how one can deal with the constraints \eqref{sumrule} in the case of the zero-momentum black brane correlator in $d=4$. 
First of all, we can test how well the asymptotic tail \eqref{eq:qnmTail} satisfies the double-trace sum rules if it is promoted to an exact spectrum.

Using the reality relation between the two towers of QNMs, 
\begin{equation}
\omega_{n}^{-}=-(\omega_{n}^{+})^{*},\qquad r_{n}^{-}=(r_{n}^{+})^{*},
\end{equation}
the sum rules 
\begin{equation}
\sum_{n>0}\sum_{\varepsilon=\pm}r_{n}^{\varepsilon}\left(i\omega_{n}^{\varepsilon}\right)^{q}=0
\end{equation}
can be rewritten as
\begin{equation}\label{sumrulesrewritten}
2\,\text{Re}\sum_{n>0}r_{n}^{+}\left(i\omega_{n}^{+}\right)^q=0,
\end{equation}
which depend on divergent series.

We start from
\begin{equation}
i\omega_{n}^{+}=r e^{i\theta}n+d_0+d_{4/3}n^{-4/3}-\frac{4}{3}\frac{d_0d_{4/3}}{r e^{i\theta}}n^{-7/3}+\dots,
\end{equation}
and use the corresponding residues
\begin{equation}
r_n^+=C\,\frac{d}{dn}\frac{i\omega_{n}^{+}}{r e^{i\theta}}\left(\frac{i\omega_{n}^{+}}{r e^{i\theta}}\right)^{2\Delta_{\mathcal O}-4}.
\end{equation}
We then compute the moments on the left-hand side of \eqref{sumrulesrewritten} by zeta regularisation.
More precisely, the sums are evaluated by expanding each summand $r_{n}^{+}(i\omega_{n}^{+})^q$ in the truncated expansion at large $n$, and assigning to every term the corresponding zeta-regularised value. To compare different values of $q$ in the same units, we report the moments obtained by replacing $i\omega_{n}^{+}$ with $i\beta\omega_{n}^{+}$.

By truncating at different orders, we find the following residuals of the first three sum rules,
\begin{equation}
\begin{array}{c|c|c|c}
q &
\omega_n^+ \text{ truncated at } d_0 &
\omega_n^+ \text{ truncated at } d_{4/3}n^{-4/3} &
\omega_n^+ \text{ truncated at } n^{-7/3}
\\ \hline
0 & -5.60218\times10^{-2} & -6.96580\times10^{-2} & -7.84542\times10^{-2}\\
1 & \phantom{-}3.59408\times10^{-1} & \phantom{-}3.40216\times10^{-1} & \phantom{-}3.07329\times10^{-1}\\
2 & -4.10174\times10^{-1} & -3.09801\times10^{-1} & -2.60329\times10^{-1}
\end{array}
\end{equation}

Although these numbers are relatively small, they are not zero. This shows that a finite asymptotic truncation of the QNM tail does not by itself solve the double-trace sum rules. We therefore turn to a more systematic way of assigning a finite meaning to the divergent moments in \eqref{sumrulesrewritten}, separating the exact low-lying QNMs from the asymptotic tail.
The strategy is inspired by the Mellin analysis of harmonic sums developed in \cite{FLAJOLET19953}. In their terminology, the QNM representation is a harmonic sum with amplitudes $r_n$, frequencies $i\omega_n$, and base function $g(x)=e^{-x}$, whose Mellin transform factorises as $\mathcal M(s)=\Gamma(s)\mathcal{Z}(s)$. This suggests a concrete way to assign a meaning to the formally divergent moments in \eqref{sumrule}: we shall subtract from the QNM summand the large-$n$ terms responsible for the divergence, sum the resulting convergent remainder, and restore the subtracted power-law contributions through their meromorphic continuation.
Sum rules of QNMs were explored also in \cite{Dodelson:2023vrw}. We first formulate this subtraction as an analytic continuation of $\mathcal{Z}(s)$, where the subtraction terms are fixed by the large-$n$ QNM data obtained in Section \ref{sec:QNMasyfromOPE}. We then implement it using a finite set of QNM data and test whether the resulting subtracted moments approach the predicted value $\mathcal{Z}(-q)=0$ as the cutoff in the QNM sum is increased.

The results of Section \ref{sec:QNMasyfromOPE} allow us to write the QNM data expansion at large-$n$ and for every non-negative integer $q$ as
\begin{equation}\label{QNMdataexpansionfull}
r_{n}^{+}(i\omega_{n}^{+})^q\sim A_0^{(q)}\left(n+e^{-i\theta}\frac{d_0}{r}\right)^{\alpha+q}+A_{7/3}^{(q)}\left(n+e^{-i\theta}\frac{d_0}{r}\right)^{\alpha+q-7/3}+\mathcal O\left(n^{\alpha+q-11/3}\right),
\end{equation}
with $\alpha=2\Delta_{\mathcal{O}}-4$ and coefficients
\begin{equation}
A_0^{(q)}=C\left(r e^{i\theta}\right)^q,\qquad A_{7/3}^{(q)}=C\left(r e^{i\theta}\right)^q\left(\alpha+q-\frac{4}{3}\right)\frac{d_{4/3}}{r e^{i\theta}},
\label{analyticMomentCoefficients}
\end{equation}
following the notations and results in \eqref{coefficientsResults1} and \eqref{coefficientsResults3}.

A contribution of degree $\xi$ in the large-$n$ expansion of the summand \eqref{QNMdataexpansionfull}  behaves as
\begin{equation}
\left(n+e^{-i\theta}\frac{d_0}{r}\right)^{\alpha+q-\xi}
\end{equation}
and is summable only if
\begin{equation}
\xi>T_q,\qquad T_q\equiv\alpha+q+1.
\end{equation}
All terms with $\xi<T_q$ must therefore be subtracted and treated by analytic continuation. A term with $\xi=T_q$ would instead behave as $n^{-1}$ and generate a pole of $\mathcal{Z}(s,\vec{0})$ at $s=-q$. Therefore, whenever this degree is present, finiteness of the double-trace sum rule requires
\begin{equation}
2\,\text{Re}\,A_{T_q}^{(q)}=0.
\end{equation}

Let $N$ denote the number of modes included from each tower. We define the subtracted moment
\begin{equation}
\mathcal{S}_q(N)=2\,\text{Re}\left[\sum_{n=1}^{N}r_{n}^{+}(i\omega_{n}^{+})^q+\sum_{\xi<T_q}A_\xi^{(q)}\zeta\left(-\alpha-q+\xi,N+1+e^{-i\theta}\frac{d_0}{r}\right)\right],
\end{equation}
where the second term, involving Hurwitz-zeta functions, is the contribution of the asymptotic QNM tail beyond the cutoff. The latter follows from the Hurwitz-zeta identity for a complex exponent $p$:
\begin{equation}
\sum_{n=1}^{N}\left(n+e^{-i\theta}\frac{d_0}{r}\right)^p+\zeta\left(-p,N+1+e^{-i\theta}\frac{d_0}{r}\right)=\zeta\left(-p,1+e^{-i\theta}\frac{d_0}{r}\right).
\end{equation}

The definition of $\mathcal{S}_q(N)$ leaves the first $N$ QNM contributions unchanged. Only the tail with $n>N$ is replaced by its large-$n$ asymptotic expansion. Among the resulting asymptotic terms, those with $\xi<T_q$ cannot be summed in the ordinary sense and are therefore restored through their Hurwitz-zeta continuation. Terms with $\xi>T_q$ are already summable and need not be included explicitly: their omitted tail converges to zero as $N\to\infty$. Moreover, after the non-summable terms have been subtracted, the difference between the exact QNM summand and the retained asymptotic approximation is $\mathcal{O}(n^{-1-\epsilon})$ for some $\epsilon>0$. Its contribution beyond the cutoff is then $\mathcal{O}(N^{-\epsilon})$ and also vanishes in the limit.
In the original region of convergence, the added and subtracted contributions cancel identically, so this prescription agrees there with $\mathcal{Z}(s,\vec{0})$ and therefore defines its continuation. It follows that
\begin{equation}
\lim_{N\to\infty}\mathcal S_q(N)=\mathcal{Z}(-q,\vec{0}).
\end{equation}
The double-trace sum rule can consequently be written in the convergent form
\begin{equation}\label{convergentDoubleTraceSumRule}
\lim_{N\to\infty}\mathcal S_q(N)=0.
\end{equation}

We tested the first three conditions in \eqref{convergentDoubleTraceSumRule} for $\Delta_{\mathcal{O}}=9/4$, using the first $101$ QNMs and residues. For this value of the external dimension,
\begin{equation}
T_0=\frac{3}{2},\qquad T_1=\frac{5}{2},\qquad T_2=\frac{7}{2}.
\end{equation}
In particular, for $q=0$, only the degree $\xi=0$ satisfies $\xi<T_0$. For both $q=1$ and $q=2$, the non-summable degrees are $\xi=0$ and $\xi=7/3$. 

At $N=101$, the exact partial moments, the analytically continued non-summable tails, and their sums are
\begin{equation}
\begin{array}{c|c|c|c}
q &
2\beta^q\,\text{Re}\displaystyle\sum_{n=1}^{101}r_{n}^{+}(i\omega_{n}^{+})^q &
2\beta^q\,\text{Re}\displaystyle\sum_{\xi<T_q}A_\xi^{(q)}
\zeta\left(-\alpha-q+\xi,102+e^{-i\theta}\frac{d_0}{r}\right) &
\beta^q\,\mathcal{S}_q(101) \\ \hline
0 & 947.727766873 & -947.727578731 & 1.8814\times10^{-4}\\
1 & 2.10393256737\times10^5 & -2.10393256688\times10^5 & 4.9697\times10^{-5}\\
2 & -1.37183459693\times10^8 & 1.37183459645\times10^8 & -4.7950\times10^{-2}
\end{array}
\end{equation}
where we introduced powers of $\beta$ so that different values of $q$ are expressed in the same units.
The two contributions are individually large and cutoff dependent, but cancel to high accuracy.

Varying the number $N$ of QNMs included exactly gives
\begin{equation}
\begin{array}{c|c|c|c}
N & \mathcal S_0(N) & \beta\,\mathcal S_1(N) & \beta^2\,\mathcal S_2(N) \\ \hline
1   & 1.1194\times10^{-2} & 2.8322\times10^{-2} & -4.2172\times10^{-1}\\
21  & 6.9930\times10^{-4} & 3.9622\times10^{-4} & -1.0824\times10^{-1}\\
51  & 3.3276\times10^{-4} & 1.2151\times10^{-4} & -6.8605\times10^{-2}\\
81  & 2.2616\times10^{-4} & 6.6237\times10^{-5} & -5.3867\times10^{-2}\\
101 & 1.8814\times10^{-4} & 4.9697\times10^{-5} & -4.7950\times10^{-2}
\end{array}
\end{equation}
The magnitude of all three subtracted moments decreases monotonically as the cutoff is increased, consistently with convergence towards the predicted value $\mathcal{Z}(-q,\vec{0})=0$.

As a further application of the double-trace sum rules, we now investigate how much information about the fundamental QNM can be obtained from the asymptotic data derived above. We isolate the first conjugate pair and approximate all the remaining modes, with $n\ge 2$, by successive truncations of their large-$n$ expansion.

For $q=0,1,2,3$, we write
\begin{equation}
0=\sum_{\varepsilon=\pm}r_{1,\varepsilon}(i\omega_{1}^{\varepsilon})^q+\sum_{n=2}^{\infty}\sum_{\varepsilon=\pm}r_{n}^{\varepsilon}(i\omega_{n}^{\varepsilon})^q.
\end{equation}
Using the reality relation between the two towers, this becomes
\begin{equation}
0=2\,\text{Re}\left[r_{1,+}(i\omega_{1}^{+})^q\right]+2\,\text{Re}\sum_{n=2}^{\infty}r_{n}^{+}(i\omega_{n}^{+})^q.
\end{equation}
Thus, the moments of the fundamental pair are fixed by the contribution of the remaining QNM tower:
\begin{equation}
m_q\equiv 2\,\text{Re}\left[r_{1,+}(i\omega_{1}^{+})^q\right]=-2\,\text{Re}\sum_{n=2}^{\infty}r_{n}^{+}(i\omega_{n}^{+})^q,
\end{equation}
where the equality is understood by replacing the right-hand side with its asymptotically subtracted and analytically continued form defined above.

The results obtained in Section \ref{sec:QNMasyfromOPE} provide three successive approximations to the asymptotic contribution to each moment.
As already described, the number of asymptotic terms required for a convergent subtraction depends on $q$ and $\Delta_{\mathcal{O}}$. For $\Delta_{\mathcal{O}}=9/4$, the moment with $q=0$ is fully subtracted once the constant correction $d_0$ is resummed into the shifted overtone number. The moments with $q=1$ and $q=2$ additionally require the fractional correction generated by $d_{4/3}$. The moment with $q=3$ would also require the next unknown coefficient. Nevertheless, we apply each available truncation to all four moments with $q=0,1,2,3$ in order to study how the estimate of the fundamental mode changes as further asymptotic information is included. The three successive approximations read
\begin{equation}
\begin{aligned}
m_q\simeq -2\text{Re}\left[C\left(r e^{i\theta}\right)^q\zeta(-\alpha-q,2)\right],
\end{aligned}
\end{equation}
when keeping only the leading asymptotic behaviour,
\begin{equation}
\begin{aligned}
m_q\simeq -2\text{Re}\left[C\left(r e^{i\theta}\right)^q\zeta\left(-\alpha-q,2+e^{-i\theta}\frac{d_0}{r}\right)\right],
\end{aligned}
\end{equation}
when including the correction $d_0$, and
\begin{equation}
\begin{aligned}
m_q\simeq -2\text{Re}\biggl\{&C\left(r e^{i\theta}\right)^q\biggl[\zeta\left(-\alpha-q,2+e^{-i\theta}\frac{d_0}{r}\right)\\
&+\left(\alpha+q-\frac{4}{3}\right)\frac{d_{4/3}}{r e^{i\theta}}\zeta\left(-\alpha-q+\frac{7}{3},2+e^{-i\theta}\frac{d_0}{r}\right)\biggr]\biggr\},
\end{aligned}
\end{equation}
when including also the fractional correction $d_{4/3}$.

Once the four moments $m_0,\dots,m_3$ have been obtained at a fixed approximation level, the fundamental frequency can be extracted by the order-two Prony method.
Indeed, the moments generated by the fundamental conjugate pair have the form
\begin{equation}
m_q=r_{1,+}(i\omega_{1}^{+})^q+r_{1,-}(i\omega_{1}^{-})^q.
\end{equation}
The two quantities $i\omega_{1}^{+}$ and $i\omega_{1}^{-}=(i\omega_{1}^{+})^*$ are the roots of the quadratic polynomial
\begin{equation}
x^2-2\,\text{Re}(i\omega_{1}^{+})x+|\omega_{1}^{+}|^2=0.
\end{equation}
Therefore, each of them satisfies
\begin{equation}
(i\omega_{1}^{\pm})^{q+2}=2\,\text{Re}(i\omega_{1}^{+})(i\omega_{1}^{\pm})^{q+1}-|\omega_{1}^{+}|^2(i\omega_{1}^{\pm})^q.
\end{equation}
Multiplying these identities by the corresponding residues and adding the two contributions gives the order-two Prony recurrence
\begin{equation}
m_{q+2}=2\,\text{Re}(i\omega_{1}^{+})m_{q+1}-|\omega_{1}^{+}|^2m_q.
\end{equation}
Applying this relation at $q=0$ and $q=1$ gives the following system of equations for $2\,\text{Re}(i\omega_{1}^{+})$ and $|\omega_{1}^{+}|^2$:
\begin{equation}
2\,\text{Re}(i\omega_{1}^{+})=\frac{m_0m_3-m_1m_2}{m_0m_2-m_1^2},\qquad |\omega_{1}^{+}|^2=\frac{m_1m_3-m_2^2}{m_0m_2-m_1^2}.
\end{equation}

At $\Delta_{\mathcal{O}}=9/4$, the numerical result for the fundamental frequency is
\begin{equation}
(\omega_{1}^{+})^{\text{exact}}=1.512919-1.050026\,i,
\end{equation}
and we find the following successive approximations:
\begin{equation}
\begin{array}{c|c|c}
\text{asymptotic data included} &
(\omega_{1}^{+})^{\text{estimate}} &
\text{relative error} \\ \hline
\text{leading term} & 1.541553-2.149204\,i & 5.9706\times10^{-1}\\
\text{leading term and }d_0 & 1.434928-1.113814\,i & 5.4710\times10^{-2}\\
\text{leading term, }d_0\text{ and }d_{4/3} & 1.447503-1.133456\,i & 5.7569\times10^{-2}
\end{array}
\end{equation}
The constant correction $d_0$ substantially improves the leading estimate. The fractional correction changes the result only moderately because the $q=3$ moment still contains additional non-summable asymptotic contributions that must be subtracted to define $m_3$ consistently.
These can be compared with the direct evaluation of \eqref{eq:qnmTail}, which gives $(\omega_{1}^{+})^{\text{tail}}=1.50874-1.04369\,i$, with a relative error of $4.13\times 10^{-3}$ compared with the exact numerical result.

In fact, we tested the effect of these missing contributions by resolving the remainder in \eqref{QNMdataexpansionfull} fitting numerically the coefficients $A_{11/3}^{(q)}$ and $A_4^{(q)}$ for each $q$ from the QNM data with $n=1,\dots,101$. Since both degrees satisfy $\xi<T_3=9/2$, both contributions must be included in the subtraction defining $m_3$.
The result then greatly improves:
\begin{equation}
\begin{array}{c|c|c}
\text{asymptotic data included} &
(\omega_{1}^{+})^{\text{estimate}} &
\text{relative error} \\ \hline
\text{up to }d_{4/3}\text{ and fitted }A_{11/3}^{(q)},A_4^{(q)}
&1.512908-1.050047\,i&1.2753\times10^{-5}
\end{array}
\end{equation}
It would be interesting to compare this estimate with the direct evaluation of \eqref{eq:qnmTail}, as done above at lower order. A consistent comparison would require the analytic QNM tail to be known to the same order as the asymptotic data used here, including the coefficients that generate $A_{11/3}^{(q)}$ and $A_4^{(q)}$ in the moment expansion.
Finally, we can compare the relative error of this moment-based estimate with the relative error $3.14\times10^{-5}$ obtained for the fundamental mode in Table \ref{table:numerics} from the analytic continuation of the OPE at $\Delta_{\mathcal O}=11/4$, indicating a comparable level of accuracy.
On the one hand, the OPE analytic-continuation method requires much less input; on the other hand, the procedure described here is analytic in principle and gives a direct use of the double-trace sum rules.

\subsection{Mellin completion using Carlson's theorem}

In Section \ref{sec:opeExtended}, we showed how recovering non-asymptotic QNM data from the OPE requires an analytic continuation beyond the domain in which the OPE is originally defined. The exactly solvable R-current correlator at zero momentum provides a setting in which the additional analytic input can be stated precisely.

Let us start from the OPE data in \eqref{Rcurrent-OPE1}-\eqref{Rcurrent-OPE2}, which imply the presence of poles in $\mathcal{M}(s,\vec{k}=\vec{0})$ at $s=3-4m$, $m\in\mathbb Z_{\geq 0}$. It is convenient to map this pole lattice to the non-positive integers by introducing
\begin{equation}
w=\frac{s-3}{4}.
\end{equation}
We consider Mellin completions of the form
\begin{equation}
\mathcal{M}(s,\vec{0})=\Gamma(w)B(w),
\end{equation}
where $B(w)$ is a holomorphic function in the left half-plane, needed to match the OPE residues. Since
\begin{equation}
\text{res}\left[\Gamma(w(s)),s=3-4m\right]=\frac{4(-1)^m}{m!},
\end{equation}
we ask
\begin{equation}
B(-m)=\frac{m!}{4(-1)^m}\,\beta^{-4m}\,b_{4m,0}.
\end{equation}
In particular, the OPE determines $B(w)$ at every non-positive integer, but does not by itself determine its analytic continuation away from these points.
To separate this interpolation problem from the factorial growth contained in $B(-m)$, we define
\begin{equation}
F(w)=\frac{B(w)}{\Gamma(1-w)}.
\end{equation}
The OPE data then fix
\begin{equation}
F(-m)=\frac{\beta^{-4m}\,b_{4m,0}}{4(-1)^m},\qquad m\in\mathbb Z_{\geq 0}.
\end{equation}
Substituting the explicit R-current OPE residues gives a direct interpolation,
\begin{equation}
\widetilde B(w)=-\frac{4^{-w}\Gamma(1-w)}{\pi}\frac{\zeta(1+4w)}{\Gamma(-4w-2)},
\end{equation}
and hence the candidate Mellin transform
\begin{equation}
\widetilde{\mathcal M}(s,\vec{0})=\Gamma(w)\widetilde B(w)=\frac{2^{\frac{3-s}{2}}\csc\left(\frac{\pi(s+1)}{4}\right)\zeta(s-2)}{\Gamma(1-s)}.
\end{equation}
By construction, $\widetilde{\mathcal M}(s,\vec{0})$ has all the required OPE poles and residues. Nevertheless, this interpolation is not uniquely selected by the OPE data. Indeed, multiplication by any entire function $h(w)$ equal to one at every nonpositive integer leaves all OPE poles and residues unchanged. The missing information is supplied by a growth condition inherited from the physical Mellin transform. We require the normalised interpolation $F(w)$ to be holomorphic in the left half-plane, of exponential type strictly smaller than $\pi$, namely $|F(w)|\le\,\text{const}\,e^{(\pi-\epsilon)|w|}$ for some $\epsilon>0$ in the left half-plane, and polynomially bounded on the imaginary axis. 

We use Carlson's theorem in the following form. If $f(w)$ is holomorphic for $\text{Re}\,w<0$, continuous for $\text{Re}\,w\le 0$, of exponential type strictly smaller than $\pi$, polynomially bounded on the imaginary axis, and satisfies $f(-m)=0$ for every $m\in\mathbb Z_{\ge 0}$, then $f(w)$ vanishes identically. 

Suppose that $F_1(w)$ and $F_2(w)$ are two normalised Mellin interpolations satisfying these conditions and reproducing the same OPE data. Their difference
\begin{equation}
f(w)=F_1(w)-F_2(w)
\end{equation}
belongs to the same growth class and obeys
\begin{equation}
f(-m)=0,\qquad m\in\mathbb Z_{\ge 0}.
\end{equation}
Carlson's theorem therefore gives $F_1(w)=F_2(w)$. 
The OPE data consequently admit at most one Mellin completion within the physical growth class.

It remains to show that such a completion exists. Defining $\widetilde F(w)=\widetilde B(w)/\Gamma(1-w)$ and using the functional equation of the Riemann zeta function gives
\begin{equation}
\widetilde F(w)=\cos(2\pi w)F_{\mathrm C}(w),
\end{equation}
where
\begin{equation}
F_{\mathrm C}(w)=-4^{1+w}\pi^{-1+4w}(8w^2+6w+1)\zeta(-4w).
\end{equation}
The apparent singularity of $\zeta(-4w)$ at $w=-1/4$ is cancelled by the polynomial prefactor, so $F_{\mathrm C}(w)$ is holomorphic in the left half-plane. Moreover, it is polynomially bounded along vertical lines and has exponential type zero. Since $\cos(2\pi m)=1$, it reproduces exactly the same interpolation data, $F_{\mathrm C}(-m)=\widetilde F(-m)$, $m\in\mathbb Z_{\ge 0}$.
By contrast, $\cos(2\pi w)$ has vertical exponential type $2\pi$, so $\widetilde F(w)$ lies outside the Carlson class. The unique admissible completion is therefore
\begin{equation}
B_{\mathrm C}(w)=\Gamma(1-w)F_{\mathrm C}(w)=\frac{\widetilde B(w)}{\cos(2\pi w)}.
\end{equation}
Substitution into $\mathcal M(s,\vec{0})=\Gamma(w)B_{\mathrm C}(w)$ yields
\begin{equation}
\mathcal M(s,\vec{0})=-\frac{2^{\frac{7-s}{2}}}{\pi}\Gamma(s)\cos\left(\frac{\pi(s+1)}{4}\right)\zeta(s-2),
\end{equation}
which reproduces the full R-current Mellin transform. 

The mechanism described above is not specific to the R-current correlator. Whenever the OPE poles lie on a regular lattice, they can be mapped to the non-positive integers and their residues interpreted as interpolation data for a suitably normalised function $F(w)$. If this function is holomorphic in the left half-plane and has exponential type strictly smaller than $\pi$, Carlson's theorem again guarantees uniqueness. From the spectral representation, the relevant vertical growth is controlled by the angular support of the QNM frequencies.
Thus, independent information constraining the extremal QNM angles, for example from the locations of the bouncing singularities, may provide the Carlson bound without requiring knowledge of the individual QNMs. The main practical obstruction beyond the R-current example is that the OPE coefficients are generally not known in a closed analytic form, making it difficult to construct an explicit interpolation. With finitely many OPE coefficients one may instead seek numerical interpolations subject to the same growth constraint, although Carlson's theorem then provides a guiding uniqueness principle rather than an exact reconstruction.

Ideally, if one has fully reconstructed $\mathcal{M}(s,\vec{k})$, it is possible to reconstruct the QNMs by looking in the large positive $\text{Re}(s)$ regime, as anticipated in Section \ref{sec:mellin}. We close this section by making this point more explicit.

Suppose first that there is a unique QNM of smallest modulus, denoted by $\omega_1$, and that all remaining QNMs satisfy $|\omega_n|\geq|\omega_2|>|\omega_1|$. Then, it is possible to extract $\omega_1$ directly from the large-positive-$s$ behaviour of the Mellin transform. Indeed, along the positive real $s$-axis
\begin{equation}
\mathcal{Z}(s,\vec{k})=r_1(i\omega_1)^{-s}\left[1+\mathcal{O}\left(\left|\frac{\omega_1}{\omega_2}\right|^s\right)\right].
\end{equation}
It follows that
\begin{equation}
\lim_{s\to+\infty}\frac{\mathcal{Z}(s,\vec{k})}{\mathcal{Z}(s+1,\vec{k})}=i\omega_1.
\end{equation}

In general, however, there can be more frequencies with the same modulus. Notably, this is the case for the black brane correlator, where QNMs always come in pairs with $\omega_{n}^{-}=-(\omega_{n}^{+})^*$. In that case, writing $i\omega_{1}^{+}=\rho_1 e^{i\theta_1}$ with residue $r_{1,+}$, and assuming that all remaining QNMs have modulus greater than some $\rho_2>\rho_1$, one finds, along the positive real $s$-axis
\begin{equation}\label{Zsasymptotics}
\mathcal{Z}(s,\vec{k})=2|r_{1,+}|\rho_1^{-s}\cos\left(\arg r_{1,+}-s\theta_1\right)+\mathcal{O}\left(\rho_2^{-s}\right).
\end{equation}
The modulus can still be obtained by the exponential rate
\begin{equation}
\frac{1}{\rho_1}=\limsup_{s\to+\infty}|\mathcal{Z}(s,\vec{k})|^{1/s},
\end{equation}
where the $\limsup$ is taken to get rid of the isolated suppressions produced by the zeros of the oscillatory cosine.

Once $\rho_1$ is known, the angle can be extracted from computing consecutive values $\mathcal{Z}(s,\vec{k}), \mathcal{Z}(s+1,\vec{k}), \mathcal{Z}(s+2,\vec{k})$, and using the following Prony recurrence relation from \eqref{Zsasymptotics}:
\begin{equation}
\rho_1^2\mathcal{Z}(s+2,\vec{k})-2\rho_1\cos\theta_1\,\mathcal{Z}(s+1,\vec{k})+\mathcal{Z}(s,\vec{k})=\mathcal{O}\left(\left(\frac{\rho_1}{\rho_2}\right)^s\right),
\end{equation}
from which
\begin{equation}
\cos\theta_1=\lim_{s\to+\infty}\frac{\rho_1^2\mathcal{Z}(s+2,\vec{k})+\mathcal{Z}(s,\vec{k})}{2\rho_1\mathcal{Z}(s+1,\vec{k})}.
\end{equation}
Once the first pair of modes has been determined and subtracted from $\mathcal{Z}(s,\vec{k})$, the same procedure can be applied iteratively to recover the subsequent pairs in order of increasing modulus.

\section{Lightcone OPE=QNM and stress tensor correlators}\label{sec:lightcone}
In this section we initiate the study of OPE$=$QNM in the limit of large spatial momentum $k$. In the OPE considered so far, the contributing operators are organised in terms of their scaling dimension $\Delta$. It is, however, interesting to further consider a different lightcone limit where the contributing operators are organised in terms of their twist $\tau=\Delta-J$. The success of the lightcone bootstrap is largely due to this simple difference and the universality of the low-twist spectrum \cite{Komargodski:2012ek,Fitzpatrick:2012yx}. Moreover, this limit is especially interesting when studying stress tensor correlators as we discuss. The lightcone bootstrap for finite-temperature correlators in holography on a spatial $S^{d-1}$ has been explored extensively, see e.g.\ \cite{Karlsson:2019qfi,Karlsson:2019dbd,Li:2019zba} in terms of long-lived modes \cite{Festuccia:2008zx,Dodelson:2022eiz}. Here, as in the rest of the paper, we consider finite temperature correlators on $\mathbb{R}^{d-1}$ and assume no low-lying scalar operators with twist $\tau<d-2$.

Let us consider the large-$k$ ($k=|\vec{k}|$) limit of the blocks in \eqref{abmap} at fixed $t$. One finds using $\frac{J_{\sigma_p}(kt)}{(kt)^{\sigma_p}}\sim \sqrt{\frac{2}{\pi}}(kt)^{-\sigma_p-\frac{1}{2}}\cos\left(kt-\frac{\pi\sigma_p}{2}-\frac{\pi}{4}\right)$ that 
\bea 
G_R(t,\vec{k})&=&\sum_{\Delta,J}\frac{a_{\Delta,J}c_{\Delta,J}}{\beta^{\Delta}}\frac{t^{\frac{1}{2} \left(d-2 \Delta_{\mathcal{O}}+2 J+\tau -2\right)} }{k^{\frac{d}{2}-\Delta_{\mathcal{O}}+\frac{\tau }{2}} }\cos \left(k t-\frac{1}{4} \pi  \left(d-2 \Delta_{\mathcal{O}}+\tau \right)\right)+\cdots,\cr
c_{\Delta,J} &=& 2^{\frac{d+\Delta+J-2\Delta_{\mathcal O}+2}{2}}\,\pi^{\frac{d-2}{2}}\,\frac{\left(\frac{d-2}{2}\right)_{J}}{J!}\,\Gamma\!\left(1+\frac{\tau}{2}-\Delta_{\mathcal O}\right)\sin\!\left[\pi\!\left(\frac{\Delta}{2}-\Delta_{\mathcal O}\right)\right], \label{eq:GRLargeK}
\eea 
where for each block we have kept only the leading term as $k\to\infty$. In the spatial integration when going from position space to fixed spatial momentum, this contribution arises from the region $r=|\vec{x}|\sim t$ in Eq.\ \eqref{eq:IInt}. In particular, we see that the suppression in $k$ is given in terms of the twist $\tau$ rather than the scaling dimension $\Delta$. On the other hand, while in the large-$k$ limit, low-twist operators become important, as $t$ increases the contribution from higher-twist operators grows. 

Let us consider the universal minimal-twist operators present in any CFT corresponding to the identity operator and the stress tensor operator $T_{\mu\nu}$. We find using \eqref{eq:GRLargeK} that their relative contribution in the large-$k$ limit is 
\be\label{eq:1Tratio}
\frac{G_R(t,\vec{k})|_{T}}{G_R(t,\vec{k})|_1}\propto b_{T_{\mu\nu}}\beta ^{-d}k^{1-\frac{d}{2}} t^{\frac{d}{2}+1}
\ee 
where the thermal stress-tensor one-point function $\langle T_{\mu\nu}\rangle_\beta \sim \beta^{-d}b_{T_{\mu\nu}}$. Note that here we have dropped the stress tensor OPE coefficient in the light-light OPE since it is fixed in terms of the scaling dimension for scalar operators and do not play an important role, for external spinning operators we will see below, however, that the OPE coefficients are crucial. In particular, as the time $t$ increases, at some point the stress tensor contribution is no longer suppressed compared to the identity contribution in \eqref{eq:1Tratio}. This happens at a timescale 
\be 
t_{\text{ST}} \sim \left(\frac{\beta^d}{b_{T_{\mu\nu}}}\right)^{\frac{2}{d+2}}k^{\frac{d-2}{d+2}},
\ee 
where the stress-tensor contribution is no longer suppressed at large-$k$ compared to the identity contribution in \eqref{eq:1Tratio}. In particular, in the context of holography where we further have multi-stress tensor operators $[T^p]_{J=2p}$, with $\tau=p(d-2)$ and $J=2p$, as $t\to t_{\text{ST}}$ they all contribute equally and needs to be resummed. Lastly, note that in order to recover the (in general non-analytic) $t$ dependence in \eqref{eq:GRLargeK}, there has to be an infinite family of these modes similar to the mechanism in Section \ref{sec:QNMasyfromOPE}, see also \cite{Festuccia:2008zx,Dodelson:2023nnr} where this follows naturally from WKB in the bulk. This will be discussed further in Section \ref{sec:LOCOPEQNMK}. 

Let us now consider how this is reproduced from the point of view of the QNM representation. In the limit where only the identity contribution contributes, the QNMs effectively approach lightcone modes with $\omega_n^\pm(k) =\pm k$, trivially reproducing $\propto \cos(kt)$. The leading correction to these modes as $k\to \infty$ are therefore expected to appear when

\be\label{eq:breakdown}
|\omega^\pm(k)\mp k| \sim \frac{1}{t_{\text{ST}}} =  \frac{b_{T_{\mu\nu}}^{\frac{2}{d+2}}}{\beta}\left(\frac{1}{\beta k}\right)^{\frac{d-2}{d+2}}.
\ee 

It is interesting to compare this to holography which in \cite{Festuccia:2008zx} (see also \cite{Dodelson:2023nnr}) was found using WKB methods. The large-$k$ modes there obey the following dispersion relation 
\be\label{eq:largekModes}
\omega^\pm_n = \pm k\pm \frac{e^{\mp i\frac{2\pi}{d+2}}}{k^{\frac{d-2}{d+2}}}\left(\frac{\sqrt{\pi}n\Gamma \left(\frac{3}{2}+\frac{1}{d}\right)}{2^{\frac{1}{d}-\frac{1}{2}}\Gamma \left(\frac{1}{d}\right)}\right)^{\frac{2 d}{d+2}}+\cdots,
\ee 
with $\beta = \frac{4\pi}{d}$ and $k\gg n\gg1$. Comparing \eqref{eq:breakdown} and \eqref{eq:largekModes}, we find that the time scale where multi-stress tensors can not be neglected correctly reproduces the $k$-scaling of the subleading correction to the lightcone modes. This time scale correspond in the bulk to a timelike geodesic that is close to the boundary with a large radius $r\sim k^{\frac{d-2}{d+2}}$ during a time $t\sim k^{\frac{d-2}{d+2}}$. \cite{Festuccia:2008zx}

It is further interesting to extend the discussion above to the context of stress tensor correlators\footnote{See e.g.\ \cite{Fuini:2016qsc} for work on large spatial momentum modes of graviton perturbations. }. Before discussing the stress tensor kinematics in detail, which we do below, let us argue how the behaviour of large-$k$ modes change for stress tensor correlators. In \cite{Kulaxizi:2010jt} it was shown that the stress-tensor contribution to thermal correlator is proportional to the conformal collider bounds and in \cite{Esper:2023jeq} (whose conventions we follow) it was argued that in a lightcone limit the same is true for minimal-twist $[T^p]_{2p}$ operators. Let us denote (in $d=4$) 
\bea
\hat{\mathcal{C}}^{(1)} &=& \frac{5\pi^2}{3}\frac{-7\hat{a}-2\hat{b}+\hat{c}}{C_T}\geq0,\cr
\hat{\mathcal{C}}^{(2)} &=&  \frac{10\pi^2}{3}\frac{16\hat{a}+5\hat{b}-4\hat{c}}{C_T}\geq0,\cr
\hat{\mathcal{C}}^{(3)} &=&15\pi^2\frac{-4\hat{a}-2\hat{b}+\hat{c}}{C_T}\geq0,\cr
C_T &=&\frac{\pi^2}{3}(14\hat{a}-2\hat{b}-5\hat{c}),
\eea
and we further define $\mathcal{C}^{(i)}\equiv \frac{b_{T_{\mu\nu}}}{C_T}\hat{\mathcal{C}}^{(i)}$ with $(\hat{a},\hat{b},\hat{c})$ are stress-tensor three-point functions\footnote{These are related to the conformal collider coefficients $t_2=\frac{30 (13 \hat{a}+4 \hat{b}-3 \hat{c})}{14\hat{a}-2\hat{b}-5\hat{c}}$ and  $t_4=-\frac{15 (81\hat{a}+32\hat{b}-20 \hat{c})}{28\hat{a}-4\hat{b}-10\hat{c}}$. } and the inequalities correspond to the conformal collider bounds \cite{Hofman:2008ar}. These follow from the more general statement of the ANEC. The stress tensor OPE and the appearance of the averaged null energy operator in the thermal state was explored in \cite{Kulaxizi:2010jt,Huang:2022vet,Esper:2023jeq}. 

In particular, for stress tensor correlators, the time scale $t_{\text{ST}}=t_{\text{ST}}^{(i)}$ will then depend on which polarisation\footnote{Here we refer to the scalar, shear or sound channel of the stress-tensor correlator, as in \cite{Kovtun:2005ev}, the details of which will be clarified below.} of the stress tensor correlator we consider. In the scalar case, the inverse time scale correctly reproduced the scale of the subleading corrections to the lightcone modes. For the stress-tensor QNMs we would therefore further expect that this subleading correction is further proportional to $\sim(\mathcal{C}^{(i)})^{1/3}$ and the polarisation-dependent time scale analogous to \eqref{eq:breakdown} to be given by 
\be\label{eq:tbrkTT}
t^{(i)}_{\text{ST}} \simeq  \beta\left(\frac{\beta k}{\mathcal{C}^{(i)}}\right)^{\frac{1}{3}}\implies |\omega^{(i),\pm}(k)\mp k| \sim \frac{1}{t_{\text{ST}}} =  \frac{1}{\beta}\left(\frac{\mathcal{C}^{(i)}}{\beta k}\right)^{\frac{1}{3}},
\ee 
where for concreteness we put $d=4$. In holographic theories where one can approach saturation continuously, we expect that subleading corrections in the large-$k$ expansion become important, such subleading terms were considered in \cite{Aniceto:2026zmc} and were found to be in powers of $k^{-4/3}$.

Let us now consider the kinematics of stress tensor correlators in more detail and show how the conformal collider bounds appear. Consider the Euclidean stress tensor correlator 
\be 
G_{\mu\nu,\rho\sigma}^E(x) = \langle T_{\mu\nu}(x) T_{\rho\sigma}(0)\rangle_\beta.
\ee 
The OPE decomposition can then be written as 
\be 
G_{\mu\nu,\rho\sigma}^E(x) = \frac{1}{(x^2)^{d}}\sum_{(\Delta,J)}\sum_{i}a_{\Delta,J}^{(i)}\beta^{-\Delta}P^{(\Delta,J),i}_{\mu\nu\rho\sigma}(x)\, (x^2)^{\frac{\Delta-J-4}{2}},
\ee 
where $P^{(\Delta,J),i}_{\mu\nu\rho\sigma}(x)$ are polynomials of degree $J+4$ and the sum over $i$ sums over different OPE structures, and thus the thermal coefficients likewise. The conformal block decomposition can be constructed using the embedding space formalism \cite{Costa:2011mg,Costa:2011dw}, described in the context of thermal correlators in detail in \cite{Karlsson:2022osn}. Integrated stress tensor commutators, relevant for the retarded correlator, were also studied in \cite{Besken:2020snx}. Here we mainly focus on the relevant part for studying the large-$k$ limit, the contribution from the identity operator and the stress tensor operator. We are going to align the spatial momentum $\vec{k}\,||\hat{z}$ and mainly consider three different components of stress tensor correlators, $G^R_{xy,xy}$, $G^R_{tx,tx}$ and $G^R_{tz,tz}$ and refer to these as different polarisations, or channels, in analogy with the common terminology in holography \cite{Kovtun:2005ev}.
 
Let us consider first the identity operator in the OPE which can be obtained from the vacuum correlator
\bea
\langle T_{\mu\nu}(x) T_{\rho\sigma}(0)\rangle_{\text{vac}}
&=& \frac{C_T}{x^{2d}}
\left( \frac{1}{2}\big(I_{\mu\rho}(x)I_{\nu\sigma}(x) + I_{\mu\sigma}(x)I_{\nu\rho}(x)\big)
- \frac{1}{d}\eta_{\mu\nu}\eta_{\rho\sigma} \right), \cr
I_{\mu\nu}(x) &=& \eta_{\mu\nu} - 2\frac{x_\mu x_\nu}{x^2},
\eea
where $C_T$ is the central charge. It is clear that this takes the form 
\be\label{eq:TTVac} 
\langle T_{\mu\nu}(x) T_{\rho\sigma}(0)\rangle_{\text{vac}}= \frac{C_TP^{\text{vac}}_{\mu \nu,\rho\sigma}(x)}{(x^2)^{d+2}},
\ee 
where $P^{\text{vac}}_{\mu \nu,\rho\sigma}$ is a polynomial of degree-$4$ ($2J_{T_{\mu\nu}}=4$). The discontinuity of \eqref{eq:TTVac} will be delta-function localized on the lightcone, and thus simplifies the Fourier transform. The result is given as a linear combination of $\sin kt$ and its derivatives, see \eqref{eq:free} with $n\leq d+2$. We can thus write the contribution from the identity operator to the retarded correlator at finite spatial momentum as
\bea 
G^R_{\mu\nu\rho\sigma}(t,\vec{k})|_{\text{vac}}&=&-i\theta(t)\int d^3x e^{-i\vec{k}\cdot\vec{x}}\,\text{disc}\,\langle T_{\mu\nu}(x) T_{\rho\sigma}(0)\rangle_{\text{vac}}\cr
&=& -i\theta(t)\sum_{n=1}^{d+2}\hat{P}^{n,\text{vac}}_{\mu\nu,\rho\sigma}(t,-i\partial_{k_i})\mathcal{I}_{n}(t,k),
\eea
where $k=|\vec{k}|$, $\hat{P}^{\text{vac}}_{\mu\nu,\rho\sigma}$ can be read off from \eqref{eq:TTVac} and $\mathcal{I}_n(t,k)$ are given in \eqref{eq:free}. 

The next contribution in the stress tensor OPE is the exchange of a stress tensor itself, which comes with three different structures labelled by the OPE coefficients $(\hat{a},\hat{b},\hat{c})$. The relevant structure takes a similar form, see Appendix C in \cite{Karlsson:2022osn} for relevant details and whose conventions we follow,
\be 
\langle T_{\mu\nu}(x) T_{\rho\sigma}(0)\rangle_{\beta}|_{T_{\mu\nu}} = \frac{P^{T}_{\mu \nu,\rho\sigma}(x;\hat{a},\hat{b},\hat{c})}{x^{d+6}},
\ee 
where $P^{T}_{\mu \nu,\rho\sigma}(x;\hat{a},\hat{b},\hat{c})$ is a degree-$6$ polynomial. By the same argument as above, we obtain
\be
G^R_{\mu\nu,\rho\sigma}(t,\vec{k}) = -i\theta(t)\sum_{n=1}^{5}\hat{P}^{n,T}_{\mu\nu,\rho\sigma}(t,-i\partial_{k_i};\hat{a},\hat{b},\hat{c})\mathcal{I}_{n}(t,k).
\ee 
The necessary expression here are computed explicitly in Appendix \ref{app:StressTensors}. E.g.\ for $G^R_{xyxy}|_T$ we find ($\vec{k}=(0,0,k)$): 
\bea 
 G^R_{xy,xy}\big|_{\text{vac}} &=&\theta(t)\frac{C_T\pi^2}{40\,t^{5}}\Big[\big((kt)^2-3\big)\cos(kt)-3kt\,\sin(kt)\Big],\cr
G^R_{xyxy}|_T &=& \theta(t)\frac{-4b_T}{9C_T\beta^{4}}(-7\hat a-2\hat b+\hat c)\,k\,\sin(kt),
\eea 
where we see that the stress tensor contribution is proportional to $\mathcal{C}^{(1)}$ as expected. Note, moreover, that this takes the same form as the stress tensor contribution to a scalar correlator with $\Delta=4$, consistent with the bulk EOM of this mode being the massless scalar wave equation. For the other polarisations, however, the dependence on $(\hat{a},\hat{b},\hat{c})$ is more complicated and only the term with the highest power of $k$ non-trivially combine into the conformal collider bound. These are computed in Appendix \ref{app:StressTensors} and the ones we need are 
\bea 
G^R_{tx,tx}\big|_{\text{vac}}
&=&\theta(t)\frac{C_T\pi^2 k^2}{160\,t^{3}}\big(kt\,\sin(kt)+\cos(kt)\big),\cr
G^R_{tx,tx}\big|_{T}&=&\theta(t)\frac{b_T\,k}{3\,C_T\,\beta^{4}}\Big[\tfrac13(16\hat a+5\hat b-4\hat c)\,kt\cos(kt)
+\tfrac12(2\hat b+\hat c)\,\sin(kt)\Big],
\eea
as well as 
\bea 
G^R_{tz,tz}\big|_{\text{vac}}&=&\theta(t)\frac{C_T\pi^2 k^2}{240\,t^{3}}\Big[(2-(kt)^2)\cos(kt)+2kt\,\sin(kt)\Big],\cr
G^R_{tz,tz}\big|_{T}&=&\frac{\theta(t)b_T\,k}{3\,C_T\,\beta^{4}}\Big[-\tfrac13(4\hat a+2\hat b-\hat c)\,(kt)^2\sin(kt)
+2(4\hat a+\hat b-\hat c)\,kt\cos(kt)\cr
&-&2(2\hat a-\hat b-\hat c)\sin(kt)\Big].
\eea

Let us denote $i=1,2,3$ the three different channels, and consider the ratio of the largest $k$ terms to find 
\be
\frac{G^R_{(i)}(t,k\gg1)\big|_T}{G^R_{(i)}(t,k\gg1)\big|_\text{vac}} \sim \frac{t^3}{k\beta^{4}}\mathcal{C}^{(i)},\label{eq:scaleTT}
\ee 
in agreement with the scaling \eqref{eq:1Tratio} and \eqref{eq:tbrkTT}. In particular, this become $\mathcal{O}(1)$ as $t\sim\beta\left(\beta k/\mathcal{C}^{(i)}\right)^{\frac{1}{3}}$ as anticipated in \eqref{eq:tbrkTT}. Note that for minimal-twist multi-stress tensor operators, the factor $\frac{t^3}{k\beta^{4}}$ is purely kinematical and will appear but a priori it is not clear that the $\mathcal{C}^{(i)}$ does. It was, however, argued in \cite{Esper:2023jeq} that this is indeed the case, and that the effective expansion parameter is precisely \eqref{eq:scaleTT}, which was verified explicitly in Gauss-Bonnet gravity. We thus expect that the thermal stress tensor correlators thermalise at a slower rate as the conformal collider bounds are approaching saturation. Note that a CFT saturating the conformal collider bounds is expected to be free \cite{Zhiboedov:2013opa} consistent with having only normal modes.

To explore briefly the physics of the stress tensor correlators and the appearance of the conformal collider bounds in practice, let us consider Gauss-Bonnet gravity with a coupling constant $\lambda_{\text{GB}}$. We will be brief as most of the discussion is not new, and refer to the literature for further details on Gauss-Bonnet gravity and the connection to causality. For the AdS$_5$ black brane with Gauss-Bonnet coupling $\lambda_{\text{GB}}$, writing $\kappa\equiv\sqrt{1-4\lambda_{\text{GB}}}$  following the notation in \cite{Esper:2023jeq},  the three polarisations of $T_{\mu\nu}$ are sensitive to the three different conformal collider bounds
\be\label{eq:Cfactors}
\mathcal{C}^{(1)}=\frac{8(5\kappa-4)}{\kappa^2(\kappa+1)^3}\geq 0,\qquad \mathcal{C}^{(2)}=\frac{8(2-\kappa)}{\kappa^2(\kappa+1)^3}\geq 0,\qquad \mathcal{C}^{(3)}=\frac{8(4-3\kappa)}{\kappa^2(\kappa+1)^3}\geq 0,
\ee
which vanish as $\kappa\to (4/5,2,4/3)$, corresponding to $\lambda_{\text{GB}}\to (9/100,-3/4,-7/36)$. This was recently explored in detail \cite{Buchel:2026rep}. Let us consider $\omega,k\gg1$ with $u=\frac{\omega}{k}$ fixed. Using e.g.\ \cite{Buchel:2026rep}, the effective Schrodinger problem in the bulk is given by 
\be 
-\psi''(z)-\left[k^2\left(u^2-1+\frac{\mathcal{C}^{(i)}}{16}z^4\right)-\frac{15}{4z^2}\right]\psi(z) = 0,
\ee 
up-to corrections that are subleading in the limit we consider, and where $z\to 0$ corresponds to the boundary limit in the bulk. Following \cite{Festuccia:2008zx}, the WKB turning point is located at 
\be 
V_{\text{eik}}  = k^2(u^2-1+\frac{\mathcal{C}^{(i)}}{16}z^4) =0\implies z_T^4 = \frac{16}{\mathcal{C}^{(i)}}(1-u^2),
\ee
and the corresponding action is
\be 
S(u,k)=2k\int_0^{z_T} dz \sqrt{u^2-1+\frac{\mathcal{C}^{(i)}}{16}z^4} \propto \frac{k (u^2-1)^{3/4}}{(\mathcal{C}^{(i)})^{1/4}}.
\ee 
Imposing that it is an $S\sim m$, for $m\gg1$ integer, as a quantization condition and solving for $\omega-k$ leads to 
\be 
|\omega_m-k| \sim \left(\frac{\mathcal{C}^{(i)}}{k}\right)^{1/3}m^{4/3},
\ee 
consistent with \eqref{eq:tbrkTT} and the OPE stress tensor expectations. We thus conclude that we recover the scaling $(\mathcal{C}^{(i)})^{1/3}$ for the $\mathcal{O}(k^{-1/3})$ correction to the lightcone modes both from the OPE and the QNM point of view. From this point of view, the time scale \eqref{eq:tbrkTT} is nothing but the bulk time delay proportional to $\left|\frac{\partial S}{\partial \omega}\right|_{\omega =k}$. 

\subsection{From lightcone OPE to weakly damped large-$k$ QNMs}\label{sec:LOCOPEQNMK}
Let us consider how \eqref{eq:GRLargeK} can be reproduced by a single family of lightcone modes labelled by $n$ in more detail. We proceed similarly to what was done in Section \ref{sec:QNMasyfromOPE}. The non-analytic in $t$ behaviour is expected to arise from the tail of QNMs. We therefore approximate the QNM sum by an integral
\be
G_R(t,k)\simeq \int^\infty dn\,  r_n^+ e^{-i\omega^+_n t}+\cdots. 
\ee
where here, and below, the dots refer to the contribution from the $\omega^-_n$ modes as well as subleading corrections in $1/k$ or $t$. Assuming $\omega_n^+ = k+e^{-i\theta}Kk^{-\frac{d-2}{d+2}}n^\gamma+\cdots$, and we further estimate the residues by the vacuum scaling 
\be 
r_n \sim \frac{\partial\omega}{\partial n}(\omega^2-k^2)^{\Delta-\frac{d}{2}}\Big|_{\omega=\omega^+_n} \sim \frac{k^{\Delta-\frac{d}{2}}}{n}\left(k^{-\frac{d-2}{d+2}}n^{\gamma}\right)^{\Delta-\frac{d}{2}+1},
\ee 
where $(\omega^2-k^2)^{\Delta-\frac{d}{2}}$ is the vacuum spectral density and $\partial\omega/\partial n$ the Jacobian from approximating a sum of delta functions with a smooth function. The integral is dominated by $|n|\sim (k^{-\frac{d-2}{d+2}}t)^{-\frac{1}{\gamma}}$ and we then find 
\be
G_R(t,k)\propto k^{\Delta-\frac{d}{2}}t^{-\Delta+\frac{d}{2}-1}e^{-ikt}+\cdots,
\ee 
consistent with the scaling of the identity contribution in \eqref{eq:GRLargeK}. Moreover, in order to be consistent with the OPE due to multi-stress tensors with $\tau=p(d-2)$ and $J=2p$ the expansion has to take the form 
\be 
G_R(t,k)\simeq \int^\infty dn\, \frac{\partial\omega}{\partial n}\Big|_{\omega=\omega_n} k^{\Delta-\frac{d}{2}}(k^{-\frac{d-2}{d+2}}n^\gamma)^{\Delta-\frac{d}{2}}\sum_{p\geq 0} c_pn^{-p\gamma\frac{d+2}{2}} e^{-i\omega^+_n t}+\cdots,\label{eq:largeKInt}
\ee 
where the coefficients $c_p$ are combinations of corrections to the residues and QNMs, and we have translated powers in $t$ into powers of $n$.  Performing the integral in \eqref{eq:largeKInt}, we end up with an expansion 
\bea
G_R(t,k)&\simeq&   \mathcal{N} k^{\Delta-\frac{d}{2}}t^{-\Delta+\frac{d}{2}-1}e^{-ikt}\sum_{p\geq 0} \tilde{c}_pe^{\frac{ip}{4}(d+2) (\pi -2 \theta )} (K^{\frac{d}{2}+1}k^{1-\frac{d}{2}} t^{\frac{d}{2}+1})^p+\cdots \,\cr
\tilde{c}_p &=&c_p \Gamma (\Delta-\frac{d-2+p(d+2)}{2}),
\label{eq:largeKExp}
\eea
where $\tilde{c}_p$ are related to the thermal coefficients by comparison to \eqref{eq:GRLargeK}. Matching the phase in \eqref{eq:largeKExp} with \eqref{eq:GRLargeK}, assuming $c_p$ is real, leads to 
\bea 
&\frac{1}{4}& i (d+2) (\pi -2 \theta )=\frac{1}{4} i \pi  (d-2)\implies\cr
&\theta& = \frac{2\pi}{d+2}. 
\eea
This is consistent with the phase found from WKB in the bulk \eqref{eq:largekModes}. We thus conclude that the phase of the large-$k$ correction in holographic theories follows naturally from the exchange of leading-twist operators in the OPE. Lastly, as discussed in the previous section, in the context of holographic stress tensor correlators, we further expect $K^{\frac{d}{2}+1}\propto \mathcal{C}^{(i)}$.

\section{Discussion}\label{sec:discussion}
In this work we provided a general framework for OPE=QNM. In order to achieve this, we argued that mixed time-spatial momentum retarded correlators, $G_R(t,\vec{k})$, serve as the perfect avenue for such a relation. In particular, the mixed nature serves as an intermediary between the two expansions and the causal nature of retarded correlators further ensures that both representations are well behaved. Moreover, we showed that the OPE and QNM expansions of the mixed correlator have a common region of overlap in the complex $t$ plane, allowing for a direct identification of OPE and QNM data. We used this constructively in at least two different ways: 1) matching at real finite $t>0$ and 2) matching at singular points whose behaviour is simultaneously dominated by the asymptotics of each representation. This enabled us to accurately predict any QNM from the OPE. 

The OPE=QNM relation we have outlined constitutes a new kind of bootstrap programme, in which fundamental constraints on OPE data (such as the spectrum) or on QNM data (such as causality and stability, see e.g. \cite{Heller:2022ejw, Heller:2023jtd}) are intertwined and constrain one another. A sharp manifestation of this relation are the relations \eqref{OPEreconstruction} or the sum rules \eqref{sumrule}, which constrains QNM data given the OPE spectrum. It would be interesting to develop this programme further, in particular the role of Mellin space thermal correlators, see also \cite{Alday:2020eua}.

The O($N$) model at large-$N$, which we present in Appendix \ref{app:ON}, illustrates the same framework applies to non-holographic models. It would be interesting to explore other models which do not rely on the existence of a holographic dual. For example, one could study the hydrodynamic quasinormal modes of CFT$_2$ which admit analytic control through conformal perturbation theory along the lines of recent works \cite{Davison:2024msq, Asplund:2025nkw, Davison:2025xdj}. In these cases, the square Penrose diagram for BTZ may be relaxed through the addition of relevant deformations, which would adjust bouncing singularity locations and asymptotic QNM angles associated to changes in the OPE data.

Tauberian theorems have been used in CFT at zero temperature to relate OPE asymptotics in different channels \cite{Mukhametzhanov:2018zja,Mukhametzhanov:2019pzy}, see also \cite{Marchetto:2023xap} for applications at finite temperature. Typically, this requires certain positivity conditions that are not obviously present in our OPE$=$QNM relation. It would be interesting to understand if one can generalise such relations to apply to the setup considered in this paper.\footnote{We thank Alexander Zhiboedov for discussions.}

It would be interesting to investigate the role of hydrodynamic QNMs as a special case of the general OPE=QNM relations we have outlined. We point out that the OPE for $G_R(t,\vec{k})$ is naturally already organised as series in $k$, see \eqref{abmap}. Thus, the small $k$ expansion of the QNM representation should also be organised in this way, resulting in a set of OPE=QNM relations for each even power of $k$. To illustrate this latter point, consider scalar correlators in BTZ case at $k\neq 0$, for which the frequency space Green's function is given by
\begin{equation}
\widetilde{G}_R(\omega,k)= \frac{\Gamma\left(\frac{\Delta_\mathcal{O}-i(\omega-k)}{2}\right)\Gamma\left(\frac{\Delta_\mathcal{O}-i(\omega+k)}{2}\right)}{\Gamma\left(1-\frac{\Delta_\mathcal{O}+i(\omega-k)}{2}\right)\Gamma\left(1-\frac{\Delta_\mathcal{O}+i(\omega+k)}{2}\right)},
\end{equation}
with QNM frequencies at $\omega_n^\pm(k)=\pm k-i(\Delta_\mathcal{O}+2n)$ for $n\in\mathbb Z_{\geq0}$. The resulting QNM representation is
\be
G_R(t,\vec{k}) = -i\sum_{n\geq 0} \sum_{\pm} \text{res}\left(\widetilde{G}_R(\omega,k) e^{-i\omega t},\; \omega = \omega_n^\pm(k)\right).
\ee
Since $\widetilde{G}_R(\omega,k)$ is invariant under $k\to -k$, each term in the $n$ sum is proportional to $\cos(kt)$ and thus admits a power series in even powers of $k$, in accordance with the structure of the OPE.

There are various further interesting directions to explore:
\begin{itemize}
    \item Charged black holes: In this case, there is also a line of purely imaginary QNMs which contribute to non-analytic terms in the OPE singularity but we do not expect them to do so at the bouncing singularities at an angle. It would be interesting to explore if there are still enough constraints to determine the asymptotic QNM expansion in this case. 
    \item It would be interesting to study the implications in other models, such as the SYK model. There, the spectrum of QNMs is much richer, as well as the bouncing singularities. Can one use similar methods to those here to explore the asymptotic density of QNMs in these cases?  
    \item Boundary signatures of shockwaves behind the horizon have been explored in \cite{Horowitz:2023ury}, it would be interesting to explore how such a setup would affect the OPE$=$QNM relations that we have explored in this work. 
\end{itemize}
    
\section*{Acknowledgements}
It is a pleasure to thank David M. Ramirez and  Alexander Zhiboedov for discussions. 
We would also like to thank Julien Barrat, Deniz N. Bozkurt, Enrico Marchetto, Alessio Miscioscia, and Elli Pomoni for sharing drafts and coordinating release.
BW would like to thank Elli Pomoni and the DESY Theory Group for hospitality while this work was being completed.
PA is supported by the Royal Society grant URF\textbackslash R\textbackslash 231002, `Dynamics of holographic field theories'. CI is partially supported by the
Swiss National Science Foundation Grant No. 185723 and the NCCR SwissMAP and acknowledges funding from the European Research Council (ERC) and the Swiss State Secretariat for Education, Research and Innovation (SERI) through the consolidator grant ProbQuant.
RK is supported by the Titchmarsh Research Fellowship at the Mathematical Institute and by the Walker Early Career Fellowship at Balliol College.
BW is supported by a Royal Society University Research Fellowship and in part by the Science and Technology Facilities Council (Consolidated Grant `New Frontiers in Particle Physics, Cosmology and Gravity').  For the purpose of open access, the authors have applied a CC BY public copyright licence to any Author Accepted Manuscript (AAM) version arising from this submission.

\appendix
\section{Fourier transform of thermal  blocks}\label{app:FTblock}

In this appendix, we evaluate $\mathcal{I}_{\Delta,J}(t,\vec{k})$ from \eqref{eq:IDeltaJtk}. Since the integrand is rotationally invariant, the result for $\mathcal{I}_{\Delta,J}(t,\vec{k})$ depends only on $k = |\vec{k}|$. We will denote with $\alpha=\frac{\Delta}{2}-\Delta_{\mathcal O}$. Performing the angular integral, we obtain
\be\label{eq:radialInt}
\mathcal{I}_{\Delta,J}(t,\vec{k})=\,\theta(t)\,(2\pi)^{\frac{d-1}{2}}k^{-\frac{d-3}{2}}\int_0^t\mathrm{d}r\,r^{\frac{d-1}{2}}J_{\frac{d-3}{2}}(kr)C_J^{\left(\frac{d-2}{2}\right)}\left(\frac{t}{\sqrt{t^2-r^2}}\right)(t^2-r^2)^{\alpha}.
\ee
Using the Gegenbauer expansion
\begin{equation}\label{eq:IInt}
C_J^{\left(\frac{d-2}{2}\right)}(z)=\sum_{p=0}^{\lfloor J/2\rfloor}(-1)^p\frac{\left(\frac{d-2}{2}\right)_{J-p}}{p!(J-2p)!}(2z)^{J-2p},
\end{equation}
this becomes
\begin{equation}
\begin{aligned}
\mathcal{I}_{\Delta,J}(t,\vec{k})=\,&\theta(t)\,(2\pi)^{\frac{d-1}{2}}k^{-\frac{d-3}{2}}\\
&\sum_{p=0}^{\lfloor J/2\rfloor}(-1)^p\frac{\left(\frac{d-2}{2}\right)_{J-p}}{p!(J-2p)!}2^{J-2p}t^{J-2p}\int_0^t\mathrm{d}r\,r^{\frac{d-1}{2}}J_{\frac{d-3}{2}}(kr)(t^2-r^2)^{\alpha_p},
\end{aligned}
\end{equation}
where we defined $\alpha_p=\frac{\Delta}{2}-\Delta_{\mathcal O}-\frac{J-2p}{2}$. The remaining radial integral is a Sonine integral,
\begin{equation}
\int_0^t\mathrm{d}r\,r^{\frac{d-1}{2}}J_{\frac{d-3}{2}}(kr)(t^2-r^2)^{\alpha_p}=2^{\alpha_p}\Gamma(\alpha_p+1)k^{-\alpha_p-1}t^{\frac{d-1}{2}+\alpha_p}J_{\frac{d-1}{2}+\alpha_p}(kt).
\end{equation}
Defining $\sigma_p=\frac{d-1}{2}+\alpha_p$, we obtain 
\be
\mathcal{I}_{\Delta,J}(t,\vec{k})=\,\theta(t)\,t^{d-1-2\Delta_{\mathcal O} + \Delta}(2\pi)^{\frac{d-1}{2}}\sum_{p=0}^{\lfloor J/2\rfloor}(-1)^p\frac{\left(\frac{d-2}{2}\right)_{J-p}}{p!(J-2p)!}2^{J-2p+\alpha_p}\Gamma(\alpha_p+1)\frac{J_{\sigma_p}(kt)}{(kt)^{\sigma_p}}. \label{FTblock}
\ee
This is the result for the spatial Fourier transform of the block. For our purposes we wish to arrange the result in powers of $t$ to get the OPE of $G_R(t,\vec{k})$ \eqref{introGRtk}, we can do this by expanding the Bessel function in powers of $t$,
\begin{equation}
\frac{J_{\sigma_p}(kt)}{(kt)^{\sigma_p}}=\sum_{\ell=0}^{\infty}\frac{(-1)^\ell}{\ell!\Gamma(\ell+\sigma_p+1)}\frac{(kt)^{2\ell}}{2^{2\ell+\sigma_p}},
\end{equation}
and collect powers of $t$, 
\begin{equation}\label{eq:IDeltaJexpansion}
\mathcal{I}_{\Delta,J}(t,\vec{k})=\theta(t)t^{d-1-2\Delta_{\mathcal O}+\Delta}\sum_{\ell=0}^{\infty}\mathcal{I}_{\Delta, J, \ell}  (kt)^{2\ell},
\end{equation}
with coefficient,
\begin{equation}
\begin{aligned}
\mathcal{I}_{\Delta, J, \ell}=&\pi^{\frac{d-1}{2}}\frac{(-1)^\ell}{4^\ell\ell!}\sum_{p=0}^{\lfloor J/2\rfloor}(-1)^{p}\frac{\left(\frac{d-2}{2}\right)_{J-p}}{p!(J-2p)!}2^{J-2p}\frac{\Gamma\left(1+\frac{\Delta-2\Delta_{\mathcal O}-J+2p}{2}\right)}{\Gamma\left(\ell+\frac{d+1}{2}+\frac{\Delta-2\Delta_{\mathcal O}-J+2p}{2}\right)},\\
=&\pi^{\frac{d-1}{2}}\frac{(-1)^\ell}{4^\ell\ell!}\frac{2^J\left(\frac{d-2}{2}\right)_J}{J!}\frac{\Gamma\left(1+\frac{\Delta-2\Delta_{\mathcal O}-J}{2}\right)}
{\Gamma\left(\ell+\frac{d+1}{2}+\frac{\Delta-2\Delta_{\mathcal O}-J}{2}\right)}\times\\
&{}_3F_2\left(
-\frac{J}{2},\ \frac{1-J}{2},\ 1+\frac{\Delta-2\Delta_{\mathcal O}-J}{2};\ 
1-\frac{d-2}{2}-J,\ \ell+\frac{d+1}{2}+\frac{\Delta-2\Delta_{\mathcal O}-J}{2};\ 1
\right).
\end{aligned}\label{IDeltaJell}
\end{equation}
This completes the spatial Fourier transform of each block as a series in small $t$.

\section{The \texorpdfstring{O($N$)}{O(N)} model} \label{app:ON}
In this section we study the O($N$) model as an example that is not holographic. 
\subsection{Large-\texorpdfstring{$N$}{N}}
As a simple non-holographic example of OPE=QNM, in which everything can be worked out in complete detail, we now consider the critical O($N$) model at large $N$. We first work in $d=3$, with external field $\phi$ of dimension $\Delta_\phi=\frac{1}{2}$ and keep the thermal circle length $\beta$ explicit. It is convenient to use the dimensionless thermal mass $m_{\text{th}}$, so that the physical mass is $m_{\text{th}}/\beta$. At large $N$, the Euclidean correlator can be written as the thermal image sum
\begin{equation}
G_E(\tau,\vec x)=\frac{1}{\beta}\sum_{n\in\mathbb Z}\frac{e^{-m_{\text{th}}\sqrt{\left(\frac{\tau}{\beta}-n\right)^2+\left(\frac{r}{\beta}\right)^2}}}{\sqrt{\left(\frac{\tau}{\beta}-n\right)^2+\left(\frac{r}{\beta}\right)^2}},
\qquad
r=|\vec x|.
\end{equation}
After analytic continuation $\tau=it+0^+$, only the $n=0$ term contributes to the discontinuity. For $r<t$ one has $\sqrt{r^2-(t-i0^+)^2}=i\sqrt{t^2-r^2}$ and therefore the retarded correlator in position space is
\begin{equation}\label{GRONmodel}
G_R(t,\vec x)=-2\theta(t-r)\frac{\cos\left(\frac{m_{\text{th}}}{\beta}\sqrt{t^2-r^2}\right)}{\sqrt{t^2-r^2}}.
\end{equation}
The Fourier transform in space gives the mixed $t,\vec{k}$ correlator,
\begin{equation}
G_R(t,\vec{k})=-4\pi\theta(t)\int_0^t dr\,r\,J_0(kr)\frac{\cos\left(\frac{m_{\text{th}}}{\beta}\sqrt{t^2-r^2}\right)}{\sqrt{t^2-r^2}}=-4\pi\theta(t)\frac{\sin\left(t\sqrt{ k^2+m_{\text{th}}^2/\beta^2}\right)}{\sqrt{k^2+m_{\text{th}}^2/\beta^2}}. \label{ONGRTk}
\end{equation}

The same result follows from the thermal OPE for $G_R(t,\vec{k})$, \eqref{abmap}. At large $N$, the exchanged operators that contribute to the discontinuity are $\sigma^m$, with
\begin{equation}
\Delta=2m,
\qquad
J=0,
\qquad
a_{\sigma^m}=\frac{m_{\text{th}}^{2m}}{\Gamma(2m+1)}.
\end{equation}
With this OPE spectrum the relevant coefficients in \eqref{abmap} become 
\be
\mathcal{I}_{2m, 0, \ell} = \frac{(-4)^{-\ell} \pi \Gamma\left(m + \frac{1}{2}\right)}{\ell! \Gamma\left(m + \ell + \frac{3}{2}\right)}
\ee
thus, through \eqref{abmap} lead to the $G_R(t,\vec{k})$ OPE coefficients, 
\begin{equation}
b_{2m,k} =-4\pi\frac{(-1)^m\left((\beta k)^2+m_{\text{th}}^2\right)^m}{(2m+1)!}, \label{ONOPEdata}
\end{equation}
and when summed over $m$ in \eqref{abmap} leads precisely to \eqref{ONGRTk}. Since \eqref{ONGRTk} has no singularities, the OPE converges for all $t$.

The finite-$k$ data also allow one to invert explicitly the relation between the coefficients $b_{2n, k}$ and the OPE coefficients $a_{\sigma^m}$. Expanding the power in \eqref{ONOPEdata} by the binomial theorem,
\begin{equation}
b_{2n,k}=-4\pi\frac{(-1)^n}{(2n+1)!}\sum_{m=0}^{n}\binom{n}{m}m_{\text{th}}^{2m}(\beta k)^{2n-2m}.
\end{equation}
and using $m_{\text{th}}^{2m}=(2m)!a_{\sigma^m}$ one obtains
\begin{equation}
b_{2n,k}=\sum_{m=0}^{n}\left[-4\pi\frac{(-1)^n(2m)!}{(2n+1)!}\binom{n}{m}\right]\left(\beta\,k\right)^{2n-2m}a_{\sigma^m}. \label{ONbn}
\end{equation}
The inverse binomial transform then gives
\begin{equation}\label{btoaON}
a_{\sigma^m}=\frac{1}{(2m)!}\sum_{r=0}^{m}\binom{m}{r}\left(-(\beta k)^2\right)^{m-r}\frac{(-1)^{r+1}(2r+1)!}{4\pi}b_{2r,k}.
\end{equation}
This explicit inversion is special to the large-$N$ O($N$) model because only spin zero contributes. In a general theory the finite-$k$ expansion provides a triangular linear system mixing the different spins at fixed dimension, and one needs the full set of powers in $k^2$ to reconstruct the spin-resolved OPE data.

Similarly, \eqref{ONGRTk} can be expressed as a QNM sum. In particular, the sine is a sum of two exponentials, corresponding to QNM frequencies and residues
\bea
\omega_\pm = \pm \sqrt{k^2+m_{\text{th}}^2/\beta^2}, \qquad r_\pm = \mp \frac{2\pi i}{\sqrt{k^2+m_{\text{th}}^2/\beta^2}}.
\eea
Hence the entire QNM spectrum consists of two normal modes, in this case. One can verify that all the relevant QNM sum rules \eqref{sumrule} are satisfied,
\be
\sum_{\pm} r_\pm (i\omega_\pm)^{2q} = 0, \qquad q \in \mathbb{Z}_{\geq 0},
\ee 
in accordance with the absence of $t^{2q}$ terms in the OPE \eqref{ONbn}.
Since there are only finitely many QNMs the QNM sum is also convergent everywhere in $t$.

The connection in Mellin space is also simple. The Mellin space correlator, as obtained through the Mellin-QNM relation \eqref{MellinQNM}, is
\be
\mathcal{M}(s, \vec{k}) = -4\pi \omega_+^{-1-s}\Gamma(s) \sin\left(\frac{\pi s}{2}\right).
\ee
Here the sine factor cancels the poles of $\Gamma(s)$ at negative even integers, leaving simple poles at
\be
s = -1 -2m, \qquad m \in \mathbb{Z}_{\geq 0}.
\ee
From the Mellin-OPE relation \eqref{MellinOPE} one then recovers the OPE data \eqref{ONOPEdata}.

\subsection{\texorpdfstring{$\epsilon$}{epsilon} expansion}
In this subsection we contrast the large-$N$ meromorphic example with the Wilson-Fisher model in $d=4-\epsilon$. At order $\epsilon$, the momentum-space OPE for the thermal two-point function of $\phi$ receives contributions from the identity and from $\phi^2$. The external dimension and the anomalous dimension of $\phi^2$ are
\begin{equation}
\Delta_\phi=1-\frac{\epsilon}{2}+O(\epsilon^2),
\end{equation}
and
\begin{equation}
\gamma_{\phi^2}=\epsilon\frac{N+2}{N+8}+O(\epsilon^2).
\end{equation}
The Euclidean momentum-space correlator takes the form
\begin{equation}
G_E(\omega_n,\vec{k})=\mathcal N_1(\epsilon)(k^2+\omega_n^2)^{-1+\frac{\epsilon}{2}}-\frac{8\pi^4}{3}\frac{N+2}{N+8}\epsilon\frac{1}{(k^2+\omega_n^2)^2}+O(\epsilon^2),
\end{equation}
where
\begin{equation}
\mathcal N_1(\epsilon)=\frac{4^{1-\epsilon/2}\pi^{2-\epsilon/2}}{\Gamma(1-\epsilon/2)^2}.
\end{equation}
The retarded correlator is obtained by the standard Matsubara continuation
\begin{equation}
\omega_n\rightarrow -i(\omega+i0^+).
\end{equation}
Hence,
\begin{equation}
G_R(\omega,\vec{k})=\mathcal N_1(\epsilon)\left(k^2-(\omega+i0^+)^2\right)^{-1+\frac{\epsilon}{2}}-\frac{8\pi^4}{3}\frac{N+2}{N+8}\epsilon\frac{1}{\left(k^2-(\omega+i0^+)^2\right)^2}+O(\epsilon^2).
\end{equation}
For any nonzero $\epsilon$, the exponent $-1+\epsilon/2$ is not an integer. Hence the points
\begin{equation}
\omega=\pm k
\end{equation}
are branch points, not isolated poles. Expanding to first order in $\epsilon$ makes this explicit:
\begin{equation}
\left[k^2-(\omega+i0^+)^2\right]^{-1+\frac{\epsilon}{2}}=\frac{1}{k^2-(\omega+i0^+)^2}\left[1+\frac{\epsilon}{2}\log\left(k^2-(\omega+i0^+)^2\right)\right]+O(\epsilon^2).
\end{equation}
The logarithm produces a branch cut starting at $\omega=\pm k$. The rational terms in $G_R(\omega,k)$ by themselves would have poles at $\omega=\pm k$, but these points are already branch points of the full correlator. Therefore the correlator is not meromorphic at order $\epsilon$, and the discrete QNM description is lost. In the strict $\epsilon=0$ limit one recovers
\begin{equation}
G_R^{(0)}(\omega,\vec{k})=\frac{4\pi^2}{k^2-(\omega+i0^+)^2},
\end{equation}
with two simple poles at $\omega=\pm k$. This example shows that the general relation is broader than an OPE to QNM relation.

\section{Bulk Mellin space}\label{app:bulkmellin}
In this section, we consider the Mellin transform at the level of an AdS Schwarzschild  wave equation. 

Consider a wave equation with redshift factor $f(r)$. Let us consider $\Phi(t,r,\phi)=e^{ik\phi}\psi(t,r)$ and Mellin transform in $t$ as
\be
\psi(s,r) = \int_0^\infty dt\,\, t^{s-1}\psi(t,r). 
\ee 
In particular, time derivatives are turned into shift operators 
\be 
\mathcal{M}_t(\partial_t\psi)(s)= -(s-1)\psi(s-1,r),\qquad \mathcal{M}_t(\partial^2_t\psi)(s)= (s-1)(s-2)\psi(s-2,r).
\ee 
The wave equation for the metrics we consider takes the form 
\be 
(\mathcal{L}_r+\frac{(s-1)(s-2)}{f(r)}\mathcal{P}_{-2})\psi(s,r) = 0,
\ee 
where $\mathcal{L}_r$ is a second-order differential operator in $r$ and $\mathcal{P}_{i} f(s)=f(s+i)$ is a shift operator. Now note that $(s-1)(s-2)..(s-k)P_{-k}\Gamma(s)=\Gamma(s)$ and thus diagonalise the action of the shift operator $\Gamma(s)$. We can therefore make the ansatz 
\be
\psi(s,r)=\Gamma(s)(i\omega)^{-s} R_{\omega,k}(r)
\ee
from which we get 
\be 
(\mathcal{L}_r-\frac{\omega^2}{f(r)})R_{\omega,k}(r) = 0,
\ee 
which is the expected radial ODE. The Mellin space more general solution is then given by  
\be 
\psi(s,r) = \Gamma(s)\int_{\mathcal{C}} \frac{d\omega }{2\pi}(i\omega)^{-s}R_{\omega,k}(r),
\ee 
where $R_{\omega,k}(r)$ are the usual solutions to the radial ODE. In particular, imposing infalling boundary conditions and normalizability at the boundary the frequencies become the quantised QNMs and reduces to a discrete sum 
\be\label{eq:psiMellin} 
\psi_k(s,r) = \Gamma(s) \sum_{n,\pm}c_{n}^\pm(i\omega_n^\pm)^{-s}R_n^{\pm}(r).
\ee 
Consider the bulk-boundary propagator $K_k(s,r)$ of the form \eqref{eq:psiMellin} with a unit-normalized delta function at the boundary. We can then apply the extrapolate dictionary $G_{R}(s,k)=\mathcal{N}_\Delta \lim_{r\to\infty} r^{\Delta}K_k(s,r)$ to obtain the expected form
\be 
G_R(s,k) = \Gamma(s)\sum_n r_n^\pm (k)(i\omega_n^\pm)^{-s},
\ee 
where the residues $r_n^\pm$ are easily related to the other quantities.

\section{Further details on stress tensor correlators}\label{app:StressTensors}
Here we include some further details in the computation for stress tensor correlators. We will use the results in \cite{Karlsson:2022osn}, take the discontinuity and perform the Fourier transform. Let us begin with a discussion about the symmetry structure for the mixed time-spatial correlator. The time direction is singled out $u^\mu = (1,\vec{0})$ and with a fixed spatial momentum $\vec{k}$ the remaining little group is $\text{SO}(2)$. We mainly consider $\vec{k}\,||\hat{z}$ and classify the correlators according to this. 

Let us consider a scalar operator with integer dimension $\Delta = n$ whose discontinuity is given by $r^2=\vec{x}^2$
\be
\text{disc}\,\frac{1}{(r^2-t^2)^{n}}
= -\frac{2\pi i}{(n-1)!}\,\delta^{(n-1)}\!\big(t^2-r^2\big),
\ee
which is delta-function supported on the lightcone. The spatial Fourier transform can then be performed by first integrating over angular coordinates, then the remaining radial integral localises at the endpoint $t=r$ due to the delta function support. Using $\int d\Omega e^{-i\vec{k}\cdot\vec{x}} = 4\pi j_0(kr)=4\pi\sin(kr)/(kr)$ we get 
\bea 
\mathcal{I}_n(t,k)=\int d^3 x e^{-i\vec{k}\cdot \vec{x}}\,\text{disc}\,\frac{1}{(r^2-t^2)^{n}} &=& \frac{-8\pi^2 i}{k(n-1)!}\int_0^t dr r\sin (kr)\delta^{(n-1)}(t^2-r^2) \cr
&=& -\frac{4\pi^2 i}{k(n-1)!}((2t)^{-1}\partial_t)^{n-1}\sin(kt),\label{eq:free}
\eea 
which is just a sum of $\sin kt$ and $\cos kt$ weighted by a polynomial in $k$. In particular, we see that the larger the $n$, the more powers of $k$. This will be the basic integral which we will need below for the stress tensor correlator. In particular, the stress tensor correlators can be written as a sum of terms $\propto (x^2)^{-n}$. Spatial factors $x_i$ in the numerator can be converted to $-i\partial_{k_i}$ on \eqref{eq:free}; even powers of $t$ can either be carried as external factors or eliminated with $t^2=-x^2+r^2$ (in Lorentzian signature); a single (odd) factor of $t$ can be converted to derivatives w.r.t.\ to $t$ acting on $I_{n-1}$.

From
$\langle T_{\mu\nu}T_{\rho\sigma}\rangle_{\rm vac}=\tfrac{C_T}{x^{2d}}\big[\tfrac12(I_{\mu\rho}I_{\nu\sigma}
+I_{\mu\sigma}I_{\nu\rho})-\tfrac1d\eta_{\mu\nu}\eta_{\rho\sigma}\big]$ one finds the single structure
\begin{equation}
\langle T_{xy}(x)T_{xy}(0)\rangle_{\rm vac}
=\frac{C_T}{2\,(x^2)^{6}}\Big[(t^2-z^2)^2-(x^2_{\perp\!,1}-x^2_{\perp\!,2})^2+4x^2_{\perp\!,1}x^2_{\perp\!,2}\Big],
\label{eq:DTxyid}
\end{equation}
where $(x_{\perp,1},x_{\perp,2},z)=(x,y,z)$. Its discontinuity and transform gives
\begin{equation}
G^R_{xy,xy}\big|_{\text{vac}}
=\theta(t)\frac{C_T\pi^2}{40\,t^{5}}\Big[\big((kt)^2-3\big)\cos(kt)-3kt\,\sin(kt)\Big].
\end{equation}

Next we consider the stress tensor exchange
\begin{equation}
\langle T_{xy}(x)T_{xy}(0)\rangle_{\beta}\big|_{T}
=\frac{b_{T}}{(x^2)^{5}}\,\mathcal P^{T}_{xy,xy}(x;\hat a,\hat b,\hat c),
\label{eq:DTxyT}
\end{equation}
with $\mathcal P^{T}_{xy,xy}$ the polynomial in Eq.~(C.24) \cite{Karlsson:2022osn} and $b_{T}$
the thermal one-point coefficient, $\langle T_{\mu\nu}\rangle_\beta\sim\beta^{-d}b_{T}$. Taking
the discontinuity and transforming, with the overall thermal normalization one finds
\be
G^R_{xy,xy}\big|_{T}=-\theta(t)\frac{4(-7\hat a-2\hat b+\hat c)}{9C_T}\frac{b_T}{\beta^{4}}\,k\,\sin(kt)
\label{eq:DGxyT}
\ee
proportional to the scalar-channel conformal-collider combination
$\mathcal{C}^{(1)}=\tfrac{b_T}{C_T}\hat{\cal{C}}^{(1)}$.

Now consider the shear channel. The identity contribution is
\begin{equation}
\langle T_{tx}T_{tx}\rangle_{\rm vac}
=\frac{C_T}{2\,(x^2)^{6}}\Big[-(x^2)^2+2(x_{\perp\!,1}^2-t^2)\,x^2+8t^2 x_{\perp\!,1}^2\Big],
\end{equation}
which transforms to 
\begin{equation}
G^R_{tx,tx}\big|_{\text{vac}}
=\theta(t)\frac{C_T\pi^2 k^2}{160\,t^{3}}\big(kt\,\sin(kt)+\cos(kt)\big)
\end{equation}
The stress-tensor exchange gives
\begin{equation}
G^R_{tx,tx}\big|_{T}
=\theta(t)\frac{b_T\,k}{3\,C_T\,\beta^{4}}\Big[\tfrac13(16\hat a+5\hat b-4\hat c)\,kt\cos(kt)
+\tfrac12(2\hat b+\hat c)\,\sin(kt)\Big],
\end{equation}
whose leading large-$k$ term is $\propto(16\hat a+5\hat b-4\hat c)\propto \mathcal{C}^{(2)}$,
the shear-channel collider combination.

We now consider the sound channel. The identity contribution to the energy-density and longitudinal correlators are
\begin{align}
G^R_{tt,tt}\big|_{\text{vac}}&=-\theta(t)\frac{C_T\pi^2 k^4}{240\,t}\cos(kt)\cr
G^R_{tz,tz}\big|_{\text{vac}}&=\theta(t)\frac{C_T\pi^2 k^2}{240\,t^{3}}\Big[(2-(kt)^2)\cos(kt)+2kt\,\sin(kt)\Big],
\end{align}
related by conservation. The stress-tensor exchange in the longitudinal (sound) polarization is
\begin{equation}
G^R_{tz,tz}\big|_{T}
=\frac{\theta(t)b_T\,k}{3\,C_T\,\beta^{4}}\Big[-\tfrac13(4\hat a+2\hat b-\hat c)\,(kt)^2\sin(kt)
+2(4\hat a+\hat b-\hat c)\,kt\cos(kt)-2(2\hat a-\hat b-\hat c)\sin(kt)\Big],
\end{equation}
with leading term $\propto -(4\hat a+2\hat b-\hat c)\propto \mathcal{C}^{(3)}$, the
sound-channel collider combination.

\bibliographystyle{ytphys}
\bibliography{refs}

\end{document}